\documentclass[12pt]{article}
\usepackage[text={16cm,23cm},centering]{geometry}
\usepackage[T1]{fontenc}
\usepackage{lmodern,textcomp}
\usepackage{graphicx}
\usepackage{amsmath,amssymb,amsfonts}
\usepackage{array,multirow,booktabs,rotating}
\newcolumntype{L}[1]{>{\raggedright\let\newline\\\arraybackslash\hspace{0pt}}m{#1}}
\newcolumntype{C}[1]{>{\centering\let\newline\\\arraybackslash\hspace{0pt}}m{#1}}
\newcolumntype{R}[1]{>{\raggedleft\let\newline\\\arraybackslash\hspace{0pt}}m{#1}}
\usepackage[font=small,labelfont=bf,tableposition=top]{caption}
\usepackage[numbers]{natbib}
\usepackage{fancyvrb}
\usepackage{longtable}
\usepackage{tikz}
\usetikzlibrary{positioning,calc,arrows}
\tikzset{
  contract/.style={draw,circle,minimum size=2.2em,black!80,thick,fill=blue!20,inner sep=0mm},
  user/.style={draw,minimum size=1.8em,fill=green!20},
  create/.style={-stealth',red,thick},
  call/.style={-stealth'},
  delegatecall/.style={-stealth',densely dashed},
  action/.style={font=\footnotesize,sloped}
}
\newcommand\action[1]{\textcircled{\tiny#1}}

\newcommand\fnurl[1]{\footnote{\url{#1}}}
\newcommand\ES[1]{\href{https://etherscan.io/address/0x#1\#code}{\texttt{ES:0x#1}}}
\newcommand\A[1]{\href{https://etherscan.io/address/0x#1}{\texttt{0x#1}}}

\usepackage{pifont}
\newcommand\YES{\ding{51}}
\newcommand\no{\textcolor{lightgray}{\ding{55}}}

\newcommand\SW[1]{\begin{turn}{-90}#1\end{turn}}

\newcommand\op[1]{{\textsc{#1}}}
\newcommand\PUSH{\op{push}}
\newcommand\ethclient{\texttt{openethereum}}
\newcommand\Dapp{dApp}
\newcommand\Dapps{\Dapp s}

\newcommand\ITEM[1]{\par\smallskip\noindent\textit{#1}}
\newcommand\ETH{Ether}
\newenvironment{Wallet}[1]{\subsection*{#1}}{}
\newenvironment{Author}{\ITEM{Author:}}{}
\newenvironment{Source}{\ITEM{Source:}}{}
\newenvironment{Description}{}{}
\newenvironment{Identification}{\ITEM{Identification:}}{}
\newenvironment{Addresses}{\ITEM{Addresses:}}{}

\newcommand\HD[1]{\texttt{#1}}
\usepackage{solidity}
\usepackage[breaklinks=true]{hyperref}
\usepackage{breakurl}
\begin{document}

\title{Wallet Contracts on Ethereum\\Identification, Types, Usage, and Profiles
\thanks{An earlier version of this work is published in \url{https://ieeexplore.ieee.org/document/9223287}}}
\author{Monika di Angelo \and Gernot Salzer}
\date{%
   TU Wien, Vienna, Austria\\%
   \{monika.di.angelo,gernot.salzer\}@tuwien.ac.at\\[2ex]%
   April 4, 2021
 }
 \maketitle

\begin{abstract}
In the area of blockchains, a wallet is anything that manages the access to cryptocurrencies and tokens.
Off-chain wallets appear in different forms, from paper wallets to hardware wallets to dedicated wallet apps, while on-chain wallets are realized as smart contracts.
Wallet contracts are supposed to increase trust and security by being transparent and by offering features like daily limits, approvals, multiple signatures, and recovery mechanisms.
The most prominent platform for smart contracts in general and the token ecosystem im particular, and thus also for wallet contracts is Ethereum.

Our work aims at a better understanding of wallet contracts on Ethereum, since they are one of the most frequently deployed smart contracts.
By analyzing source code, bytecode, and execution traces, we derive usage scenarios and patterns.
We discuss methods for identifying wallet contracts in a semi-automatic manner by looking at the deployed bytecodes and the on-chain interaction patterns.
We extract blueprints for wallets and compile a ground truth.
Furthermore, we differentiate characteristics of wallets in use, and group them into six types.
We provide numbers and temporal perspectives regarding the creation and use of wallets.
For the 40 identified blueprints, we compile detailed profiles.
We analyze the data of the Ethereum main chain up to block 11\,500\,000, mined on December 22, 2020.
\end{abstract}

\newpage
\tableofcontents
\newpage
\section{Introduction}
Wallets keep valuables, credentials, and items for access rights (like cash, licenses, credit cards, key cards) in one place, for ease of access and use.
On the blockchain, cryptocurrencies play a role similar to cash, while cryptographic tokens are a universal tool for handling rights and assets.
Software wallets manage the cryptographic keys required for authorization and implement the protocols for interacting with blockchains in general and smart contracts (on-chain programs) in particular.

On-chain wallets are smart contracts that hold cryptocurrencies and access to tokens and that may offer advanced methods for manipulating the assets.
Simply by introducing the role of an `owner' it becomes possible to transfer all assets of an on-chain wallet transparently and securely in a single transaction.
More refined methods include multi-signature wallets, which grant access only if sufficiently many owners sign.

Regarding the number of transactions and public availability of data, Ethereum is the major platform for smart contracts, and thus also for tokens and on-chain wallets.
This paper investigates the usage and purpose of on-chain wallets on the main chain of Ethereum qualitatively as well as quantitatively.
In particular, we address the following questions.
\begin{enumerate}
\item How can deployed wallets be identified from transaction data?
\item Regarding functionality, which types of wallets are deployed?
\item When and in which quantities are wallets created, and how many are actually used?
\item What is the role of wallets in the overall Ethereum smart contract landscape?
\end{enumerate}
Methodologically, we start from the source code of wallets and determine characteristic functions.
Then we search the deployed bytecode for variants of the wallets with the same profile.
Some wallets can also be detected by their creation history or by the way they interact with other contracts.
We group the wallets according to their functionality and collect creation and usage statistics from the blockchain data.
Finally, we relate wallets to other frequently occurring contract types.

Regarding their number, on-chain wallets form a substantial part of the smart contracts on the chain.
This work thus contributes to a better understanding of what smart contracts on Ethereum are actually used for.
Moreover, the collection of wallet features and blueprints may serve as a resource when designing further decentralized trading apps.
Our methods for detecting wallets and analyzing their activities may help in assessing the liveliness of on-chain projects.
E.g., a temporal view on the use of wallets is more informative than just the number of wallets initially deployed.

\subsubsection*{Roadmap}
Section~\ref{data} clarifies terms and presents our methods for bytecode analysis.
Section~\ref{id} discusses methods for the identification of potential on-chain wallets.
Section~\ref{types} describes characteristic features of wallets and categorizes them into types.
Section~\ref{analysis} analyzes interactions of wallets.
Section~\ref{sec:comparison} compares our approach to related work.
Finally, section~\ref{conclusions} concludes with a summary of our results.

\section{Terms, Bytecode Analysis, and Tools}\label{data}
We assume the reader to be familiar with blockchain essentials.
For Ethereum specifics, we refer to~\cite{Ethereumwiki,Buterin2017,Wood2018}.

\subsection{Terms and Data}\label{sec:terms}
Ethereum distinguishes between externally owned accounts, often called \emph{users}, and contract accounts or simply \emph{contracts}.
Accounts are uniquely identified by addresses of 20 bytes.
Users can issue \emph{transactions} (signed data packages) that transfer value to users and contracts, or that call or create contracts.
These transactions are recorded on the blockchain.
Contracts need to be triggered to become active, either by a transaction from a user or by a call (a \emph{message}) from another contract. 
Messages are not recorded on the blockchain, since they are deterministic consequences of the initial transaction.
They only exist in the execution environment of the Ethereum Virtual Machine (EVM) and are reflected in the execution trace and potential state changes.
We use `message' as a collective term for any (external) transaction or (internal) message.

Unless stated otherwise, statistics refer to the Ethereum main chain up to block 11\,500\,000 (mined on Dec 22, 2020).
We abbreviate factors of 1\,000 and 1\,000\,000 by the letters k and M, respectively.

To a large extent, our analysis is based on the EVM bytecode of deployed contracts.
If available, we use verified source code from \texttt{etherscan.io}.
However, relying solely on such contracts would bias the results: in contrast to 36.7\,M successful create operations, there are verified source codes for 125\,k addresses (0.34\,\%) only.

\subsection{Code Skeletons}\label{sec:skeletons}
To detect functional similarities between contracts we compare their \emph{skeletons}.
They are obtained from the bytecodes of contracts by replacing meta-data, constructor arguments, and the arguments of $\PUSH$ operations uniformly by zeros and by stripping trailing zeros.
The rationale is to remove variability that has little impact on the functional behavior, like the swarm hash added by the Solidity compiler or hard-coded addresses of companion contracts.

Skeletons allow us to transfer knowledge about a contract to others with the same skeleton.
Note that the 36.7\,M contract deployments correspond to 364\,k distinct bytecodes and just 170\,k distinct skeletons.
This is still a large but manageable number of bytecodes to consider when aiming for full coverage.

In addition, we are drastically increasing the proportion of contracts for which we can associate a source code.
This in turn helps us understand the semantics of a large part of the deployments.
In table~\ref{tab:source}, we list the number of deployments (in thousands) that can be associated with source code.
The first line indicates the number of addresses with verified source code on \texttt{etherscan.io} as well as the percentage iin relation to all deployed contracts.
When we relate contract addresses via the source code of factories, we arrive at 17.5\,M deployments (47.6\,\%) that can be associated with the verified source codes.
Considering contract addresses that have identical bytecode, we can associate 17.8\,M (48.4\,\%) deployments with the verified source codes.
Considering contract addresses that have identical skeletons, we can associate 20.8\,M (56.6\,\%) deployments with the verified source codes.

\begin{table}[!htb]
\caption{Share of Contracts with associated source code [in thousands]}\label{tab:source}
\centering
\begin{tabular}{l|rr}
contracts    & total & percentage  \\
\midrule
with verified source	&       125 &   0.3 \\
related via factories 	& 17\,480 & 47.6 \\
related via bytecodes	& 17\,770 & 48.4 \\
related via skeletons	& 20\,772 & 56.6 \\
\end{tabular}
\end{table}

\subsection{Contract Interfaces}
Most contracts in the Ethereum universe adhere to the ABI standard~\cite{ABI}, which identifies functions by signatures that consist of the first four bytes of the Keccak-256 hash of the function names and the parameter types.
Thus, the bytecode of a contract contains instructions to compare the first four bytes of the call data to the signatures of its functions.
To understand the implemented interface of a contract, we extract it from the bytecode, and then try to restore the corresponding function headers.

\subsubsection{Interface Extraction}
We developed a pattern-based tool to extract the interface contained in the bytecode.
We compiled a ground truth for validation with data up to block height 8.45\,M, for which we used the combination of verified source code, corresponding bytecode, and ABI provided by Etherscan.
The signatures extracted by our tool differed from the ground truth in 42 cases.
Manual inspection revealed that our tool was correct also in these cases, whereas the ABIs did not faithfully reflect the signatures in the bytecode (e.g.\ due to compiler optimization or library code).

Before applying the tool to all deployed bytecodes, a few considerations regarding the validation of the tool are due.
Apart from very few LLL or Vyper contracts, the validation set consists almost exclusively of bytecode generated by the Solidity compiler, covering virtually all its releases (including early versions).
Regarding the large group of 9.6\,M deployed contracts (220\,k codes, 107\,k skeletons) generated by the Solidity compiler, it is thus representative.\footnote{%
Deployed code generated by \texttt{solc} can be identified by the first few instructions.
It starts with one of the sequences \texttt{0x6060604052}, \texttt{0x6080604052}, \texttt{0x60806040818152},
\texttt{0x60806040819052}, or \texttt{0x60806040908152}.
In the case of a library, this sequence is prefixed by a \texttt{PUSH} instruction followed by \texttt{0x50} or \texttt{0x3014}.
}
Another interesting group of deployed contracts consists of 5.2\,M short contracts (18\,k codes, only 271 skeletons) without entry points.
They are mainly contracts for storing gas (gasToken), but also proxies (contracts redirecting calls elsewhere) and contracts involved in attacks.
As a last group of deployed contracts, we are left with remaining 595 codes.
For them, our tool shows an error rate of 8\,\%, estimated from a random sample of 60 codes that we manually checked.

\subsubsection{Interface Restoration}
To understand the purpose of contracts we try to recover the function headers from the signatures.
As the signatures are partial hashes of the headers, we use a dictionary of function headers with their 4-byte signatures (collected from various sources), which allows us to obtain a function header for 61.2\,\% of the 385\,k distinct signatures on the mainchain.%
\footnote{An infinity of possible function headers is mapped to a finite number of signatures, so there is no guarantee that we have recovered the original header.
The probability of collisions is low, however.
For example, of the 471\,k signatures in our dictionary only 33 appear with a second function header.}
Since signatures occur with varying frequencies and bytecodes are deployed in different numbers, this ratio increases to 91\,\% (or 89\,\%) when picking a code (or a deployed contract) at random.

\subsection{Third Party Tools}\label{sec:tools}
We employ the Ethereum client \ethclient\ in archive mode to obtain the execution traces.
\texttt{PostgreSQL} serves as our primary database that stores the messages extracted from the traces as well as information on the contracts.
For analyzing contract interactions as graphs, we use the graph database \texttt{Neo4j}. 
Furthermore, we utilize \texttt{etherscan.io} for information on deployed contracts and \texttt{matplotlib} for plotting.

\section{Identifying Potential Wallets}\label{id}
We define a proper wallet to be a contract whose sole purpose is to manage assets.
In contrast, contracts that serve other purposes as well beyond managing assets are termed non-wallet contracts.
The latter group contains all applications that require \ETH{} or tokens.

Our approach first identifies potential wallet contracts and then checks if the bytecode actually implements a proper wallet.
In this section, we discuss four methods to identify potential wallets, of which we utilize the last three in combination for the first step.
In section~\ref{types} we elaborate on the second step, the check whether they implement a proper wallet. 

\subsection{Wallets as Recipients of \ETH{} or Tokens}\label{ssec:recipients}
In a broad sense, any address that has sent or received \ETH{} or tokens at some point in time may be called a wallet: on-chain wallet in case of a contract address and off-chain wallet otherwise.
Addresses transferring \ETH{} can be easily identified by looking at the senders and receivers of successful messages with an \ETH{} value greater than zero.
Token transfers are harder to detect, as the addresses receiving tokens are not directly involved in the transfer.

{\sloppy
We identify token holders as the addresses that appear in calls of the methods \texttt{transfer(address,uint256)}, \texttt{transferFrom(address,address,uint256)}, \texttt{mint(address,uint256)}, \texttt{balanceOf(address)} or in events of the type \texttt{Transfer(address,address,uint256)}.

}

\begin{table}[!htb]
\caption{Accounts having held \ETH{} or tokens [in millions]}\label{tab:recipients}
\centering
\begin{tabular}{lR{1.4cm}R{1.4cm}R{1.4cm}R{1.4cm}}
& only \ETH{} & only tokens & both & neither \\
\midrule
contracts &   2.3 &  6.8 &   0.5 & 27.1 \\ 
users      & 49.4 & 21.9 & 43.5 & 
\end{tabular}
\end{table}
Table~\ref{tab:recipients} lists the number of accounts that ever held \ETH{} or tokens.
The number of user accounts that never held \ETH{} or tokens is difficult to assess.
According to \texttt{etherscan.io}, the number of addresses in the state was almost 130\,M.
However, as accounts are removed from the state space when no longer in use, this number is smaller than the sum of the numbers in the table above.

\ITEM{Limitations.}
The liberal definition of wallets as senders or receivers of assets gives a first idea of the quantities involved.
As we will see below, many on-chain wallets have not yet been used and thus cannot be detected by the above method.
Inherently, this method yields many non-wallets, while it misses the high amount of unused proper wallets.
Thus, we did not use this method for further analyses.

\subsection{Identifying Wallets by their Interface}
Given the source code of a wallet contract, we can use its bytecode and partially restored ABI to identify \emph{similar} contracts on the chain.
Employing the methods described in section~\ref{data}, we first locate all deployed contracts with identical bytecode or skeleton.

In order to capture variants of the already found wallets, we then look for contracts that implement the same characteristic functions as a given wallet.
This is achieved by allowing for some fuzziness regarding additional signatures.
For this, the choice of signatures is crucial. 
One has to avoid using unspecific signatures for the search or being too liberal with additional signatures.
This can be achieved by checking the functionality of contracts, e.g.\ by reading the bytecode or by looking at the interaction patterns of deployed contracts.
As we will see, the number of wallet blueprints is small enough to actually read its code.

\ITEM{Limitations.}
This approach misses contracts that do not adhere to the ABI specification.
Moreover, contracts with similar signatures may implement different functionality, and thus may not be related.

\subsection{Identifying Wallets by their Factory}
Wallets appearing in large quantities are usually deployed by a small number of contracts (so-called factories) or external addresses.
Factories can be located either by the same methods as wallets, or by specifically looking for addresses that create many other contracts and by verifying that the latter are indeed wallets.
Once the factories are identified, a database query is sufficient to select all wallets created by them.

\ITEM{Limitations.}
When looking for factories, we may encounter the same problems as for the wallet interfaces.
Otherwise, this method is robust as the signatures and the interaction patterns of factories are distinctive.
This approach misses wallets not deployed by factories.

\subsection{Identifying Wallets by their Name}
To detect wallets in a more systematic fashion and to include also less popular ones, we scanned the 128\,k source codes on \texttt{etherscan.io} for contracts containing the string `wallet` or `Wallet' in their name.

\ITEM{Limitations.}
This approach is a heuristic that yields false positives and misses wallets named differently.
Again, reading the bytecode or looking at the interaction patterns of the deployed contracts is indispensable.

\subsection{Combination of Identification Methods}
Since we aim at finding as many wallets as possible, we combine the three aforementioned methods (that complement each other, albeit with overlaps) and add a fuzzing step.
We would like to mention, that -- in retrospect -- most wallets can be identified uniquely by a small set of functions they implement, often just one function.

\subsubsection{Blueprint Fuzzing}
Even when lacking standardized interfaces, contracts can form classes based on their purpose.
For example, wallet contracts manage the access to Ether and tokens, but implement the functionality diversely.
In such situations, we employ a method we call \emph{blueprint fuzzing}, consisting of the following steps.
\begin{enumerate}
\item Identify contracts (blueprints) that belong to the class under consideration.
Samples can be found as source code on public repositories like Github or Etherscan, but also as bytecode that attracts one's attention, e.g.\ by having been frequently deployed or by exhibiting characteristic interaction patterns.
\item Analyze the code to understand its functionality.
\item Identify idiosyncratic function signatures and behavior.
E.g., functions for multi-signature transactions are characteristic of a small group of contracts.
As another example, controlled wallets show characteristic interaction patterns involving so-called controllers and sweepers.
\item Collect all bytecodes sharing the idiosyncratic features.
By concentrating on a selection of features, we capture different versions of the same blueprint.
\item Check whether the collected codes belong indeed to the same class, e.g.\ by checking for further similarities.
This step prevents over-generalization.
\item Repeat the steps for contracts that create contracts (`factories').
Contracts deployed by variants of a factory usually also belong to the same class.
\end{enumerate}

\subsubsection{Establishing a Ground Truth of Wallets}
For several reasons, it is indispensable to have a set of contracts that are classified with certainty (a gound truth): we need it to get started with blueprint fuzzing, to understand new bytecode without known relatives, or to evaluate heuristics and algorithms.
We achieve certainty in classifying a contract by inspecting its code and its deployed instances manually, drawing on all methods mentioned above as well as on disassembling and decompilation.
By analyzing 40 wallet contracts manually, we were able to classify 6\,512 bytecodes (deployed about 7.3 million times) via blueprint fuzzing as being also wallets.
Clearly, manual inspection is laborious, but its benefits are greatly amplified by the other more automated methods.

\ITEM{Limitations.}
The combination of the three methods (interface, factory, name) might still miss some wallets when they do not adhere to the ABI specification, are not deployed by a factory, and have non-descript contract names.
Moreover, the fuzzing step depends on a manual classifiaction of function headers as idiosyncratic.

\section{Classification and Comparison of Wallets}\label{types}
In this section, we discuss wallets regarding the functionalities they provide.
First, we detail our definition of a proper wallet.
Then, we determine six types of proper wallets and compare them.

\subsection{Proper Wallet}
The main functionality of a wallet consists in funding it as well as in submitting, confirming, and executing transactions to transfer \ETH{} and tokens.
Some wallets offer additional features.
To distinguish proper wallet contracts from non-wallet contracts, we define that any \emph{optional} wallet function beyond the transfer of assets must fall into one of these categories: administration and control, security mechanisms, lifecycle functions, or extensions.

Whether an implemented function belongs to these optional categories was decided upon reading the code.
This is feasible due to the heavy code reuse in wallets.

\subsection{Wallet Blueprints}
Applying the identification methods from sections~\ref{data} and \ref{id}, we could find 40 blueprints for wallets (see table~\ref{tab:characteristics1}).
When fuzzing the blueprints and employing the technique of code skeletons, we can collect all instances of these blueprints by automatically checking every bytecode deployed, thus yielding all addresses of proper wallets.
To ensure we did not over-generalize and to exclude false positives, we had to examine 835 skeletons manually.

\subsection{Types of Wallets}\label{groups}
The identified 40 variants of proper wallets (blueprints) differ in functionality.
Based on their features, we assign them to one of the six types that we define below.

\subsubsection{Simple Wallets} provide little extra functionality beyond handling \ETH{} and tokens.

\subsubsection{MultiSig Wallets} require that $m$ out of $n$ owners sign a transaction before it is executed.
Usually the required number of signatures ($m$) is smaller than the total number of owners ($n$), meaning that not all owners have to sign.
In most cases, the set of owners and the number of required signatures can be updated.

\subsubsection{Forwarder Wallets} forward the assets they receive to some main wallet.
They may include owner management.

\subsubsection{Controlled Wallets} can be compared to traditional bank accounts.
They are assigned to customers, who can use them as target of transfers, but the control over the account remains with the bank.
Withdrawals are executed by the bank on behalf of the customer.
This construction allows to comply with legal regulations that may restrict transactions.

\subsubsection{Update Wallets} provide a mechanism to update their main features at the discretion of the owner.

\subsubsection{Smart Wallets} offer enhanced features like authorization mechanism for arbitrary transactions, recovery mechanisms for lost keys, modular extension of features, or advanced token standards.

\begin{table}
\caption{Characteristics of Wallet Contracts}
\label{tab:characteristics1}
\centering\scriptsize
\makebox[0pt]{%
\begin{tabular}{@{}ll|ccc|cccccc|ccc|cc|cc@{}} 
&& \multicolumn{3}{c|}{handles} & \multicolumn{6}{c|}{control} & \multicolumn{3}{c|}{security} & \multicolumn{2}{c|}{life} & \multicolumn{2}{c}{extra} \\
     Type     & Blueprint Name                          &\SW{\ETH}
                                                                          &\SW{ERC-20}
                                                                                 &\SW{advanced tokens}
                                                                                        &\SW{owner administration}
                                                                                               &\SW{cosigner (2-of-2)}
                                                                                                      &\SW{multi signature}
                                                                                                             &\SW{third party control}
                                                                                                                    &\SW{forwarding of assets}
                                                                                                                           &\SW{flexible transaction}
                                                                                                                                  &\SW{daily limit}
                                                                                                                                         &\SW{time lock}
                                                                                                                                                &\SW{recovery mechanism}
                                                                                                                                                       &\SW{safe mode, pause, halt}
                                                                                                                                                              &\SW{destroy}
                                                                                                                                                                     &\SW{update logic}
                                                                                                                                                                            &\SW{module administration}  \\
\midrule
   Simple  
   	   & AutoWallet~\cite{auto}                   & \YES & \YES & \YES & \YES & \no  & \no  & \no  & \no  & \no  & \no  & \no  & \no  & \no  & \no  & \no  & \no  \\
           & BasicWallet~\cite{basic}                 & \YES & \YES & \YES & \YES & \no  & \no  & \no  & \no  & \no  & \no  & \no  & \no  & \no  & \no  & \no  & \no  \\
           & ConsumerWallet~\cite{consumer}           & \YES & \YES & \no  & \YES & \no  & \no  & \no  & \no  & \no  & \YES & \no  & \no  & \no  & \no  & \no  & \no  \\
           & EtherWallet1~\cite{EtherWallet1}         & \YES & \YES & \no  & \YES & \no  & \no  & \no  & \no  & \no  & \no  & \no  & \no  & \no  & \no  & \no  & \no  \\
           & EtherWallet2~\cite{EtherWallet2}         & \YES & \YES & \no  & \YES & \no  & \no  & \no  & \no  & \no  & \no  & \no  & \no  & \no  & \no  & \no  & \no  \\
           & SimpleWallet~\cite{SimpleWallet}       & \YES & \no & \no  & \no & \no  & \no  & \no  & \no  & \no  & \no  & \no  & \no  & \no  & \YES  & \no  & \no  \\
           & SimpleWallet2~\cite{SimpleWallet2}        & \YES & \no  & \no  & (\YES) & \no  & \no  & \no  & \no & \no  & \no  & \no  & \no  & \no  & \no  & \no  & \no  \\
           & SimpleWallet3~\cite{SimpleWallet3}       & \YES & \YES & \no  & \YES & \no  & \no  & \no  & \no  & \no  & \no  & \no  & \no  & \no  & \no  & \no  & \no  \\
           & SmartWallet~\cite{smartWallet}           & \no  & \YES & \no  & \YES & \no  & \no  & \no  & \no  & \no  & \no  & \YES & \no  & \no  & \no  & \no  & \no  \\
           & SpendableWallet~\cite{spendable}         & \no  & \YES &(\YES)& \YES & \no  & \no  & \no  & \no  & \no  & \no  & \no  & \no  & \no  & \no  & \no  & \no  \\
           & TimelockedWallet~\cite{timelocked}       &(\YES)& \YES & \no  & \no  & \no  & \no  & \no  & \no  & \no  & \no  & \YES & \no  & \no  & \no  & \no  & \no  \\
           & Wallet1~\cite{Wallet1}                   & \YES & \YES & \no  & \no  & \no  & \no  & \no  & \no  & \no  & \no  & \no  & \no  & \no  & \no  & \no  & \no  \\
           & Wallet3~\cite{Wallet3}                         & \YES  & \YES & \YES& \YES & \no  & \no  & \no  & \no  & \no  & \no  & \no  & \no  & \no  & \no  & \no  & \no  \\
           & Wallet4~\cite{Wallet4}                     & \YES  & \YES & \no  & \no & \no  & \no  & \no  & \no  & \no  & \no  & \no  & \no  & \no  & \no  & \no  & \no  \\
           & Wallet5~\cite{Wallet5}               & \YES  & \YES & \no  & \no & \no  & \no  & \no  & \no  & \no  & \no  & \no  & \no  & \no  & \no  & \no  & \no  \\
           & Wallet6~\cite{Wallet6}                    & \no  & \YES & \no  & \YES & \no  & \no  & \no  & \no  & \no  & \no  & \no  & \no  & \no  & \no  & \no  & \no  \\
           & Wallet7~\cite{Wallet7}                    & \no  & \YES & \no  & \YES & \no  & \no  & \no  & \no  & \no  & \no  & \no  & \no  & \no  & \no  & \no  & \no  \\
\midrule
MultiSig 
           & Ambi2~\cite{ambi2}                           & \YES  & \YES & \no  & \YES& \YES & \no  & \no  & \no  & \no  & \no  & \no  & \no  & \no  & \no  & \no  & \no \\
           & Argent~\cite{msigA}                      & \YES & \YES & \no  & \YES & \YES & \YES & \no  & \no  & \YES & \no  & \no  & \no  & \no  & \no  & \no  & \no  \\
           & BitGo~\cite{MSigBitGo}                   & \YES & \YES & \no  & \YES & \YES & \YES & \no  & \no  & \no  & \no  & \no  & \no  & \YES & \no  & \no  & \no  \\
           & Gnosis/ConsenSys~\cite{MSigGnosis}       & \YES & \YES & \no  & \YES & \YES & \YES & \no  & \no  & \YES & \YES & \no  & \no  & \no  & \no  & \no  & \no  \\
           & Ivt~\cite{msigI}                         & \YES & \YES & \no  & \no  & \no  & \YES & \no  & \no  & \YES & \no  & \no  & \no  & \YES & \no  & \no  & \no  \\
           & Lundkvist~\cite{msigL}                   & \YES & \YES & \no  & \no  & \YES & \YES & \no  & \no  & \YES & \no  & \no  & \no  & \no  & \no  & \no  & \no  \\
           & NiftyWallet~\cite{msigN}                 & \YES & \YES & \YES & \YES & \YES & \YES & \no  & \no  & \YES & \YES & \no  & \no  & \no  & \no  & \no  & \no  \\
           & Parity/Eth/Wood~\cite{MSigParity}        & \YES & \YES & \no  & \YES & \YES & \YES & \no  & \no  & \YES & \YES & \no  & \no  & \no  & \YES & \no  & \no  \\
           & TeambrellaWallet~\cite{msigT}            & \YES & \no  & \no  & \YES & \YES & \YES & \no  & \no  & \no  & \no  & \no  &(\YES)& \no  & \no  & \no  & \no  \\
           & Unchained Capital~\cite{MSigTrezor}      & \YES & \no  & \no  & \YES & \YES & \YES & \no  & \no  & \YES & \no  & \no  & \no  & \no  & \no  & \no  & \no  \\
\midrule
 Forwarder 
 	   & BitGo~\cite{forwardB}                    & \YES & \YES & \no  & \no  & \no  & \no  & \no  & \YES & \no  & \no  & \no  & \no  & \no  & \no  & \no  & \no  \\
           & IntermediateWallet~\cite{forwardI}       & \YES & \YES &(\YES)& \no  & \no  & \no  & \no  & \YES & \no  & \no  & \no  & \no  & \no  & \no  & \no  & \no  \\
	   & Poloniex2~\cite{Poloniex2}                    & \YES  & \YES & \no  & \YES& \no  & \no  & \no  & \YES & \YES & \no  & \no  & \no  & \no  & \no  & \no  & \no  \\
           & Wallet2~\cite{Wallet2}                       & \YES  & \YES & \no  & \no & \no  & \no  & \no  & \YES & \no  & \no  & \no  & \no  & \no  & \no  & \no  & \no  \\
\midrule
Controlled 
	   & Bittrex~\cite{ControlledWallet}          & \YES & \YES &(\YES)& \YES & \no  & \no  & \YES & \no  & \no  & \no  & \no  & \no  & \no  & \no  & \no  & \no  \\
           & ICTlock~\cite{ictlock}                            & \no   & \YES & \no  & \YES& \no  & \no  & \YES & \no  & \no  & \no  & \YES & \no  & \no  & \no  & \no  & \no  \\
	   & SimpleWallet4~\cite{SimpleWallet4}     & \YES  & \YES & \no  & \no & \no  & \no  & \YES & \no  & \no  & \no  & \no  & \no  & \no  & \no  & \no  & \no  \\
\midrule
Update 
    	  &  Eidoo~\cite{Eidoo}                       & \YES & \YES & \no  & \no  & \no  & \no  & \no  & \no  & \no  & \no  & \no  & \no  & \no  & \no  & \YES & \no  \\
          &  LogicProxyWallet~\cite{logicproxywallet} & \YES & \YES & \no  & \YES & \no  & \no  & \no  & \no  & \YES & \no  & \no  & \no  & \no  & \no  & \YES & \no  \\
          &  LoopringWallet~\cite{Loopringwallet} & \YES & \YES & \no  & \YES & \no  & \no  & \no  & \no  & \YES & \no  & \no  & \no  & \no  & \no  & \YES & \YES \\
\midrule
Smart  
    	  &  Argent~\cite{SmartArgent}                & \YES & \YES & \no  & \YES & \YES & \no  & \no  & \no  & \YES & \no  & \no  & \no  & \no  & \no  & \YES & \YES \\
          &  Dapper~\cite{SmartDapper}               & \YES & \YES & \YES & \YES & \YES & \no  & \no  & \no  & \YES & \no  & \no  & \YES & \no  & \no  & \YES & \YES \\
          &  Gnosis~\cite{SmartGnosis}                & \YES & \YES & \YES & \YES & \YES & \YES & \no  & \no  & \YES & \YES & \no  & \YES & \no  & \no  & \YES & \YES \\
\end{tabular}
}
\end{table}

\subsection{Features of Wallets}\label{functionalities}
Table~\ref{tab:characteristics1} shows for each identified wallet blueprint the type, name, reference to the source or bytecode, as well as an overview of the implemented features as detailed below.

\ITEM{\ETH{}.}
To handle Ether a wallet has to be able to receive and transfer it.
Some wallets are intended for tokens only and thus refuse Ether.
But even then well designed wallets provide a withdraw method, since some Ether transfers (like mining rewards and self-destructs) cannot be blocked.

\ITEM{ERC-20 tokens.}
For ERC-20 token transfers, the holding address initiates the transfer by sending to the respective token contract the address of the new holder who is not informed about this change.
No provisions have to be made to receive such tokens.
However, to s(p)end the tokens a wallet needs to provide a way to call a transfer method of the token contract. 

\ITEM{Advanced tokens} require the recipient to provide particular functions that are called before the actual transfer of the token.

\ITEM{Owner administration} enables the transfer of all assets in a wallet to a new owner in one sweep without revealing private keys.
With an off-chain wallet, one has to transfer each asset separately or share the private key with the new owner.

\ITEM{MultiSig} wallets face a trade-off between flexibility, transaction costs, and transparency.
Each transaction carries only one signature.
To supply more, the wallet either has to be called several times by different owners (incurring multiple transaction fees), or the signatures have to be computed off-chain by a trusted app and supplied as data of a single transaction.
A wallet may offer a few fixed multSig actions that are selected transparently via the entry point, or there may be a single entry point that admits the execution of arbitrary calls.
The latter case requires a trusted app that composes the low-level call data off-chain in a manner transparent for the owners who are supposed to approve it.

\ITEM{Cosigner} is a form of MultiSig where exactly two signatures are required.
Moreover, it can be employed for implementing further functionality via another contract that acts as cosigner.
This may include multiSig or off-chain signing.

\ITEM{Third party control} means that the actual control over the wallet stays with a central authority.

\ITEM{Forwarding} wallets are additional addresses for receiving assets that are transferred to a main wallet.

\ITEM{Flexible transactions} means that the wallet is able to execute arbitrary calls after adequate authorization.

\ITEM{Daily limits and time locking} restrict the access to the assets based on time.
Spending more than a daily limit may e.g.\ require additional authorization.
Time locks are useful if assets should be used only at a later point in time, after a vesting period, like after an ICO, for a smart will, or a trust fund.

\ITEM{Recovery mechanisms} provision against lost or compromised credentials.

\ITEM{Life cycle management} enable wallets to be put into safe mode, paused, or halted.
Early wallets were able to self-destruct, which results in the loss of assets sent thereafter.
When put on hold, a wallet can still reject transfer attempts.

\ITEM{Update logic} enables to switch to a newer version of the wallet logic.
This is implemented by means of proxy wallets, which derive their functionality from library code stored elsewhere, and keeps the wallets small and cheap to deploy.

\ITEM{Module administration} offers the in- or exclusion of modules to customize the wallet to user needs. 
This represents a modular and more fine-grained version of update logic.

\subsection{Code Variety of Wallets}
To discuss code variety in wallets, we first look at the six types of wallets before we discuss the 40 blueprints individually.

\subsubsection{Variety of the Types}
In table~\ref{tab:types}, we indicate for each type of wallet the number of distinct bytecodes, skeletons, and creators.
The total of 7.3\,M wallets correspond to just 6\,512 distinct deployed bytecodes (835 distinct code skeletons)\footnote{It should be noted that the total of skeletons is lower (by 4) than the sum over the rows because 3 skeletons (of proxies) are related to more than one wallet type. Also the creators are lower in total by 29 than the sum of the rows entries.}. 
This homogeneity results from the small number of on- or off-chain factories that generate most of the wallets.
\begin{table}[!h]
\caption{Code Variety of Wallet Types}\label{tab:types}
\centering
\begin{tabular}{l|rrrr}
type & deployments & bytecodes & skeletons & creators \\
\midrule
simple	& 1\,336\,246	& 4\,570	&   78	& 96 \\
multiSig	&     914\,005	& 1\,371	& 602	& 20\,668 \\
forwarder	& 2\,419\,996	&     135	&   69	& 975 \\
controlled	& 2\,553\,996	&     352	&   33	& 119 \\
update	&      16\,911	&      18	&   15	& 14 \\
smart	&      94\,690	&      66	&   42	& 737 \\
\midrule
all wallets	   	  &   7\,335\,844 	&    6\,512 	&        835 	&   22\,584 \\
all contracts & 36\,703\,122	& 363\,883	& 170\,209	& 184\,877 \\
\end{tabular}
\end{table}

Regarding the two types with the highest number of deplyoments in table~\ref{tab:types}, the forwarder (2.4\,M) and controlled wallets (2.6\,M), we notice comparably low numbers of bytecodes (135 and 352) indicating the highest code reuse of all types.
For the third most commonly deployed wallet type, the simple wallets, we find the highest number of bytecodes (4\,570). 
However, they reduce to just 78 skeletons, indicating that the code variety is only on a superficial level, i.e.\ in functionally less relevant details.
As this concerns a type with inherently little functionality, low code variety is to be expected.

The type with the highest code variety in terms of skeletons, are the multiSig wallets.
The relatively high variety is accompanied by the highest number of creators.
Still, it should be borne in mind that the overall code variety for wallets (7.3\,M deployments with 6\,512 bytecodes and 835 skeletons) is far below the average on Ethereum (36.7\,M contracts with 364\,k bytecodes and 170\,k skeletons).

\begin{table}[!hp]
\centering
\small
\caption{Code Variety of Wallet Blueprints}
\label{tab:variety}
\begin{tabular}{lrlrrr}
Type & Deployments & Blueprint Name & Bytecodes & Skeletons & Creators \\
\midrule 
   Simple  
   	   &       9\,250 & AutoWallet~\cite{auto}                   & 1 & 1 & 3 \\
           &       4\,822 & BasicWallet~\cite{basic}                 & 2 & 2 & 2 \\
           &     11\,911 & ConsumerWallet~\cite{consumer}          & 11 & 9 & 9 \\
           &     84\,656 & EtherWallet1~\cite{EtherWallet1}         & 2 & 1 & 3 \\
           &   112\,987 & EtherWallet2~\cite{EtherWallet2}         & 2 & 1 & 3 \\
           &              2 & SimpleWallet~\cite{SimpleWallet}       & 2 & 2 & 2 \\
           &           540 & SimpleWallet2~\cite{SimpleWallet2}        & 13 & 10 & 15 \\
           &      1\,288 & SimpleWallet3~\cite{SimpleWallet3}       & 4 & 2 & 4 \\
           &    46\,832 & SmartWallet~\cite{smartWallet}          & 15 & 5 & 6 \\
           &      6\,430 & SpendableWallet~\cite{spendable}        & 6 & 5 & 6 \\
           &          226 & TimelockedWallet~\cite{timelocked}       & 40 & 16 & 25 \\
           &  229\,861 & Wallet1~\cite{Wallet1}                  & 4\,447 & 5 & 4 \\
           &         645 & Wallet3~\cite{Wallet3}                         & 15 & 13 & 5 \\
           &  306\,613 & Wallet4~\cite{Wallet4}                     & 4 & 3 & 3 \\
           &  292\,285 & Wallet5~\cite{Wallet5}               & 2 & 2 & 2 \\
           &  224\,591 & Wallet6~\cite{Wallet6}                   & 3 & 1 & 3 \\
           &     3\,307 & Wallet7~\cite{Wallet7}                   & 2 & 2 & 2 \\
\midrule
MultiSig 
           &  604\,435 & Ambi2~\cite{ambi2}                          &13 & 6 & 16 \\
           &           16 & Argent~\cite{msigA}                     & 7 & 5 & 5 \\
           &  239\,839 & BitGo~\cite{MSigBitGo}                   & 136 & 95 & 1\,392 \\
           &    13\,370 & Gnosis/ConsenSys~\cite{MSigGnosis}       & 975 & 321 & 1\,280 \\
           &           96 & Ivt~\cite{msigI}                        & 2 & 2 & 8 \\
           &     3\,843 & Lundkvist~\cite{msigL}                   & 53 & 42 & 50 \\
           &         996 & NiftyWallet~\cite{msigN}                 & 3 & 3 & 2 \\
           &   50\,424 & Parity/Eth/Wood~\cite{MSigParity}        & 171 & 119 & 17\,786 \\
           &         852 & TeambrellaWallet~\cite{msigT}            & 7 & 7 & 167 \\
           &         134 & Unchained Capital~\cite{MSigTrezor}      & 4 & 3 & 9 \\
\midrule
 Forwarder 
 	   & 2\,003\,428 & BitGo~\cite{forwardB}                    & 122 & 61 & 966 \\
           &        2\,520 & IntermediateWallet~\cite{forwardI}       & 6 & 4 & 2 \\
	   &     401\,549 & Poloniex2~\cite{Poloniex2}                    & 1 & 1 & 2 \\
           &      12\,499 & Wallet2~\cite{Wallet2}                       & 6 & 3 & 5 \\
\midrule
Controlled 
	   & 2\,553\,686 & Bittrex~\cite{ControlledWallet}          & 73 & 31 & 116 \\
           &            309 & ICTlock~\cite{ictlock}                            & 278 & 1 & 2 \\
	   &                1 & SimpleWallet4~\cite{SimpleWallet4}     & 1 & 1 & 1 \\
\midrule
Update 
    	  &     3\,916 & Eidoo~\cite{Eidoo}                       & 3 & 3 & 2 \\
          &     9\,359 & LogicProxyWallet~\cite{logicproxywallet} & 3 & 3 & 3 \\
          &     3\,636 & LoopringWallet~\cite{Loopringwallet} & 12 & 10 & 8 \\
\midrule
Smart  
    	  &    36\,138 & Argent~\cite{SmartArgent}                & 17 & 13 & 15 \\
          &    46\,474 & Dapper~\cite{SmartDapper}              & 14 & 7 & 10 \\
          &    12\,078 & Gnosis~\cite{SmartGnosis}                & 35 & 22 & 712  \\ 
\end{tabular}
\end{table}

\subsubsection{Variety of the Blueprints}
Next, we look at the blueprints individually.
In table~\ref{tab:variety}, we indicate for each wallet blueprint the number of distinct bytecodes, skeletons, and creators.
We can make the following observations:

Regarding bytecodes, the simple wallet Wallet1 shows the highest number of different bytecodes with 4\,447.
However, the variety is not reflected in the number of corresponding code skeletons, which is only 5.
Thus, the code variety lies in minor details.

For non-superficial variety, the number of skeletons is a better indicator.
We find the highest number of 321 skeletons in the multiSig wallet Gnosis/ConsensSys. 
With 975 bytecodes, it also shows the second highest number of distinct bytecodes.
Other wallets with a high number of skeletons are the mutisig wallets Parity/Eth/Wood (119) and BitGo (95).
Thus, the code variety in the wallet type multiSig can be attributed largely to those three blueprints.

They also show the highest numbers of creators, with the multisisg wallet Parity/Eth/Wood in the lead.
It seems a popular wallet to copy and use, especially since the ratio of creators to deployments is about one third.

Among the wallets with high variety, the multiSig wallet Gnosis/ConsensSys sticks out with an interesting ratio of bytecodes to creators (975:1280).
This looks like a copy and customize phenomenon.

At the other end of the spectrum, we find low variety, which is indicated by a high ratio of deployments to skeletons.
Here, the forwarder wallet Poloniex2 is in the lead with about 400\,000:1, followed by the simple wallet wallet6 with roughly 225\,000:1. 
Both represent mass deployments without any code variety.

\subsubsection{Verified Source Code}
On \texttt{etherscan.io}, we find a total of 125\,251 different verified source codes, 706 of which are wallet source codes (that reduce to 597 bytecodes and 274 skeletons).
Thus, 0.01\,\% of all wallets directly provide verified source code. 
However, also on \texttt{etherscan.io}, identical bytecode is marked as such, thus increasing the number of bytecodes with related source code.
Counting all deployments with bytecode identical to the 706 ones with verified source code, we arrive at 2\,705\,976 or 36.9\,\% wallets.

Taking into account the fact that most wallets are created by factories whose code may be found, this number rises to 42.5\,\%.
By exploiting the similarity of code skeletons, we can relate even 56.5\,\% of the wallets to verified source code.

Finally, it should be noted that over 1.2\,M wallets (16.9\,\%) are just proxies that have extra short bytecode and no verified source code.

\section{Interaction Analysis}\label{analysis}
In this section, we examine the wallet contracts regarding their creation and usage.
All figures in this section depict the number of wallets in bins per 100\,k blocks (mined in about two weeks) as a stack plot on a timeline with the date on the upper horizontal axis and the mined block number on the lower one. 
The colors code different aspects of the wallet contracts.

The 11.5\,M blocks of the Ethereum main chain contain 945\,M transactions.
Adding the messages triggered by the transactions, the data comprises almost 2\,611\,M messages.
In this data, we find 36.7\,M successful create messages, of which 7.3\,M create wallet contracts.
The wallets received 44.9\,M and sent 70.9\,M messages.
Discounting the inter-wallet calls, the wallets were involved in 115.7\,M messages corresponding to 4.4\,\% of all messages.

\subsection{Creation of Wallet Types}
\begin{figure}[!htb]
\centering
\includegraphics[width=.9\columnwidth]{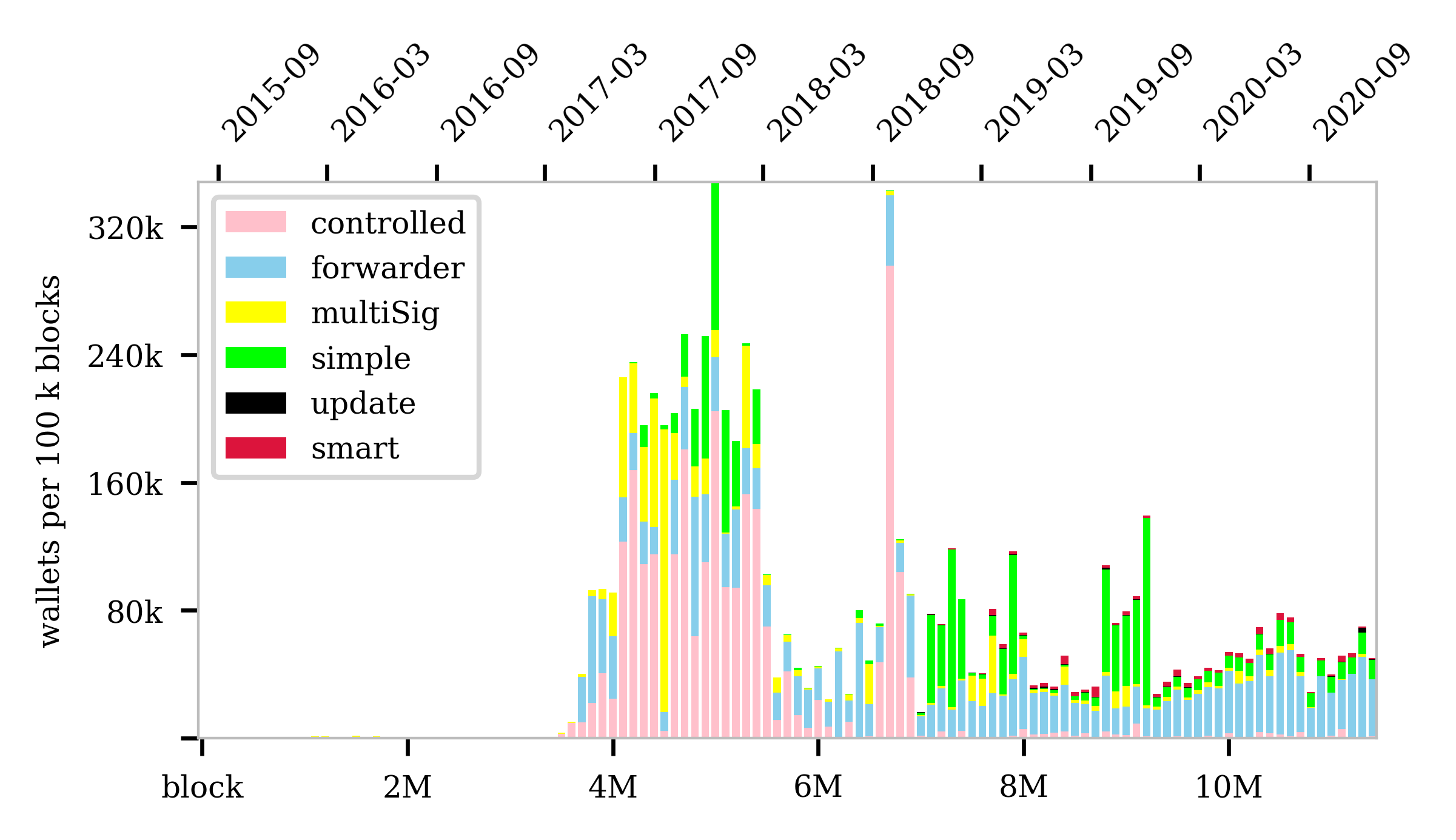}
\includegraphics[width=.86\columnwidth]{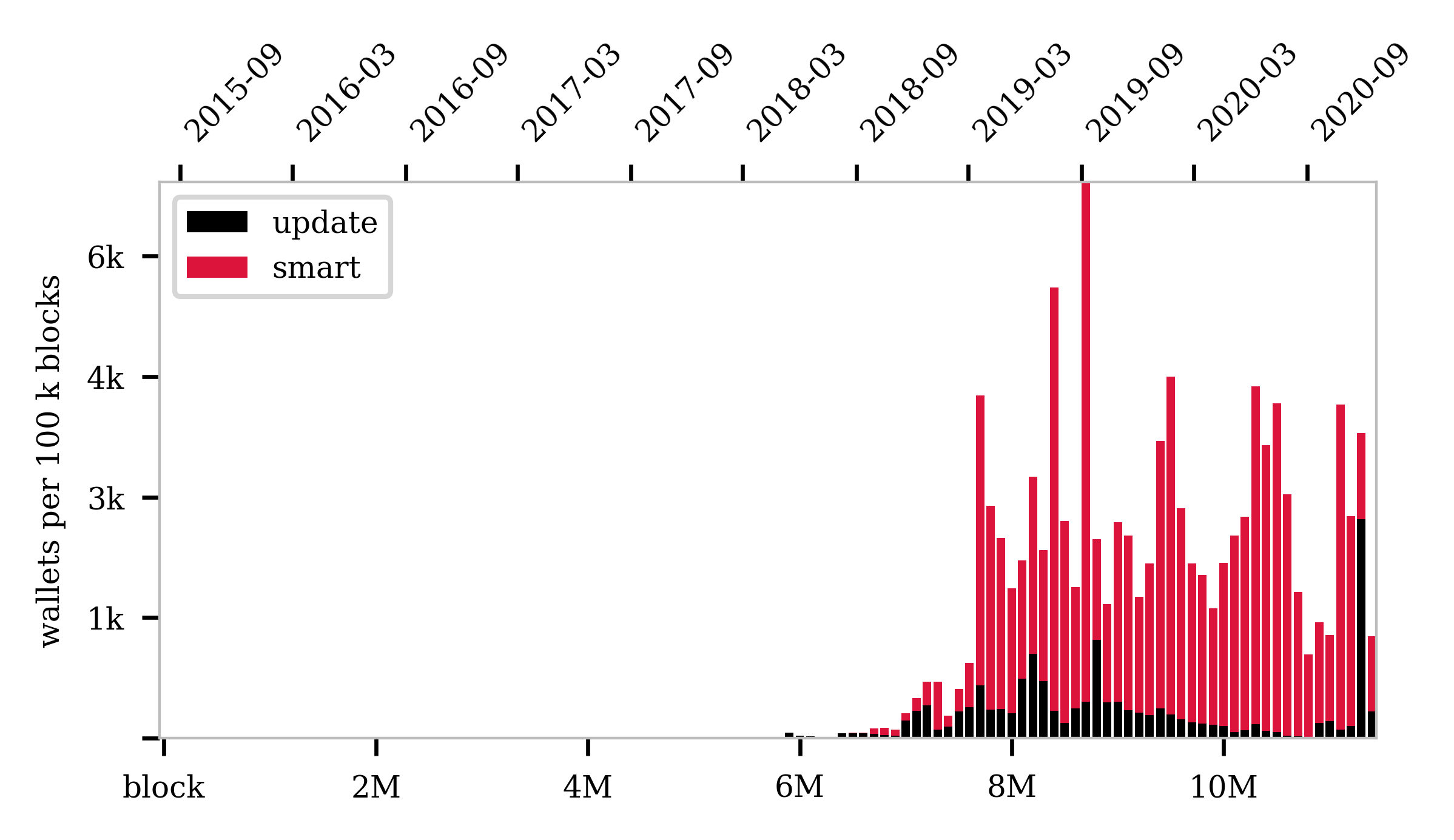}
\caption{Creation of wallets with respect to the six types. The upper plot shows all wallets with colors indicating the type, while the lower one only displays the two rarer types.} \label{fig:alltypes}\label{fig:su}
\end{figure}

In the upper part of figure~\ref{fig:alltypes}, we differentiate the wallets with respect to the six functional types (c.f.\ section~\ref{types}).
The first wallets deployed are the multiSig wallets (yellow) right from the start of Ethereum, albeit in small numbers (not visible) for about two years.
Next to come were the controlled wallets (pink) around block 3.5\,M, followed shortly after by the forwarder wallets (blue).
Simple wallets (green) started to appear in the second half of 2017 after block 4.3\,M.
Update wallets (black, hardly discernible) are as recent as block 5.9\,M (mid 2018), while smart wallets (red) are being deployed since block 6.5\,M (end of 2018).
All of them are still being created.

Since the update and smart wallets are deployed in much smaller numbers, we spotlight them in the lower part of figure~\ref{fig:su}.

\begin{figure}[!htb]
\centering
\includegraphics[width=.9\columnwidth]{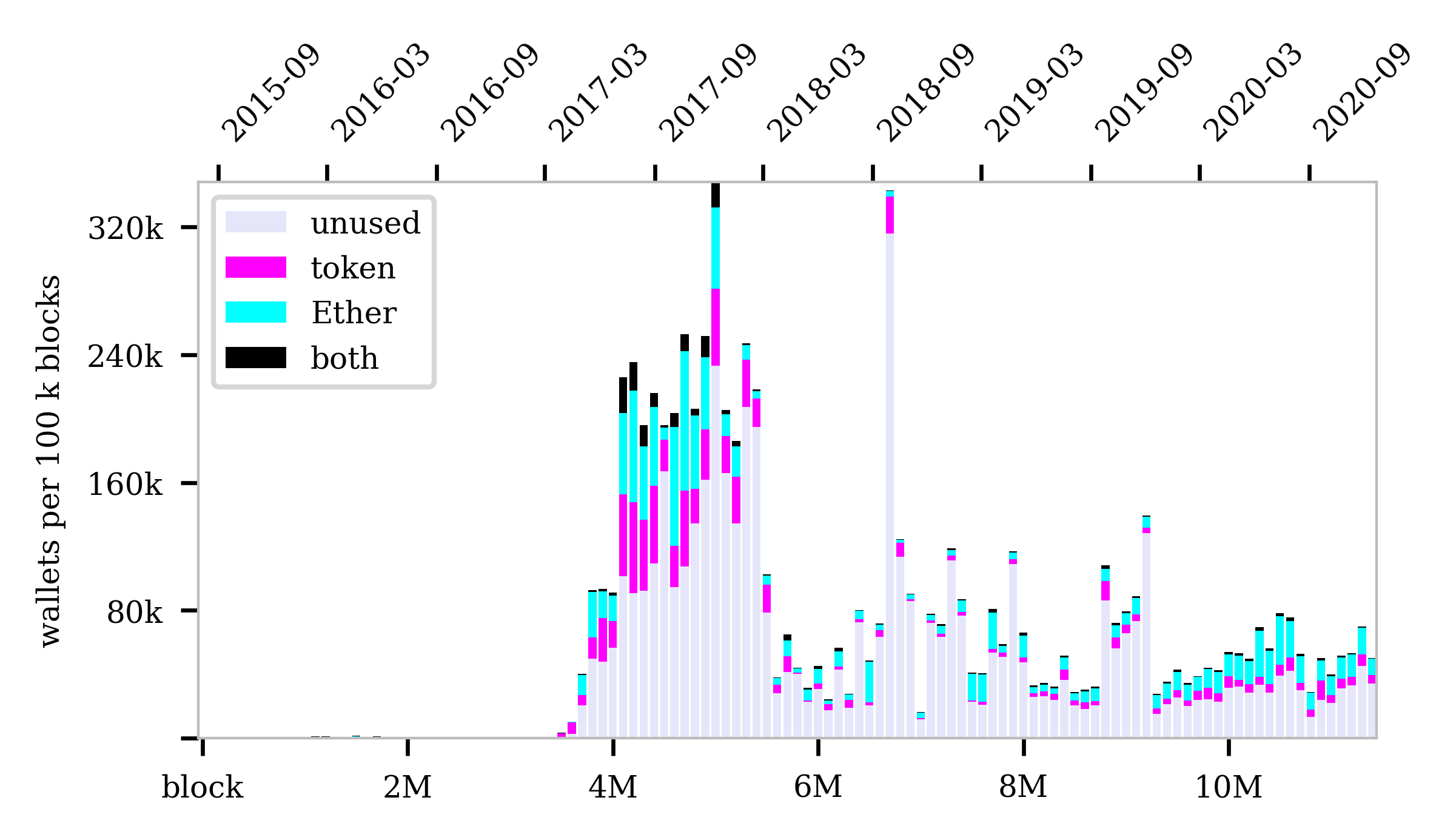}
\caption{All wallets -- creation and usage. The time line shows the creation time of the wallets, while the colors indicate the later usage.} \label{fig:allusage}
\end{figure}
\subsection{Usage of Wallets}
Next, we differentiate the wallets according to their later usage: wallets used for tokens as well as \ETH{}, wallets used for only one of them, and unused wallets.
(See section~\ref{ssec:recipients} for a discussion of how to identify token and \ETH{} holders.)

Figure~\ref{fig:allusage} gives an overview of all wallets on the timeline of creation, while the colors indicate the later usage.
Of the 7.3\,M wallets, 68.4\,\% (5.0\,M, grey) have not been used so far.
The other wallets are either used for tokens (864\,k, magenta) or for \ETH{} (1.249\,M, cyan), but only a few wallets are used for both (203\,k, black).

\begin{figure}[hbt!]
\centering
\includegraphics[width=.9\columnwidth]{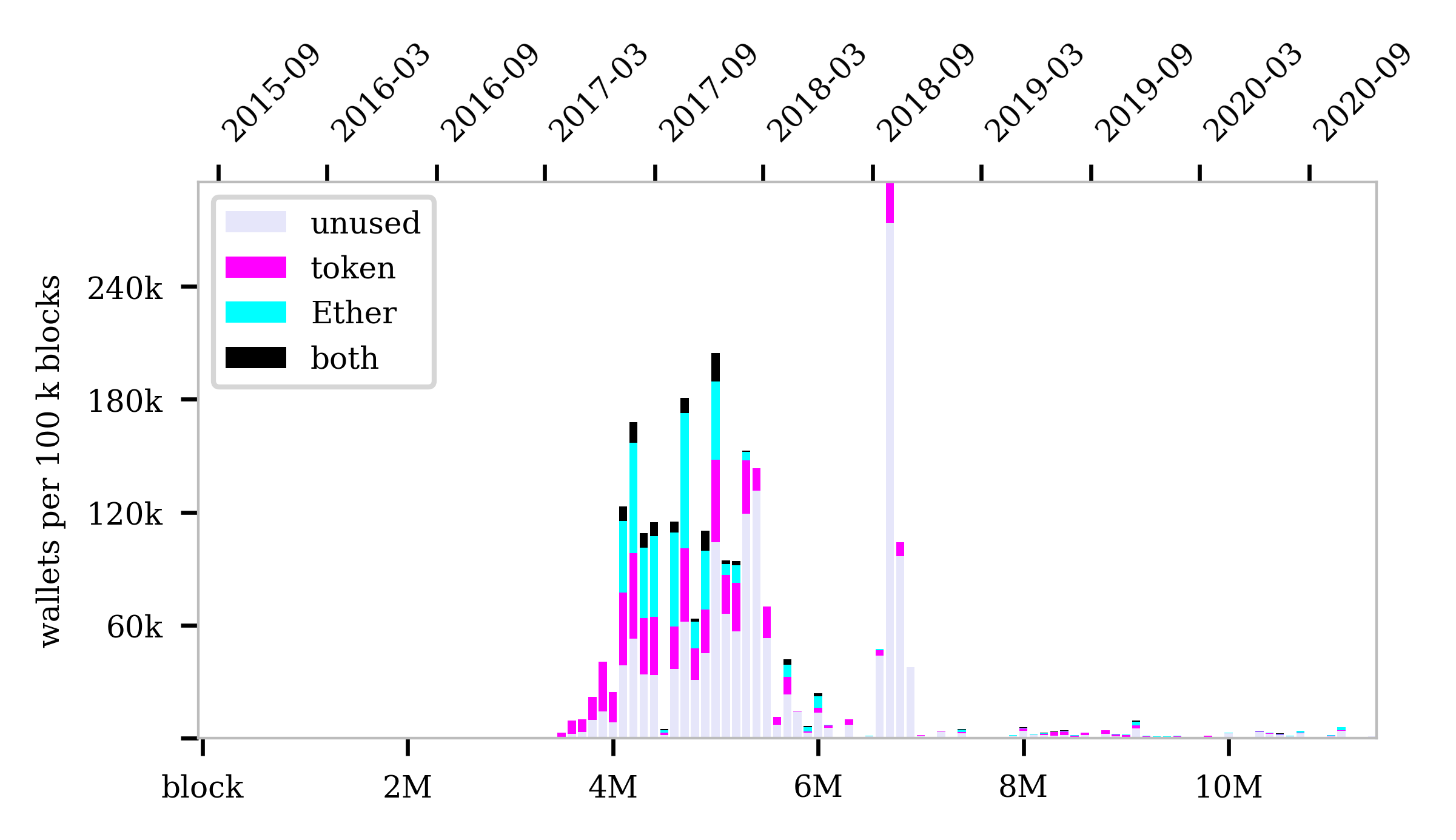}\\
\caption{Controlled wallets -- creation and usage.} \label{fig:controlled}
\end{figure}
\subsubsection{Controlled Wallets}
Figure~\ref{fig:controlled} depicts the usage of the most frequently deployed type of controlled wallets. 
From mid 2017 to early 2018, we notice a significantly higher volume of deployment than later on, while the rate of unused wallets remains more or less at the same high level.
Controlled wallets are used quite equally for tokens or \ETH{}, but rarely for both.
This wallet type is still deployed more often than all others.

\subsubsection{Forwarder Wallets}
\begin{figure}[hbt!]
\centering
\includegraphics[width=.9\columnwidth]{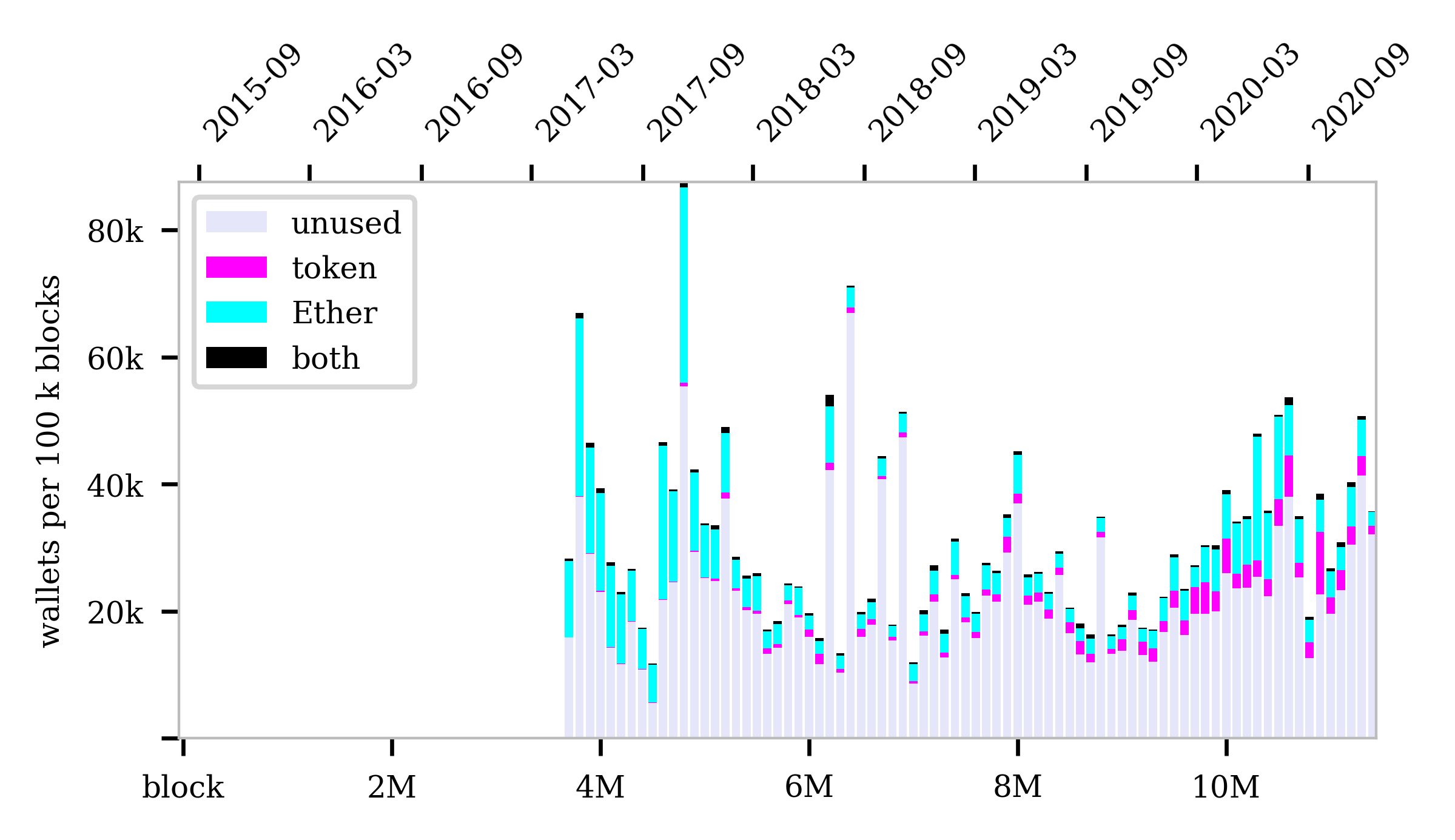}
\caption{Forwarder wallets -- creation and usage. } \label{fig:forwarder}
\end{figure}

Figure~\ref{fig:forwarder} depicts the usage of the second most frequently deployed type of forwarder wallets. 
Since its onset, this type shows a rather steady deployment history as well as a high rate of unused wallets.
While the earlier wallets are rather used for \ETH{}, the later ones are increasingly used for tokens.
Throughout the timeline, forwarder wallets are rarely used for both.

\subsubsection{Simple Wallets}
\begin{figure}[hbt!]
\centering
\includegraphics[width=.9\columnwidth]{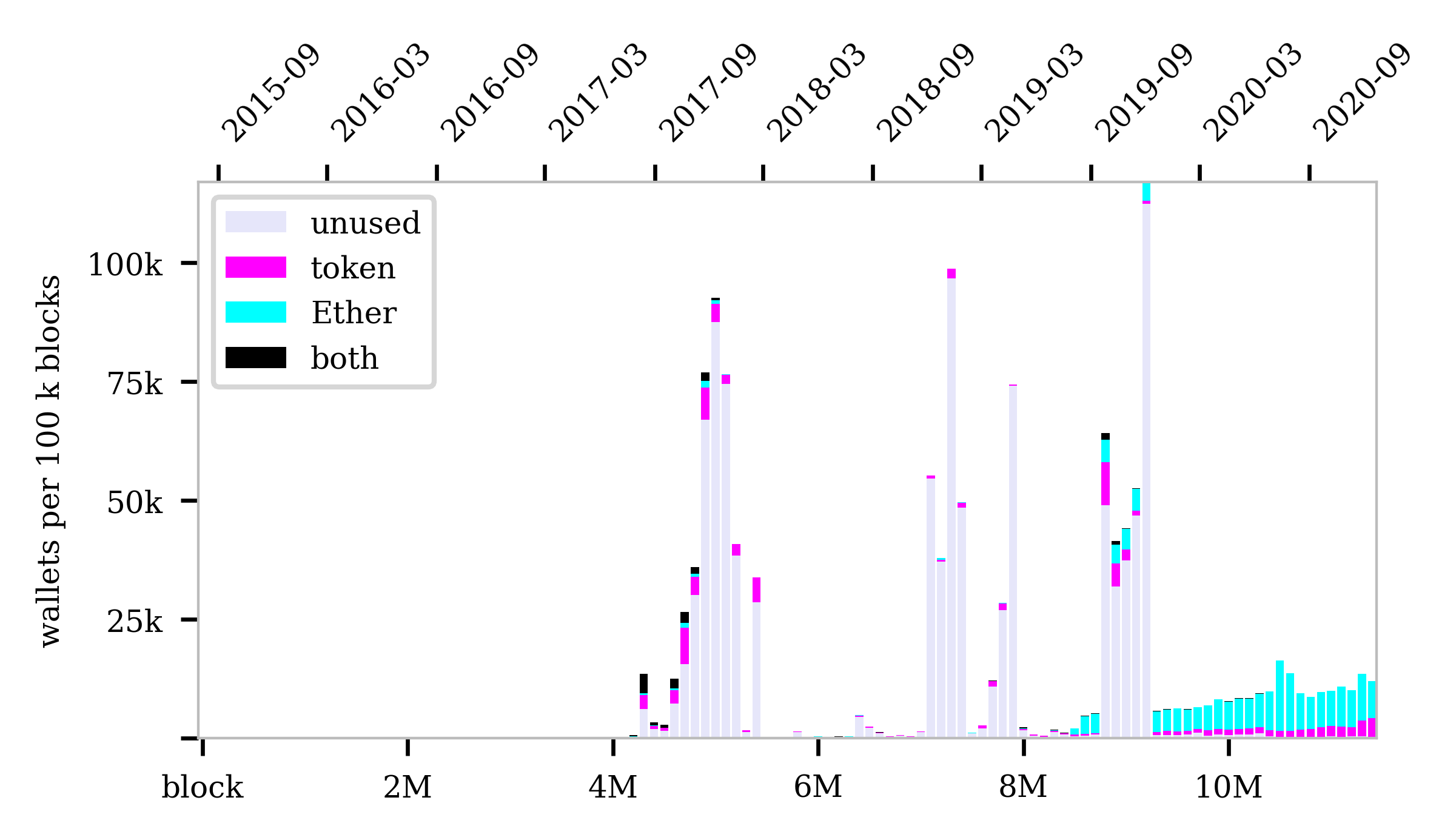}
\caption{Simple wallets -- creation and usage.} \label{fig:simple}
\end{figure}

Figure~\ref{fig:simple} depicts the usage of the third most frequently deployed type of simple wallets. 
This type consists of several diverse blueprints, which seems to be reflected in the plot.
We notice a few peaks in deployment with a low rate of different usage patterns: 
the first peak around the end of 2019 shows a low usage for token and both;
the second peak(s) in early 2019 show an even higher rate of unused wallets and a few ones used for tokens only;
the third peak towards the end of 2019 shows a rate of unused wallet similar to the first peak, while the usage is fairly distributed between token only or \ETH{} only;
most strikingly, we also see a constant flow of deployments since late 2019 with hardly any unused wallets, predominately for \ETH{}, but also a few for tokens and even less for both.

\begin{figure}[hbt!]
\centering
\includegraphics[width=.9\columnwidth]{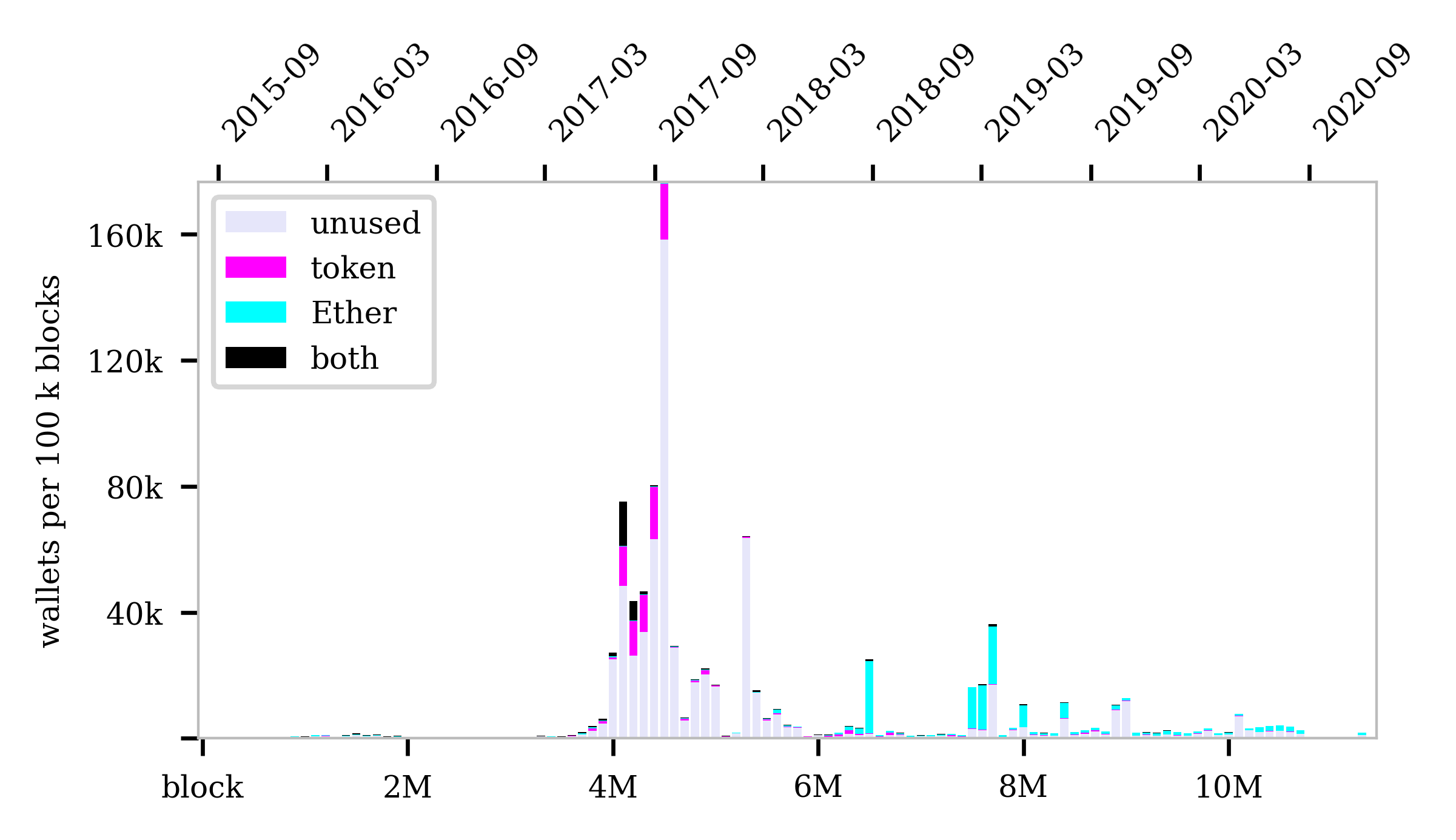}
\caption{MultiSig wallets -- creation and usage.} \label{fig:multisig}
\end{figure}
\subsubsection{MultiSig Wallets}
Figure~\ref{fig:multisig} depicts the usage of the fourth most frequently deployed type of multiSig wallets. 
We notice a pronounced Peak of deployment around mid 2017.
During this period, the multiSig wallets were hardly ever used for \ETH{} only, but rather for tokens or both, while the rate of unused wallets is high.
The picture changes, as since block 6\,M they are solely used for \ETH{} with a significant lower number of deployments and unused wallets.

\subsubsection{Update and Smart Wallets}
\begin{figure}[hbt!]
\centering
\includegraphics[width=.9\columnwidth]{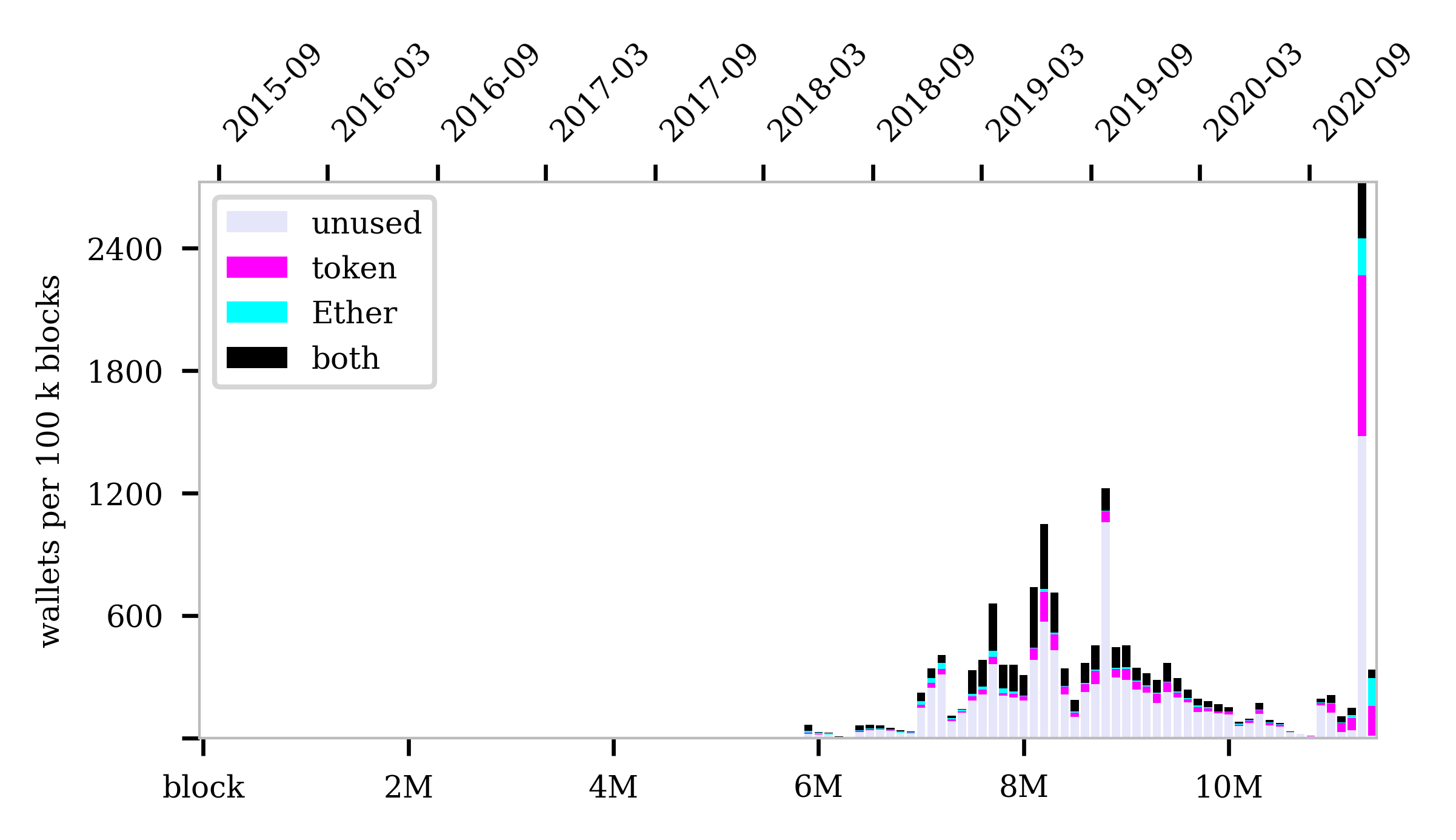}
\includegraphics[width=.87\columnwidth]{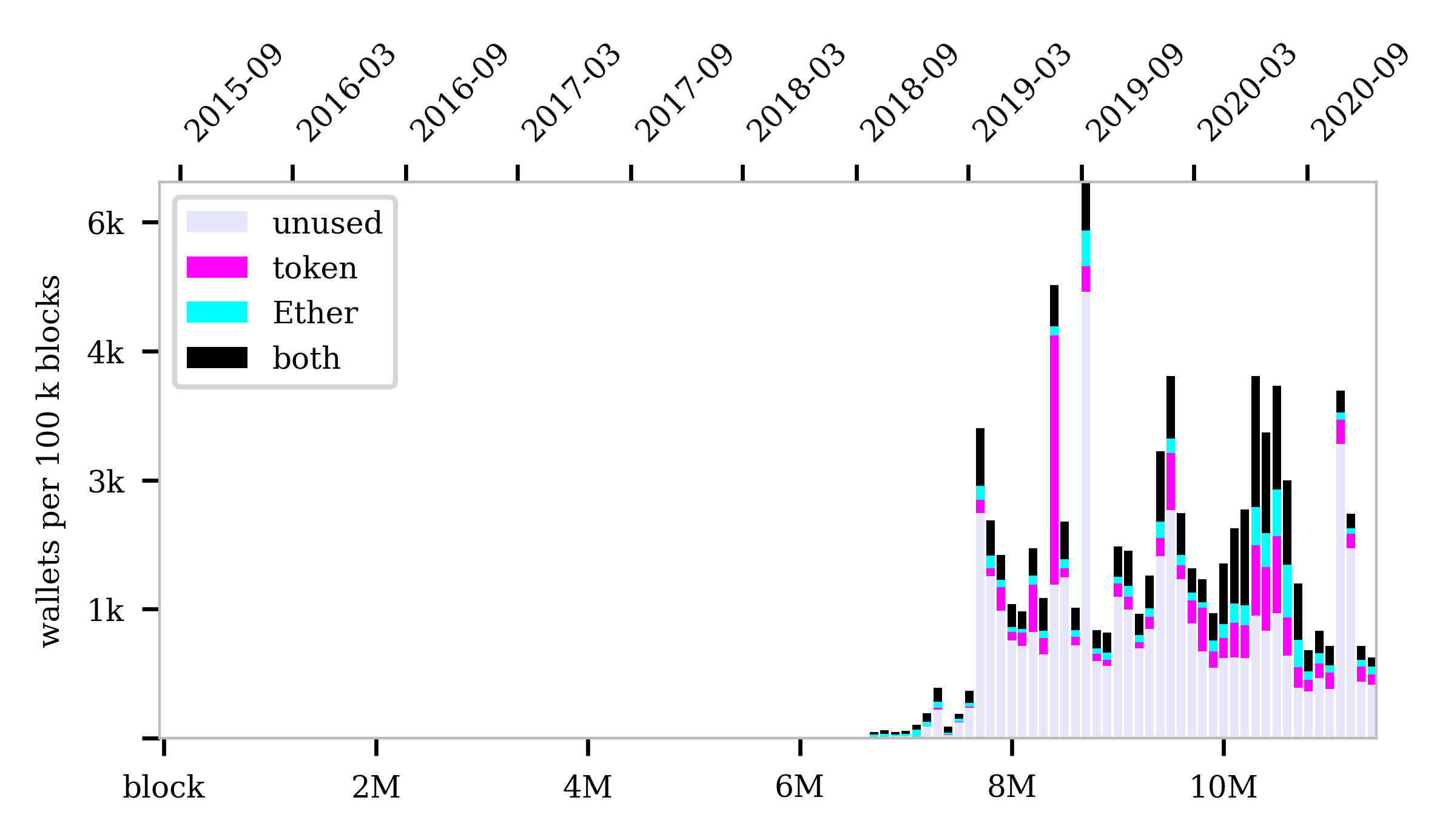}
\caption{Update (upper plot) and smart (lower plot) wallets -- creation and usage.} \label{fig:update_smart}
\end{figure}

These wallets are deployed in comparably small numbers.
If used, they are more often used for both \ETH\ and tokens than the other wallet types.
Similar to the other types, more than half of the wallets are still unused.

\subsection{Token Holdings}
Most wallets are designed for token management.
Still, only 1.1\,M wallets (14.6\,\%) have so far received at least one token, while 5.0\,M (68.4\,\%) did not.
Even though the percentage of wallets without a single token varies with the type, it is always more than 60\,\%.
If wallets do hold tokens, the number of different tokens is small for the majority of them.
Of the 1.1\,M wallets holding tokens, 219\,k hold more than one type of token.
Just 8\,055 wallets each held more than 10 different tokens.
Only single wallets held substantial amounts of different tokens over time, the maximum being 834.

\section{Wallets in the Landscape of Contracts}
Smart contracts perform their tasks stand-alone or in cooperation with companion contracts, with the number of contracts belonging to a single application going up to millions in extreme cases (like for gasTokens or wallets).
The landscape of smart contracts is as diverse as their purpose.
We find exchanges, markets, wallets, tokens, games, attackers, and all kinds of \Dapps\ implementing part of their logic on-chain.

When we want to place the large group of wallet contracts within the landscape of smart contracts, we face several challenges.
Distinguishing between any kind of groups of smart contracts is proving difficult as some groups transition smoothly in to each other, partially overlap, or are simply indistinguishable.
With \Dapps\ on the other hand, it is particularly difficult to decide which contracts are part of the app and which ones use the \Dapp\ or just interact with it.

Regarding the overall landscape, we assume that most of the applications are connected mainly over general services like wallets.
To test this assumption, we analyze the calls to smart contracts as a graph.

\subsection{Contracts as Call Graph}
In order to analyze the landscape of smart contracts, their invocations provide useful information.
For this, we build a call graph where the contracts serve as nodes that are connected by an edge whenever the trace lists a call from one contract to another.
After removing singletons (i.e.\ contracts that are called just by users or not all) and reducing multiple edges between two contracts to just one edge, we are left with 23.6\,M contracts as nodes (2.8\,M of which are wallets) and 29.9\,M calls as edges.
This graph consists of 24.6\,k connected components, with the largest one containing virtually all nodes (22.9\,M, 97\,\%), while most components have less than 100 nodes.
The high interconnection indicates that applications do not separate naturally.

\ITEM{Proper Wallets.}
To test the assumption that wallets act as a major connecting element, we additionally remove the  2.8\,M contracts from the graph that we identified as wallets. 
This further reduces the graph to 20.8\,M nodes and 24.7\,M edges.
However, the largest component still consists of 20.1\,M nodes (96.8\,\%), while the number of components slightly decreases to 23.1\,k.
We conclude that wallets likely contribute to the cohesion of the graph, but are not solely responsible for it.

\ITEM{Token holders.}
If we take any token holder to be a wallet and repeat the analysis by removing further 6.1\,M contracts that ever held a token, the graph falls apart.
The remaining 14.7\,M nodes yield 13.6\,M graph components, with the largest one containing 441\,k nodes (3\,\% of all nodes).
Thus, for a liberal definition of wallets, the assumption holds true that wallets serve as cohesive in the call graph.
However, removing all token holders is too coarse, as we remove contracts for e.g.\ exchanges, markets, applications that employ their own token, and token contracts in general.

\section{Comparison to Related Work}\label{sec:comparison}
\emph{Off-chain wallets} are compared extensively in~\cite{Haigh2019}.
In our work, we focus on wallet contracts on Ethereum.
Therefore, we consider work from the area of wallet contracts, analysis of the Ethereum transaction graph or trace, EVM bytecode analysis, as well as overview areas like applications, activity patterns or the landscape of SCs.

\subsubsection*{Wallet Contracts}
In their analysis of ERC20 token trading, the authors of~\cite{Somin2018} take any address holding tokens to be a wallet.
They demonstrate that the token trading network shows power-law properties and that it is decentralized, diverse, and mature.

The authors of~\cite{Homoliak2018} focus on 2-factor authentication for wallets where a SC handeles the passwords, but do not discuss wallet contracts in detail.

The broad analysis of smart contracts by~\cite{Pinna2019} does not focus on wallets, but concludes that most contracts that collect substantial amounts of \ETH{} are wallet contracts. 

In their taxonomy~\cite{Karantias2020}, the authors distinguish cryptocurrency wallets ``based on whether they work for transparent or private cryptocurrencies, what trust assumptions they require, their performance and their communication overhead''.
For the corresponding wallet protocols, they evaluate performance and security characteristics.
However, smart contracts are deliberately disregarded in their model.

In their security evaluation~\cite{Praitheeshan2020}, the authors investigate the safenesss of Ethereum wallet contracts using the tools Oyente, Osiris, Maian, and Mythril.
They distinguish the four wallet categories Multisig, Smart, Retailer, and Controller wallets.
The proposed security analysis framework focuses on the components Solidity, EVM, and external sources.

\ITEM{Our Approach}.
We are not interested in a broad definition of wallets being any holder of tokens or Ether like~\cite{Somin2018} or helper contracts like~\cite{Homoliak2018}.
In this paper, we focus on wallet contracts that implement characteristic functionality.
A major challenge is to identify such proper on-chain wallets.
The authors of~\cite{Praitheeshan2020} use an astonishingly similar list of wallets (in regards to Ethereum addresses and source code), but do not detail how they identify the wallets.
This paper is an extended and updated version of our previous work~\cite{MdAGS2020wallets}.

\subsubsection*{Ethereum Graph and Trace Analysis} 
The authors of~\cite{Chan2017} focus on de-anonymizing addresses by analyzing the transaction graph.

Applying network science theory, the authors of~\cite{Guo2019} ``find that several transaction features, such as transaction volume, transaction relation, and component structure, exhibit a heavy-tailed property and can be approximated by the power law function.''

The authors of~\cite{Gupta2019} investigate ``contracts of importance'', which are defined by high duplicity, Ether balance, Ether moved, and activity.

The authors of~\cite{Lee2020a} construct four interaction graphs based on Ethereum transactions, one graph each for user-to-user, user-to-contract, contract-to-user, and contract-to-contract. 
Then they investigate local and global graph properties and discuss similarities with social networks.

In a graph analysis, the authors of~\cite{Chen2020} study Ether transfer, contract creation, and contract calls.
They compute metrics like degree distribution, clustering, degree correlation, node importance, assortativity, and strongly/weakly connected components, based on which they list 35 major observations.
Beside several power law observations, they state that exchanges are important regarding money flow, while token contracts are responsible for a high transaction volume.
Wallets are mentioned only as one instance sticks out in the call graph.

Regarding ERC20 tokens on Ethereum, the authors of~\cite{Somin2018} study the tokens trading network in its entirety with graph analysis and show power-law properties for the degree distribution.

Similarly, the authors of~\cite{Victor2019} measure token networks, which they define as the network of addresses that have owned a specific type of token at any point in time, connected by the transfers of the respective token. They do not mention wallets, though.

\ITEM{Our Approach}.
Instead of examining the trading of assets like~\cite{Somin2018,Victor2019}, our investigation focuses on contracts that manage the access to the traded assets, namely wallet contracts.
In contrast to~\cite{Lee2020a}, we do not distinguish between user accounts and contracts since we focus in wallet contract.
The approach in~\cite{Gupta2019} is similar to ours~\cite{MdAGS2019e,MdAGS2020landscape} as we also look at duplicity and activity.
However, we disregard Ether in our analyses and generally aim at a more fine-grained distinction of mass phenomena.
We use high creation activity as one of the pointers to potential wallets.

\subsubsection*{EVM Bytecode Analysis}
To detect code clones, the authors of~\cite{He2020} first deduplicate contracts by ``removing function unrelated code (e.g., creation code and Swarm code), and tokenizing the code to keep opcodes only''.
Then they generate fingerprints of the deduplicated contracts by a customized version of fuzzy hashing and compute pair-wise similarity scores.

In their tool for clone detection, the authors of \cite{Liu2019} characterize each smart contract by a set of critical high-level semantic properties. 
A two-step approach of symbolic transaction sketch and syntactic feature extraction yields a numeric vector as representative of a contract and its semantic properties, which are based on the control flow graph, path conditions as well as storage and call operations.
Then they detect clones by computing the statistical similarity between the respective vectors.

The authors of~\cite{Kondo2020} investigate clones based on \emph{Solidity source code} applying a tree-based clone detector (Deckard).
At the level of s ource code, wallets do not stick out as clones.

In~\cite{Pinna2019}, the authors analyze Ethereum SCs based on code metrics. 
They define wallets as SCs that ``securely collect Ethers and could implements some functions such as the `multiple ownership' or the `escrow'.''
Unsurprisingly, they also confirm that wallets are the group of contracts that in total hold most Ether.

Using deep learning, the authors of~\cite{Kim2020} assign attribute tags (extracted from source code and metadata) to unknown bytecode, and determine the category (application area) of a bytecode, like markets and exchanges. 
As for wallets, the only mention the Parity wallet hack.

To detect token systems automatically, the authors of \cite{Froewis2019} compare the effectiveness of a behavior-based method combining symbolic execution and taint analysis, to a signature-based approach limited to ERC20-compliant tokens.
They demonstrated that the latter approach detects 99\,\% of the tokens in their ground truth data set.
For all deployed bytecode, though, it bears a ``false positive risk in case of factory contracts or dead code''.

\ITEM{Our Approach}.
Our method of computing code skeletons is comparable to the first step for detecting similarities by~\cite{He2020}.
Instead of their second step of fuzzy hashing though, we rely on the set of function signatures extracted from the bytecode and manual analysis, as our purpose is to identify wallets reliably. 
Relying on the interface is in line with the results in~\cite{Froewis2019,MdAGS2019e,MdAGS2020tokens}.
However, we base our identification of wallets on interface blueprints for several types of wallets. Additionally, we fuzz the interfaces to include variants.

\subsubsection*{Applications, Activity and Landscape of SCs}
In the empirical analysis~\cite{Bartoletti2017a}, the authors investigate platforms for SCs, cluster SC applications on Bitcoin and Ethereum into six categories, and evaluate design patterns for SCs of Solidity code until January 2017.
On Ethereum, they find 17 wallet contracts involved in 1\,342 transactions. All wallets use authorization and termination design patterns, and most of them also use time constraints.

In their empirical study~\cite{Kiffer2018}, the authors investigate Ethereum smart contracts by looking at contract creation, interaction, and code reuse.
They find that wallet contracts are a group with highly similar code.

In their graph analysis~\cite{Chen2020}, the authors conclude that financial applications dominate Ethereum. Regarding wallets, they just mention one type of wallet contract.

In the exploratory study of Ethereum SCs~\cite{Oliva2020a}, the goal is ``a broader understanding of all contracts that are currently deployed in Ethereum''.
However, the authors do not include internal transactions (i.e.\ messages) in the data set and thus miss a substantial part of the activity on Ethereum including contracts creations, calls, and destructions. This is especially true for wallet contracts, which are deployed frequently by other contracts (i.e. wallet factories).

\ITEM{Our Approach}.
In our previous work, we identify tokens~\cite{MdAGS2020tokens}, wallets~\cite{MdAGS2020wallets} and other groups of smart contracts~\cite{MdAGS2019e,MdAGS2020landscape}, and provide quantitative and qualitative characteristics for each identified type.
In this work, we aim at understanding the landscape of smart contract with a focus on the specific group of wallet contracts.
We argue that wallets are one of the backbones of the SC landscape keeping the call graph connected.

\section{Conclusions}\label{conclusions}
We examined smart contracts that provide a wallet functionality on the Ethereum main chain up to block 11\,500\,000, mined on Dec 22, 2020.
For a semi-automatic identification of wallet contracts, we discussed methods based on deployed bytecode and interactions.
By analyzing source code, bytecode, and execution traces, we derived features and types of wallets in use, and compared their characteristics.
Moreover, we provided a quantitative and temporal perspective on the creation and use of identified types of wallets, and discussed their role in the smart contracts landscape.

\ITEM{Identification of wallets.}
The identification of wallets as recipients of tokens or Ether can be done automatically, but includes many contracts beyond proper wallets. 
Our method of identifying wallets by name, interface, and ancestry yields blueprints for wallets, which then are used to locate contracts with similar implementations or same deployers.
This approach is only semi-automatic, but more reliable.

\ITEM{Blueprints for wallets.}
Since we manually verify the Solidity source code, our work yields a ground truth of wallets that can be used for evaluating automated tools.

\ITEM{Wallet Features.}
Features of wallets in use beyond the transfer of assets can be grouped into administration and control, security mechanisms, life cycle functions, and extensions.
By distilling a comprehensive list of features for pure wallets, we are able to separate wallet contracts from non-wallets.
Moreover, we could depict actual use cases via the extracted features.

\ITEM{Wallet Types.}
Wallets can be categorized into the six types simple, multiSig, controlled, forwarder, update, and smart wallet according to the features they provide.
MultiSig wallets were the first to appear shortly after the launch of Ethereum, while controlled and forwarder wallets followed in 2017.
Update wallets and smart wallets with a modular design are a recent phenomenon starting at the end of 2018.
We observe an evolution of features in the wallet types.
Still, the multiSig wallet seems popular, either as it is or incorporated into smart wallets.

\ITEM{Usage of Wallets.}
On-chain wallets are numerous, amounting to 7.3\,M contracts (20\,\% of all contracts).
However, most wallet contracts (68.4\,\%) are not in use yet. They may have been produced on stock for later use.
Interestingly, the wallets are used either for tokens or for \ETH{}, but rarely for both.
Even though most wallets are designed for token management, only 1.1\,M wallets (14.6\,\%) have so far received at least one token.
Of the few wallets holding tokens, 79,5\,\% hold just one type of tokens, while 99.2\,\% hold at most 10 different types.

\ITEM{Code Reuse.}
Due to mass deployments, the code reuse in wallets is much higher than the average on Ethereum.
The 7.3\,M wallets have only 6\,512 distinct bytecodes and 835 functionally equivalent skeletons.
The average deployment factor of 1\,126 for wallet bytecodes is more than 10 times higher than the average of 100 for all contracts on Ethereum.

\ITEM{Landscape.}
Our assumption that wallets act as a cohesive in the graph of executed calls between contracts holds only for a very broad definition of wallets.
To dissect the landscape of smart contracts effectively into applications, we may have to identify further contract groups that handle assets.

\subsection{Future Work}
When aiming at a deeper understanding of the role of \Dapps\ and smart contracts, there are still some pieces of the puzzle missing.
Our contribution to understanding on-chain wallets may serve as a basis for further research in this direction, as wallets are a major application type.
Moreover, as wallets link many \Dapps, removing them from the overall picture may let other applications stand out clearer.
Examples of such applications are markets and exchanges, which also act as a cohesive in the call graph and which still need thorough investigation.
Additionally, we can use the number of calls as weights on the edges and apply neighborhood algorithms.
First experiments with the latter approach show that it may be effective if we manage to remove some of the major connecting applications.

To determine reliably what smart contracts actually implement, it is still indispensable to analyze bytecode.
Adequate tool support for a massive automated semantic code analysis would be helpful to obtain a comprehensive picture of the smart contract ecosystem.

\sloppy
\newpage
\addcontentsline{toc}{section}{References}
\bibliographystyle{IEEEtran}
\bibliography{IEEEabrv,wallets}
\newpage
\appendix
\section{Wallet Profiles}\label{app:wallets}
In this appendix, we detail each identified wallet blueprint with regards to 
\begin{itemize}
\item a brief description of features
\item the author(s) if known
\item the location of the Solidity source code that it is based on (if any)
\item the identification method: function headers, creation history, or subtleties of the detection procedure (if any)
\item the addresses of exemplary bytecodes and creators, linked to \texttt{etherscan.io}.
\end{itemize}

\subsection{Simple Wallets}\label{app:simple}
\begin{Wallet}{AutoWallet}
  \begin{Description}
This is a simple wallet for \ETH , ERC-20 tokens and non-fungible tokens (ERC-721).
It provides owner management.
Received \ETH\ is forwarded automatically to the owner of the wallet, but the wallet provides also a sweep function to access \ETH\ that was deposited e.g.\ as a mining reward or by a self-destruct.
  \end{Description}
  \begin{Identification}
    The wallet can be uniquely identified by the following function:
\begin{verbatim}
transferNonFungibleToken(address,address,uint256)
\end{verbatim}
  \end{Identification}
  \begin{Addresses}
    The wallet is deployed 9\,250 times. All but two instances were deployed by the externally owned account
    \A{13d0c7ada3f98eec232ed7e57fefc4c300f25095}.
    We find one bytecode deployed e.g.\ at address
    \A{1991af53e07b548a062a66e8ff3fac5cc9e63b22}. The wallet provides verified source code.
  \end{Addresses}
\end{Wallet}

\begin{Wallet}{BasicWallet}
  \begin{Description}
This is a basic wallet with owner management for \ETH, ERC-20 and ERC-223 tokens.
  \end{Description}
  \begin{Identification}
    The wallet can be uniquely identified by the four functions:
\begin{verbatim}
changeOwner(address)
transfer(address,uint256)
transferToken(address,address,uint256)
tokenFallback(address,uint256,bytes)
\end{verbatim}
  \end{Identification}
  \begin{Addresses}
    This wallet was deployed 4\,822 times with verified source code and two versions of the bytecode (due to different versions of the Solidity compiler) by the externally owned account \A{ff3249da62ca5286997f31f458959de9ae2f4dad}.
    \begin{center}
    \begin{tabular}{rll}
      wallets & version & \multicolumn{1}{c}{deployed e.g.\ at} \\
      \midrule
      3\,367    & newer   & \A{a4db5156d3c581da8ac95632facee7905bc32885}\\
      1\,455    & older   & \A{850c3beae3766e3efcf76ade7cbd6e3e0aec517e}
    \end{tabular}
  \end{center}
  \end{Addresses}
\end{Wallet}

\begin{Wallet}{ConsumerWallet}
  \begin{Description}
This is a wallet for \ETH\ and ERC-20 tokens with various security features, like white-listing of receivers, daily limits (using an oracle for converting tokens to \ETH), two factor authentication, and gas management.
  \end{Description}
  \begin{Source}
    \url{https://github.com/tokencard/contracts/}
  \end{Source}
  \begin{Identification}
    We identify the wallets by checking the presence of one or two of the following functions:
\begin{verbatim}
topUpAvailable()
bulkTransfer(address,address[])
\end{verbatim}
    This allows us to identify seven variants of deployed bytecode, which we check manually to make sure that they are indeed the same type of wallet.
    Most wallets are deployed by factory contracts containing the characteristic function
\begin{verbatim}
deployWallet(address)
\end{verbatim}
  \end{Identification}
  \begin{Addresses}
    This wallet is deployed 11\,911 times.
    Of the 11 bytecodes of this wallet, the five most frequent ones were deployed by six factories, while most of the less frequent ones were deployed by one particular externally owned account. All but one provide verified source code.
    \begin{center}
      \begin{tabular}{rccc}
        wallets & code deployed e.g.\ at& creator & source\\
        \midrule
6\,853	& \A{a8e7213d64e29f6e5e81cb5d6cd48bcdcf722dc4} & A & yes \\
2\,395	& \A{15e0c211c9a221d2e9dead41ceb596755d7b5b66} & B & yes \\
1\,380	& \A{20ab867160e73788e0db311f445e67bc596e0ec0} & C & yes \\	
531	& \A{d883f8a6080ea6c473efc05c8ff3238255ad0e02} & D,E & yes \\
273	& \A{b112e2ede29fa9fd9485aab8d637336aaa020ebb} & F & yes \\
255	& \A{613e20da62058aa4f8bc2c8b6fddc03e43b89b5a} & G & yes \\	
144	& \A{e6510c19c7768ca0937e8f4daf0b16859af9c271} & G & yes \\	
53	& \A{5c76fb5fb117d190beac217bc3568e70f2b6b71d} & G & yes \\	
22	& \A{e69ac6adf65c682f3809e8273b123c2d6ec7726a} & H & yes \\	
4	& \A{1e7d250a2ac2646125be3823290fbb5d61d57c13} & G & no \\
1	& \A{13d66821f5dc94b242bd91768546b5adb1273b45} & I & yes 	
      \end{tabular}
      \smallskip

      \begin{tabular}{ccc}
        & creator address & user? \\
        \midrule
        A & \A{85bb8a852c29d8f100cb97ecdf4589086d1be2dd} & no \\
        B & \A{a678cad8f13c0f8b88ed5fee227dfae1e6fe218e} & no \\
        C & \A{95bebe7bfc6acc186c13d055d0aacc2de5f81502} & no \\
        D & \A{5e7a685ed8bd3e9dc24bfd67813e9c26b5891308} & no \\
        E & \A{b24d47364163f909e64cf2cf7788d22c51cea851} & no\\
        F & \A{5b47acb25073234421f1f8b66d2c8056620d41ff} & no \\
        G & \A{e0731c1a30e6ed0c6e9162eb87fc85e831caf382} & yes \\
        H & \A{e0731c1a30e6ed0c6e9162eb87fc85e831caf382} & no \\
        I & \A{9837103896199878CaD21A42015ce0513fC2e80e} & yes \\
      \end{tabular}
    \end{center}
  \end{Addresses}
\end{Wallet}

\begin{Wallet}{EtherWallet1}
  \begin{Description}
This is a simple wallet for \ETH\ and ERC-20 tokens with owner management.
  \end{Description}
  \begin{Identification}
    The wallet was identified via the creation history of the factories.
  \end{Identification}
  \begin{Addresses}
This wallet is deployed 84\,656 times with two versions of the bytecode by three factory contracts.
It does not provide verified source code.
    \begin{center}
    \begin{tabular}{rllc}
      wallets & \multicolumn{1}{c}{deployed e.g.\ at} & factory\\
      \midrule
      84\,653   & \A{be1390c5bbc48f731511bcba0dc7052ceea0a48a} & A, B\\
               3   & \A{6F6C95f1a3f4430840e3E7162Af55CA567D34b61} & C
    \end{tabular}
    \smallskip
    
    \begin{tabular}{c@{\quad}c}
        & factory address \\
        \midrule
        A & \A{D02C52f828a35b808Ce8335E7F02805dcc380b35} \\ 
        B & \A{67cfdc0a3a8af28d4691a94fb15029e1086c8498} \\
        C & \A{7635CF79776738c77d529cAE99BEA7283b05d52f} \\
    \end{tabular}
    \end{center}
  \end{Addresses}
\end{Wallet}

\begin{Wallet}{EtherWallet2}
  \begin{Description}
This is a simple wallet for \ETH\ and ERC-20 tokens with owner management.
  \end{Description}
  \begin{Identification}
    The wallet was identified via the creation history of the factories.
  \end{Identification}
  \begin{Addresses}
This wallet is deployed 112\,987 times with two versions of the bytecode by three factory contracts.
It does not provide verified source code.
    \begin{center}
    \begin{tabular}{rllc}
      wallets & \multicolumn{1}{c}{deployed e.g.\ at} & factory \\
      \midrule
      112\,986   & \A{cd8a5fa00ebc0efe813f1952755984497d5184f3} & A, B \\
                1   & \A{9f227c3e79ed447d2c6f4c5dca47a4dbdb786f2a} & C
    \end{tabular}
    \smallskip
    
    \begin{tabular}{c@{\quad}c}
        & factory address \\
        \midrule
        A & \A{e0af95f7e6dd07a4268064de2cf7c8f044fd53fa} \\
        B & \A{de14e7548a46bbbf1f3f50e98a04e58c0efc60eb} \\
        C & \A{bddf790106fe05d5b5b96e666e4c873f7afec24d} \\ 
    \end{tabular}
    \end{center}
  \end{Addresses}
\end{Wallet}

\begin{Wallet}{SimpleWallet}
  \begin{Description}
This is a very basic wallet for \ETH\ only that can self-destruct.
  \end{Description}
  \begin{Identification}
    The wallet was identified via the creation history of the factories.
  \end{Identification}
  \begin{Addresses}
The wallet was deployed twice, each by a different externally owned account.
It does not provide verified source code.
    \begin{center}
    \begin{tabular}{rllcc}
      wallets & \multicolumn{1}{c}{deployed e.g.\ at} & creator & destroyed?\\
      \midrule
      1   & \A{f3ea4b7d340ac35ef26365a7c4aab0ab983a56f7} & A & yes \\
      1   & \A{66da495abc70233473cbd7419e2eda6588d03d5e} & B & no 
    \end{tabular}
    \smallskip
    
    \begin{tabular}{cc}
        & deploying user address \\
        \midrule
        A & \A{4B9fbfcd85Dd84E30154DB2BB2D25CEfFB53e5B7} \\ 
        B & \A{51a6073079AE78e22Fa313aB2D9414d6Cf02e34c} \\
    \end{tabular}
    \end{center}
  \end{Addresses}
\end{Wallet}

\begin{Wallet}{SimpleWallet2}
  \begin{Description}
This is a basic wallet for \ETH\ only.
\ETH\ is received via the fallback function, queried with \HD{weiBalance()}, and transferred to a specified destination via \HD{claim(address)}.
    All but 12 wallets have the ability to transfer ownership.
    All but 5 wallets where created by externally owned accounts.
  \end{Description}
  \begin{Identification}
    The wallet can be uniquely identified by the two following functions:
    \begin{verbatim}
weiBalance()
claim(address)
    \end{verbatim}
  \end{Identification}
  \begin{Addresses}
  The wallet is deployed 540 times. We find 13 bytecodes and 10 skeletons.
    In the following table, we list for each bytecode the number of deployments, the corresponding skeleton, an exemplary address and some notes that are explained below the table.
    \begin{center}
      \begin{tabular}{rccl}
        count & skel & sample address & notes \\
        \midrule
        439 &   A & \A{7c443e7ede89665fe64f749a51993b476d5e41c7} & \\
         31 &   B & \A{c2f21e3d752834a38a0f8a68c4cef84df1eb4541} & source\\
         21 &   C & \A{17b54df06b3186cbc3a7591c9b9b36dc587dc9d6} & \\
         17 &   B & \A{9c86825280b1d6c7db043d4cc86e1549990149f9} & source\\
         14 &   B & \A{073a014b8a67af5153a693fa7365814f4822d4cb} & \\
          5 &   D & \A{02bdd07fbea5316091f9083202198f2e74eb0e07} & nom, factory \\
          3 &   A & \A{19b4bd9792a5916859e3aaa71280c1d4044ef154} & \\
          2 &   E & \A{37f08291d4a921a662ca221b985e9e2dc63cdad8} & nom, sd \\
          2 &   F & \A{c66b8ed8d23f83ead73064c2a24aa424bc7e965d} & \\
          2 &   G & \A{36f6e5a6b22b8a47c1b59f3270b8294d2e104efb} & nom\\
          2 &   H & \A{90229c3b0d73f5158251f3b80d77c673f5e16793} & nom\\
          1 &   I & \A{57b86a80f1a5a8c3bd913dd7785b0bfd3b912c3e} & \\
          1 &   J & \A{bd4fecbb8f54a8e0ab772abd4f555d31f68f32a8} & nom
      \end{tabular}
    \end{center}
    \begin{flushleft}
      source \dots\ verified source code available\\
      nom \dots\ no owner management\\
      sd \dots\ ability to self-destruct, one wallet already self-destructed\\
      factory \dots\ the contract at \A{f1458a0324876064de8a3817c8b27f8d0ef2880e} deployed the wallets.
      The bytecode of the factory is also deployed at four other addresses, which remained inactive so far.
    \end{flushleft}
\end{Addresses}
\end{Wallet}

\begin{Wallet}{SmartWallet}
  \begin{Description}
This is a wallet for ERC-20 tokens only. 
The older version allows the user to configure a backup account where to transfer the tokens, the newer version implements a time lock (cooling period) for withdrawals.
  \end{Description}
  \begin{Identification}
    The wallet can be identified by the following function:
\begin{verbatim}
transferToUserWithdrawalAccount(address,uint256,address,uint256)
\end{verbatim}
    The older version additionally contains the function
\begin{verbatim}
transferToBackupAccount(address,uint256)
\end{verbatim}
    whereas the newer version can be identified e.g.\ by the additional function
\begin{verbatim}
requestWithdraw()
\end{verbatim}
    The newer version of the wallet keeps the cooling period in a separate contract called \texttt{WithdrawalConfigurations}, which can be found by looking for the function
\begin{verbatim}
withdrawalCoolingPeriod()
\end{verbatim}
It provides verified source code at \ES{efc7de761ae038b3bb3080ecfb98cea51fd442ea} for the newer version.
  \end{Identification}
  \begin{Addresses}
    The wallet is deployed 46\,832 times by six externally owned accounts.
    We find 15 bytecodes and five skeletons.
    In the table below, we list for each skeleton the number of deployments, the code version, an exemplary address, and the creators.
    \begin{center}
      \begin{tabular}{rccc}
        wallets & version & deployed e.g.\ at& creator\\
        \midrule
   37\,629 & older & \A{5e63e5f352fa0167b241a9824165b5cf38ee2973} & A, B, C, D, E \\
     9\,192 & newer & \A{69667ce8641ed03abec2627866543b1126240044} & A, B, D, E \\ 
            7 & older & \A{0d341e13f9f9bd6e02fa97e249a0a27522b0efb1} & F \\
            2 & older & \A{0c3e4f2961f6b8d62be9353ac2376e6438a9cf20} & F \\
            2 & older & \A{86acc9df62926e62bb6b5dd0e46409be37ffea36} & F
      \end{tabular}
      \smallskip

      \begin{tabular}{cc}
         & deploying user address  \\
        \midrule
        A & \A{0fa4be05d6c7accdeb1e59f355315ec61c9a6dbb}  \\
        B & \A{6bdd15d26ee026c0cb952e3d7fd7535cf7e1ef20}  \\
        C & \A{866f649cd9280d3dfa282372a3f5828839944959}  \\
        D & \A{aa1aeffe8bf1a7470558b31f35cb6ec7faf0679f}  \\
        E & \A{b642f8e816f64cf7b022d7521be162cbb7193dd5}  \\
        F & \A{00da955d3e1f31726c0f7511f0593452925a9acd} 
      \end{tabular}
    \end{center}
  \end{Addresses}
\end{Wallet}

\begin{Wallet}{SpendableWallet}
  \begin{Description}
This is a wallet with owner administration for a single ERC-20 token as specified on deployment.
Additionally, it provides an emergency function to withdraw \ETH\ and ERC-20 tokens other than the one the wallet is intended for.
A small number of wallets contain the \texttt{tokenFallback} function required by ERC-223.
  \end{Description}
  \begin{Identification}
    The wallets can be uniquely identified by the following two functions:
\begin{verbatim}
spend(address,uint256)
claimTokens(address)
\end{verbatim}
Depending on the version, the wallets implement three or four further, less distinctive functions.

    All instances of this wallet have been deployed by contracts.
    These factories can be uniquely identified by looking for bytecode implementing the function 
\begin{verbatim}
newPaymentAddress(address,address)
\end{verbatim}
  \end{Identification}
  \begin{Addresses}
The wallet is deployed 6\,430 times.
We find six versions of bytecode, of which the last two in the table below implement the \texttt{tokenFallback} function.
The upper three provide verified source code.
    \begin{center}\scriptsize
      \begin{tabular}{@{}rcc@{}}
        wallets & code deployed e.g.\ at & factory \\
        \midrule
  5\,555 & \A{35a1700ac75f6e9e096d9a5c90e3221b658096e0} & \A{16e73d276b2163c49db410350bfefd9f48898821} \\
      452 & \A{87f53784494c693d1ea80aafffaf53b547b87df5} & \A{2048360fef3fcb5858f8e2dbae2d4b8ad2e94c23} \\
      416 & \A{2254f46dedafa2a03f59008456a7400cfadcaf73} & \A{365443b7b06b86f1f560873de723b6d4fb3927ce} \\
         3 & \A{6bb75a4cfdcee5befd65d898198f43c280631913} & \A{25f93d0a70c3a9054a9adbae43a9bd67238f0f03} \\
         2 & \A{0a9b4b39a0e8da2855f5f8cc0aa6c203963f17e7} & \A{cac914780920e425518131557beb856f0ed9688d} \\
         2 & \A{409fd08a982826c2e7c0d7d90c673c7eec0cf922} & \A{4dafe9f324816a7f5bc8e05befb2579c30cc83d6}
      \end{tabular}
    \end{center}
  \end{Addresses}
\end{Wallet}

\begin{Wallet}{TimelockedWallet}
  \begin{Description}
This is a wallet for ERC-20 tokens, and in most cases also for \ETH.
    During deployment, the owner and a lock period is fixed.
    The wallet may receive assets at any time, but only after the lock period the owner may withdraw the assets.
  \end{Description}
  \begin{Identification}
    This wallet can be uniquely identified by the following two functions:
\begin{verbatim}
info()
unlockDate()
\end{verbatim}
Most wallets are deployed by factories containing the function:
\begin{verbatim}
newTimeLockedWallet(address,uint256)
\end{verbatim}
  \end{Identification}
  \begin{Addresses}
The wallet is deployed 226 times.
We find 33 distinct bytecodes and nine code skeletons that were deployed mostly by factory contracts.
In the following table, we list for each skeleton the number of deployed wallets, an exemplary address, whether also user addresses deployed some wallets, and a reference to the deploying contract(s).
    \begin{center}
      \begin{tabular}{rccc}
        wallets & deployed e.g.\ at & user & contract\\
        \midrule
55 & \A{1b95d18bda9b74cf7f1e74d5929bafef8ec896f7} & yes & A \\
45 & \A{009dd2460f8b84dde574c429ee4c94a7b65fc143}  & yes & B \\
27 & \A{0eff527de5ab1fdd5c9d61a36b7e564e5f7fab06}  & yes & C, D \\
10 & \A{87397ec0bc821a8c10cd35913ccd9e83c8c14f09}  & no & E, F, G \\
 4 & \A{6c42f7fc947035d1e04502d20aee5de089a437dc}  & no & H \\
 2 & \A{df934b2a8ece13041cb4e062ee0b275e0fb388dd}  & no & I \\
 1 & \A{23e012c683624d4ca63410840c01bfb02ee41585}  & yes & no \\
 1 & \A{217edc621b3c8f70e55286f56dc75eb70326ec16}  & no & J \\
 1 & \A{b5b85d88f0c5f1d5f87e47cc1cbe27ab1a7f6e51}  & yes & no
      \end{tabular}
            \medskip

      \begin{tabular}{c@{\quad}c}
         & address of factory contract  \\
        \midrule
        A & \A{2174e3cea45be4b1d74811ee70f7de16119b7c67}  \\      
        B & \A{2648c3e6727595e163b9d2f8dd9b1f1ade292ec9}  \\      
        C & \A{4b448951de99285467820aa222ac5e807285e094}  \\      
        D & \A{64f4863dee2ecfa60b6857795b048b5b3a4889a4}  \\      
        E & \A{c81cb9209a13cdeaff72987a01959714f76de564}  \\      
        F & \A{e9ca4c1b0af2cda6d49fded6d5f8cd901eb531da}  \\      
        G & \A{d0a6623e7f9830003bbb651a96fca0160fc13c9e}  \\      
        H & \A{29ad150d547ae7684bb44ce22025a57c3893b18d}  \\      
        I & \A{9dd698ffcedd642140b98685ed8def2d266031c7}  \\      
        j & \A{cb172d4602643ecece8382b4e35e152b0597a707}  \\      
      \end{tabular}
    \end{center}
  \end{Addresses}
\end{Wallet}

\begin{Wallet}{Wallet1}
  \begin{Description}
    The wallet manages \ETH\ and ERC-20 tokens, which are forwarded to a fixed destination address on request.
    Some wallets are deployed as singletons by users.
    The vast majority of wallets, however, were deployed by factories.

    The wallets come in two flavors: base wallet and proxies (clones).
    Each base wallet is associated with a destination address fixed during deployment.
    The clones of a base wallet delegate each incoming call to the base wallet (using the instruction \op{delegatecall}).
    Immediately after deployment, a call to the function \HD{init} initializes the clone with the destination address of the base wallet making it functionally identical to the latter.
    To minimize the size of base wallets, the main functionality is outsourced to a library contract.
    Figures~\ref{fig:Wallet1} and \ref{fig:Wallet1fac} display pseudo-code for the factory, the library, and the base wallet.

    The factories use the instruction \op{create2} to deploy base wallets and clones.
    Thus, with the salt chosen by the user of a factory, the deployment addresses of future wallets can be pre-computed and may receive \ETH\ and tokens prior to the creation of a wallet.
    This is also reflected in the factory code that forwards assets immediately after wallet creation.
    \begin{figure}
\begin{lstlisting}
contract Wallet1 {
    address public destination;

    constructor (address _destination) { destination = _destination; }

    fallback () payable {
        Wallet1Lib.dc426881(destination);
    }

    function flushETH() {
        Wallet1Lib.dc426881(destination);
    }

    function flushERC20(address _token) {
        Wallet1Lib.e73b9609(destination, _token);
    }

    function init(address _destination) {
        require (_destination == 0);
        destination = _destination;
    }
}

library Wallet1Lib {
    function dc426881(address _destination) external {
        _destination.send(this.balance);
    }

    function e73b9609(address _destination, address _token) external {
        uint256 amount = ERC20(_token).balanceOf(this);
        ERC20(_token).@transfer@(_destination, amount);
    }
}
\end{lstlisting}
      \caption{Wallet1~-- specification of the base wallet and the library as pseudo-code.}
      \label{fig:Wallet1}
    \end{figure}
    \begin{figure}
\begin{lstlisting}
contract Wallet1Factory {
    // create base wallet
    function cbcc054a(address _destination, uint256 _salt)
        returns (address) {
        address wallet = create2(_salt,
            code of contract Wallet1 with constructor arg _destination);
        return wallet;
    }

    // create clone
    function 6ac42711(address _base, uint256 _salt) returns (address) {
        address proxy = create2(_salt, clone code delegating to _base);
        Wallet1(proxy).init(_base.destination);
        return proxy;
    }

    // create clone and call flushETH()
    function e5ab4da2(address _base, uint256 _salt) returns (address) {
        address proxy = 6ac42711(_base, _salt);
        Wallet1(proxy).flushETH();
        return proxy;
    }

    // create clone and call flushERC20()
    function b0287e81(address _base, uint256 _salt, address _token)
        address proxy = 6ac42711(_base, _salt);
        Wallet1(proxy).flushERC20(_token);
        return proxy;
    }

    // flush tokens of wallet
    function e73b9609(address _wallet, address_token) {
        Wallet1(_wallet).flushERC20(_token);
    }
}
\end{lstlisting}
      \caption{%
        Wallet1~-- specification of the wallet factory as pseudo-code.
        The listed functions are the union of the entry points encountered with factories; the actually deployed ones implement a subset thereof.
        See figure~\ref{fig:Wallet1} for the contract \HD{Wallet1}.%
      }
      \label{fig:Wallet1fac}
    \end{figure}
  \end{Description}
  \begin{Identification}
    The base wallets can be identified by the following two functions:
\begin{verbatim}
flushERC20(address)
flushETH()
\end{verbatim}
    The interface of a clone is empty and cannot be distinguished from the interface of other contracts without entry points.
    However, clones can be identified by first determining the factories behind the base wallets, and then collecting the addresses where they deployed the clones.
  \end{Identification}
  \begin{Addresses}
  The wallet is deployed 229\,861 times, albeit in the vast majority as proxy.
  In the following table, we list the four user-created wallets that are base wallets without proxies.
  The fist one provides verified source code.
    \begin{center}
      \begin{tabular}{rcccl}
        wallets & skel & sample address                              & creator & remarks\\
        \midrule
          2   &   A  & \A{e2deeb5c889adccf2a4d902cff0b02d883972231}&   [2] \\ 
          1   &   B  & \A{0dabb48a78e2216a1caa44839fb433699eb4700d}&   [1]   & src code \\ 
          1   &   C  & \A{2dd27d3597c137c94c99d868167bbd1e06919bfa}&   [2] \\ 
      \end{tabular}
    \end{center}
    Factory-created wallets form a more complex ecosystem involving proxies, libraries and factories.
    The libraries and factories are user-created contracts, while the proxies and base wallets are deployed by the factory contracts.
    In the following table, we list the factory-created wallets grouped by the two skeletons and indicate for each skeleton the number of proxies and base wallets as well as an exemplary address and the corresponding library, factory and acting user address.
    \begin{center}
      \setlength\tabcolsep{3pt}%
      \begin{tabular}{rrccccc}
        proxies & bases & skel & deployed at e.g.\                               &  lib & fac & actor \\
		\midrule
              91&       9 &  C   & \A{bab3303a58d17ee8d335c35fb92152b765948d11} &  [4] & [6] & [2] \\
        225\,323&  4\,434 &  D   & \A{0f2ba21dc2360929dfa3dfe8dfe72b630d463b32} &  [5] & [7] & [3]
      \end{tabular}
    \end{center}
    User and contract addresses:
    \begin{center}
      \begin{tabular}{cc}
        \multicolumn{2}{c}{user accounts} \\
        {[1]} & \A{4693622053773ddb76015659fe5be7078abd054e} \\ 
        {[2]} & \A{798aba8798f32d0ce3a9ccd13c3885cbe771f941} \\ 
        {[3]} & \A{60f7f36fc9c823fd25fdf00feefc6d39bed8b53b} \\ 
        \midrule
        \multicolumn{2}{c}{library contracts} \\
        {[4]} & \A{22259db14a1a123531d74e7216ca04326be41c21} \\ 
        {[5]} & \A{ca5faf4c4134bc4f43e910c857723a744b0d2d68} \\ 
        \midrule
        \multicolumn{2}{c}{factory contracts} \\
        {[6]} & \A{92a1d964b8fc79c5694343cc943c27a94a3be131} \\ 
        {[7]} & \A{f6874c88757721a02f47592140905c4336dfbc61} \\ 
      \end{tabular}
    \end{center}
  \end{Addresses}
\end{Wallet}

\begin{Wallet}{Wallet3}
  \begin{Description}
    This is a wallet for \ETH\ and ERC-20 tokens with ownership management.
    Even though `just' a token, the stablecoin USDT is handled explicitly by storing its address as a constant in the contract.
    The contract stores lists of token contracts and corresponding receivers, and allows the owner to transfer multiple tokens, USDT and \ETH\ in one call.
    As there is no verified source code publicly available, figure~\ref{fig:Wallet3} describes the functionality as Solidity pseudo-code.
    \begin{figure}
\begin{lstlisting}
contract Wallet3 is Ownable {
    address[] recs, toks;   // list of token receivers/contracts 
    address public usdt, usdtRec; // usdt token contract, usdt receiver
    string public name, symbol, version;
    event @Transfer@(address @from@, address to, uint256 val); 
    event xc1c40e9d(address @from@, address to, uint256 val);
    constructor (address _usdtRec, address[] _recs, address[] _toks) {
        usdt = 0xdac17f958d2ee523a2206206994597c13d831ec7;
        recs = _recs;      toks = _toks;        usdtRec = _usdtRec;
        name = 'contract'; symbol = 'contract'; version = '1.0.2';
    }
    fallback () payable { }
    function sendEth (address _to, uint256 _val) payable onlyOwner() {
        _to.transfer(_val);
        emit @Transfer@(this, _to, _val); emit xc1c4039d(this, _to, _val);
    }
    function sendUSDT(address _to, uint256 _val) payable onlyOwner() {
        if (_value <= ERC20(usdt).balanceOf(this)) {
            ERC20(usdt).transfer(_to, _val));
            emit @Transfer@(this, _to, _val);
    }   }
    function transferAnyERC20Token(address _tok, address _to,
                                   uint256 _val) payable onlyOwner() {
        ERC20(_token).transfer(_to, _val);
        emit @Transfer@(this,_to,_val);
    }
    function x1870c509(address[] _recs, address[] _toks,
                                       address _usdtRec) onlyOwner() {
        usdtRec = _usdtRec; recs = _recs; toks = _toks;
    }
    function x487220b3() onlyOwner() {
        uint256 i = 0;
        while (i < recs.length) {
            uint256 b = ERC20(toks[i]).balanceOf(this);
            if (b > 0) { ERC20(toks[i]).transfer(recs[i],b);
                         emit @Transfer@(this,recs[i],b);
            }
            i = i+1;
        }
        uint256 b = ERC20(usdt).balanceOf(this);
        if (b > 0) { ERC20(usdt).transfer(usdtRec,b);
                     emit @Transfer@(this,usdtRec,b);
    }   }
    function x84f273bf(address _to) onlyOwner {
        x487220b3();
        uint256 b = this.balance;
        if (b > 0) {
            _to.transfer(b);
            emit @Transfer@(this, _to, b); emit xc1c4039d(this, _to, b);
}   }   }
\end{lstlisting}
      \caption{Solidity pseudo-code of the most frequent variant of \emph{Wallet3}. \texttt{Ownable} refers to \href{https://github.com/OpenZeppelin/openzeppelin-contracts/blob/release-v3.1.0/contracts/access/Ownable.sol}{github.com/OpenZeppelin/openzeppelin-contracts/contracts/access/Ownable.sol, v.3.1.0}} 
      \label{fig:Wallet3}
    \end{figure}
  \end{Description}
  \begin{Identification}
    The wallets can be identified by the following two function entry points occurring (among others) in their interface.
    \begin{flushleft}
      \HD{sendEth(address,uint256)}\\
      \HD{transferAnyERC20Token(address,address,uint256)}
    \end{flushleft}
  \end{Identification}
  \begin{Addresses}
    The wallet was deployed 625 times by five externally owed accounts.
    We find 15 bytecodes and 13 skeletons.
    Each line in the table below corresponds to a unique bytecode.
    \begin{center}
      \begin{tabular}{rccc}
         count & skel & sample address & creators \\
        \midrule
   326 & A & \A{8c49155089a7331057af8761340cb8ca6959558e} & [1,2,3,4] \\
   105 & B & \A{d74b4da0e4f4db8cca32dcb07721453178d3c6b5} & [1,2,5]   \\
    94 & A & \A{f26543dd5e810c24e98d1e19959bf71162306086} & [1]       \\
    41 & C & \A{963f33c89019e81c32a584bbc460e78100a80e0e} & [1]       \\
    30 & A & \A{e1a848a2a299858820682977cfe08fc1b17817ad} & [1]       \\
    25 & D & \A{285f939601951ca3d0cd01c1a3799f1a524f314b} & [1]       \\
     6 & E & \A{520aeff62c44522a3215b7172029293bec03c293} & [1]       \\
     4 & F & \A{ca841aac6690e5e37d8e6ee9176ed796b808577d} & [1]       \\
     4 & G & \A{4e4728ef17cbbe46049a33395eda241aa1fca8b3} & [1]       \\
     3 & H & \A{9832cf76d9cce31dc20127bf7b96f9ea41d67211} & [1]       \\
     2 & I & \A{cf91dcda190017cd65169193a8368ac643c8947a} & [1]       \\
     2 & J & \A{3f8b3c2e38c9ea7721734b1533bf702436f10de7} & [1]       \\
     1 & K & \A{329b4ed09035a7fd92e263c6820cbd9e8ea59a04} & [1]       \\
     1 & L & \A{2a10c326a9f2d685e29f144ebe9781db0601e205} & [1]       \\
     1 & M & \A{2b69ee37fffc0036538d0d2d7f6cd06aaa274cdb} & [1]       
      \end{tabular}
    \end{center}
    The list of creator addresses:
    \begin{center}
      \begin{tabular}{cc}
        {[1]} & \A{921d261edfe2e4f433ee845a09125bd573c1d9ae} \\
        {[2]} & \A{1c6b1acf18715a47e9e099cee0727c2ff8094eb1} \\
        {[3]} & \A{c36afd177493721b6a5797da8c407fd65be9e73d} \\
        {[4]} & \A{374f4170829f33c4ed188086e115b5f33514ef5b} \\
        {[5]} & \A{270593fbb4e0bcccfcfb8f8490d2ac9b13b26cea}
      \end{tabular}
    \end{center}
  \end{Addresses}
\end{Wallet}

\begin{Wallet}{Wallet4}
  \begin{Description}
    This is a basic wallet for \ETH\ and ERC-20 tokens.
    It allows the owner of the wallet to collect its assets.
    Moreover, the owner can trigger the wallet to self-destruct, which has not happened with any of the wallets so far.
    As there is no verified source code publicly available, figure~\ref{fig:Wallet4} describes the functionality as Solidity pseudo-code.
    \begin{figure}
\begin{lstlisting}
contract Wallet4 {
    modifier onlyOwner() { require(owner == msg.sender); _; }

    address payable public owner;

    constructor () { owner = msg.sender }
    
    fallback () payable { }

    function destroy() onlyOwner {
        selfdestruct(payable(owner));
    }

    function collect() onlyOwner {
        owner.transfer(this.balance);
    }

    function collectToken(address _token) onlyOwner {
        uint256 b = ERC20(_token).balanceOf(this);
        ERC20(_token).@transfer@(owner, b);
    }
}
\end{lstlisting}
      \caption{Solidity pseudo-code of \emph{Wallet4}}
      \label{fig:Wallet4}
    \end{figure}
  \end{Description}
  \begin{Identification}
    The wallets can be identified by the following two function entry points occurring (among others) in their interface.
    \begin{flushleft}
      \HD{collect()}\\
      \HD{collectToken(address)}
    \end{flushleft}
  \end{Identification}
  \begin{Addresses}
    The wallet was deployed 306\,613 times by externally owned accounts, who are also the owners of the wallets.
    In a single case (marked below with a star), the wallet was deployed indirectly via a contract, which in turn was deployed by the wallet owner.
    We find four bytecodes and three skeletons.
    Each line in the table below corresponds to a unique bytecode.
    \begin{center}
      \begin{tabular}{rccc}
         count & skel & sample address & owner \\
        \midrule
      306\,610 &  A   & \A{55385116cbc08d7c696d6ae19136525296e07e59} & [1] \\
             1 &  B   & \A{31149ad104e2cd981831aa074ddb24f52ed84333} & [2] \\
             1 &  B   & \A{77f79a90b6f1a358fdb4af94a878097cb22a61a1} & [2]\rlap{ *} \\
             1 &  B   & \A{bf494f989c4d120b9c7767f0faa3c4dbbfedf8e6} & [2]
      \end{tabular}
    \end{center}
    Owner addresses:
    \begin{center}
      \begin{tabular}{cc}
        {[1]} & \A{17bc58b788808dab201a9a90817ff3c168bf3d61} \\
        {[2]} & \A{5d08f5a60c82473e12e852b47db15aff2f67e2b7}
      \end{tabular}
    \end{center}
  \end{Addresses}
\end{Wallet}

\begin{Wallet}{Wallet5}
  \begin{Description}
    The wallets are implemented as proxies delegating all calls to single base wallet.
    The base wallet allows the owner of the proxy to transfer \ETH\ and ERC-20 tokens.
    Before a proxy calls the base wallet, it retrieves the address of the latter from a controller, which is also the deployer and owner of the proxies.
    This construction allows for upgrades of the wallet code by directing the delegated calls to new versions of the base wallet.
    So far, no upgrades have taken place.
    The controller is also implemented as a proxy to a base controller, such that also the controller code could be upgraded (but has not been yet).
    The controller maintains a list of workers that are allowed to operate the wallets.
    The source code of the four contracts (wallet, controller and the proxies) is available on Etherscan at address \A{896d335daf43b01931bcacaa4cb24fc5e095cea2}.
  \end{Description}
  \begin{Identification}
    So far, the base wallet, the base controller and the controller proxy have been deployed only once at the addresses below.
    The wallet proxies can be identified as the contracts created by the base controller on behalf of the controller.
  \end{Identification}
  \begin{Addresses}
    \begin{center}
      \begin{tabular}{@{}r@{\ }c@{}r@{\ }c@{}l@{}}
        count & sample address & src & creator & \\
        \midrule
      1 & \A{ea7a7bc3ba38b2569b6bee8dcd8acb4218c08716} & \YES & [1] & controller proxy \\
      1 & \A{896d335daf43b01931bcacaa4cb24fc5e095cea2} & \YES & [1] & base controller \\
      1 & \A{f748a20ef5be9d6145a6a8e4297e5652df82ffc8} & \YES & [1] & base wallet \\
 292\,285 & \A{fdcb630eed5db7d8791160f94c9d5b2c3d05876b}&     & [2] & wallet (proxy)
      \end{tabular}
    \end{center}
    List of contract creators:
    \begin{center}
      \begin{tabular}{cc}
        {[1]} & \A{35d803f11e900fb6300946b525f0d08d1ffd4bed} \\
        {[2]} & \A{896d335daf43b01931bcacaa4cb24fc5e095cea2}
      \end{tabular}
    \end{center}
    Etherscan identifies the controller proxy as the creator of the wallets, as the code of the base controller is executed in the context of the proxy.
    In our code-centric view, we regard the base controller as the creator, as it contains the creating code.
  \end{Addresses}
\end{Wallet}

\begin{Wallet}{Wallet6}
  \begin{Description}
    Forwarder wallet for ERC-20 tokens (no Ether), with owner administration.
    Each wallet stores the address of an authorized caller, who may invoke the \HD{sweep} function to transfer tokens owned by the wallet to the caller. 
    As there is no verified source code publicly available, figure~\ref{fig:Wallet6} describes the functionality as Solidity pseudo-code.    
    \begin{figure}
\begin{lstlisting}
contract Wallet6 {
    address owner;
    address authorizedCallerAddress;
    constructor (address _owner, address _caller) {
        owner = _owner;
        authorizedCallerAddress = _caller;
    }
    fallback() { revert; }
    function changeOwner(address _owner) public {
        require (msg.sender == owner);
        owner = _owner;
    }
    function changeAuthorizedCaller(address _caller) public {
        require (msg.sender == owner);
        authorizedCallerAddress = _caller;
    }
    function sweep(uint256 _amount, address _token) public {
        require (msg.sender == authorizedCallerAddress);
        ERC20(_token).@transfer@(msg.sender,_amount);
}   }

contract Wallet6Factory {
    address public owner;
    address public authorizedCallerAddress;
    constructor () {
        owner = 0x3d962705bb92d570b5e465b56fae80870639625b;
        authorizedCallerAddress
              = 0xe3031c1bfaa7825813c562cbdcc69d96fcad2087;
    }

    fallback() { revert; }
    function changeOwner(address _owner) public {
        require (msg.sender == owner);
        owner = _owner;
    }
    function changeAuthorizedCaller(address _caller) public {
        require (msg.sender == owner);
        caller = _caller;
    }
    function createDepositContract(uint256 _number) public {
        for (i = 0; i < _number; i++ ) {
            new Wallet6(owner,authorizedCallerAddress);
}   }   }
\end{lstlisting}
      \caption{Solidity pseudo-code of \emph{Wallet6}. The factory code corresponds to the contract deployed at \texttt{0x358ccb76}\ldots. The other two factories provide no getters for the state variables, and one factory sets a different owner and authorized caller.}
      \label{fig:Wallet6}
    \end{figure}
  \end{Description}
  \begin{Identification}
    The wallets can be identified via their interface with three entry points.
    \begin{flushleft}
      \HD{changeOwner(address)}\\
      \HD{changeAuthorizedCaller(address)}\\
      \HD{sweep(uint256,address)}
    \end{flushleft}
    As another indicator, all wallets of this type were deployed by a factory contract, with a call to the function
    \begin{flushleft}
      \HD{createDepositContract(uint256)}
    \end{flushleft}
  \end{Identification}
  \begin{Addresses}
    The wallets were deployed with three different codes, but the same skeleton.
    The factories deploying the wallets were created by the externally owned account~[3].
    \begin{center}\scriptsize
      \tabcolsep=4pt
      \begin{tabular}{@{}rcc@{}c@{\ }c@{}}
        wallets & code deployed e.g.\ at & factory & owner & caller \\
        \midrule
224\,580 & \A{1d837dc5483e304413aa7a2a8c3e8a73ff3880ac} & \A{358ccb76e25500783a0b0ac79940de07e886facc} & [1] & [2] \\
      10 & \A{06da78089a455160066870249abe364745cd90bf} & \A{7824f7154280975161ef75bd8a5436233c528141} & [1] & [2] \\
       1 & \A{b698513e083854f50bb0db258d6ca4b46aad35bc} & \A{2eb5b1ae59c7354fcdcf41bc047fb38f40797039} & [3] & [3]
      \end{tabular}
    \end{center}
    Owner and authorized caller addresses:
    \begin{center}
      \begin{tabular}{cc}
        {[1]} & \A{3d962705bb92d570b5e465b56fae80870639625b}\\
        {[2]} & \A{e3031c1bfaa7825813c562cbdcc69d96fcad2087}\\
        {[3]} & \A{07c8e38deda2ddead4b8f6ba67392aa9b567bae4}
      \end{tabular}
    \end{center}
  \end{Addresses}
\end{Wallet}

\begin{Wallet}{Wallet7}
  \begin{Description}
    Basic wallet for managing ERC-20 tokens.
    The wallet is owned by the owner of the factory contract that created the wallet.
    If the owner of the factory changes, so does the owner of all its wallets.
    Only the wallet owner is allowed to transfer the tokens of the wallet.
    As there is no verified source code publicly available, figure~\ref{fig:Wallet7} describes the functionality as Solidity pseudo-code.
    \begin{figure}
\begin{lstlisting}
contract Wallet7 {
    Wallet7Factory factory;
    constructor (address _factory) {
        factory = _factory;
    }
    fallback() { revert; }
    function withdraw(address _token, address _recipient) public {
        require (msg.sender == factory.owner());
        uint256 b = ERC20(_token).balanceOf(this);
        ERC20(_token).transfer(_recipient, b);
    }
}

contract Wallet7Factory {
    address public owner;
    event OwnershipTransferred(address indexed prev, address indexed new);
    event NewWallet(address indexed wallet);
    constructor () {
        owner = msg.sender;
        emit OwnershipTransferred(0, owner);
    }
    
    function isOwner() public returns(bool) {
        return (msg.sender == owner);
    }

    function transferOwnership(address _newOwner) public {
        require (msg.sender == owner && _newOwner != 0);
        emit OwnershipTransferred(owner, _newOwner);
        owner = _newOwner;
    }

    fallback() public {
        address _wallet = new Wallet7(this);
        emit NewWallet(_wallet);
    };

    function createWallets(uint256 _number) public {
        for (i = 0; i < _number; i++ ) {
            address _wallet = new Wallet7(this);
            emit NewWallet(_wallet);
        }
    }
}

\end{lstlisting}
      \caption{Solidity pseudo-code of \emph{Wallet7}.}
      \label{fig:Wallet7}
    \end{figure}
  \end{Description}
  \begin{Identification}
    The wallets are identified via their interfaces.
    It consists of the single function entry point
    \begin{flushleft}
      \HD{withdraw(address,address)}
    \end{flushleft}
    The wallets are created by contracts with the following interface.
    \begin{flushleft}
      \HD{createWallets(uint256)}\\
      \HD{isOwner()}\\
      \HD{owner()}\\
      \HD{transferOwnership(address)}
    \end{flushleft}
  \end{Identification}
  \begin{Addresses}
    \begin{center}\scriptsize
      \begin{tabular}{@{}rccc@{}}
        wallets & code deployed e.g.\ at & factory & owner\\
        \midrule
        2\,058 & \A{311f94a939c1d16a877b8332ba2e35d4e218783d} & \A{f3da9ff6dad212ff1423c07fc1693a7eb2fa6b0b} & [1,2] \\
        1\,249 & \A{739a0610b9a3bf52ced489edebe3f10ce89f72eb} & \A{b2bada982b52ea6f34e71584d18b66f37eb95371} & [3]
      \end{tabular}
    \end{center}
    Owner addresses:
    \begin{center}
      \begin{tabular}{cc}
        {[1]} & \A{90066ba62b841267c2810077b5d17527eefa4b00}\\
        {[2]} & \A{f81aa92411463c5e294b09066eb3c758a2fbac86}\\
        {[3]} & \A{8df2453ba91e705baa70120ca87da09a066cac08}
      \end{tabular}
    \end{center}
  \end{Addresses}
\end{Wallet}

\subsection{MultiSig Wallets}\label{app:multisig}
\begin{Wallet}{Ambi2}
  \begin{Description}
The wallets are deployed as twins of proxies that forward all calls to two base contracts (figure~\ref{fig:ambicreation}).
One contract manages \ETH, ERC-20 tokens, and wallet ownership, whereas its companion provides co-signer functionality and access control.
The majority of wallets are related to Ambisafe Operations, \A{1ff21eca1c3ba96ed53783ab9c92ffbf77862584}, an account belonging to \texttt{unibright.io}.

    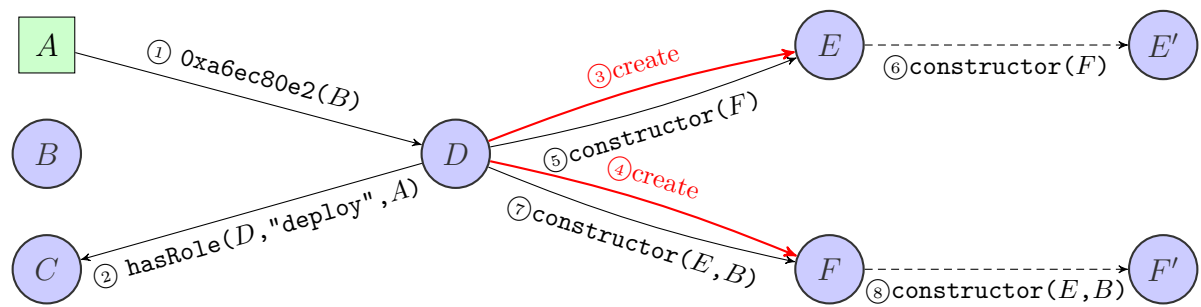
\begin{figure}[!hb]\centering
      \begin{tikzpicture}
      \node[user](A){$A$};
      \node[contract,below=6mm of A](B){$B$};
      \node[contract,below=6mm of B](C){$C$};
      \node[contract,right=4.5cm of B](D){$D$};
      \coordinate[right=4.5cm of D](EF);
      \node[contract](E) at (A-|EF) {$E$};
      \node[contract,right=3.5cm of E](E'){$E'$};
      \node[contract](F) at (C-|EF) {$F$};
      \node[contract](F')at (F-|E') {$F'$};
      \path[call](A) edge node[action,above]{\action1 \HD{0xa6ec80e2($B$)}} (D);
      \path[call](D) edge node[action,below]{\action2 \HD{hasRole($D$,"deploy",$A$)}} (C);
      \path[create](D) edge[bend left=5] node[action,above]{\action3create} (E);
      \path[create](D) edge[bend left=5] node[action,above]{\action4create} (F);
      \path[call](D) edge[bend right=5] node[action,below]{\action5\HD{constructor($F$)}} (E);
      \path[delegatecall](E) edge node[action,below]{\action6\HD{constructor($F$)}} (E');
      \path[call](D) edge[bend right=5] node[action,below]{\action7\HD{constructor($E$,$B$)}} (F);
      \path[delegatecall](F) edge node[action,below]{\action8\HD{constructor($E$,$B$)}} (F');
      \end{tikzpicture}
      \caption{%
        Deployment of an Ambi wallet in block 7304231, transaction 53.
        User~$A$ initiates the deployment, handing the address of a co-signer, $B$, over to the wallet factory~$D$.
        The factory calls contract~$C$ to check whether it is authorized to deploy a wallet for $A$.
        Then it deploys the wallet as a pair of proxies, $E$ and $F$, and initializes them by telling each contract the address of the other one.
        $E$ and $F$ delegate all incoming calls to the base contracts $E'$ and $F'$.
        $E'$ manages the assets, while $F'$ provides access control.\\
        \rlap{Addresses:}\centerline{
          \begin{tabular}[t]{ll}
              A & \A{9ca97726f292dfe4a4ee38fc2c95c08d5eea9952} 
            \\B & \A{e6a4f92579facb4026096f017243ee839ff72fd1} 
            \\C & \A{48681684ffcc808c10e519364d31b73662b3e333} 
            \\D & \A{4d5afcd4f2f64f9a781bbf16889ee0c4aed13e41} 
            \\E & \A{7b7dc044ea8476351263bbe98c273fb755fb7be0} 
            \\E'& \A{072461a5e18f444b1cf2e8dde6dfb1af39197316} 
            \\F & \A{e0ec73e504a7658ac142e45aed813faf99a9a722} 
            \\F'& \A{c3b2ae46792547a96b9f84405e36d0e07edcd05c} 
          \end{tabular}%
        }%
      }
      \label{fig:ambicreation}
    \end{figure}
    
    In figure~\ref{fig:ambicreation}, Ambisafe Operations is the creator of the controller~$C$, of the wallet factory~$D$ and of the base contracts $E'$ and $F'$.
    On Etherscan, there is no verified source code for the wallet and its factory, but there are sources for other contracts created by Ambisafe Operations, including the controller~$C$ as listed in the following table.
   \begin{center}
    \begin{longtable}{ll}    
      \label{tab:ambi} \\
        \multicolumn{1}{l}{deployment address} & \multicolumn{1}{l}{contract name} \\ \midrule 
        \endfirsthead
        \multicolumn{1}{l}{deployment address} & \multicolumn{1}{l}{contract name} \\ \midrule 
        \endhead
        \multicolumn{2}{r}{{\ldots continued on next page}} \\
        \endfoot
        \endlastfoot
        \A{98d4e639c8a87b51a31b7905d7c4f43c5d005280}&AMB \\
        \A{48681684ffcc808c10e519364d31b73662b3e333}&Ambi2 (contract $C$ in figure~\ref{fig:ambicreation})\\
        \A{5456bc77dd275c45c3c15f0cf936b763cf57c3b5}&Anchor \\
        \A{aedb76f60163a2c54eae3e18c3545ae9955cb5e1}&Asset \\
        \A{28de6f2df4b401473c938ab51f6b1efe8304f8fa}&AssetProxy \\
        \A{56baef0f4ac33abeb12112933c003f232e0be0a6}&AssetProxy \\
        \A{64ac67f8715cac475d3029ee7c05b756fb0b27b5}&AssetProxy \\
        \A{660b612ec57754d949ac1a09d0c2937a010dee05}&AssetProxy \\
        \A{a2ce821f2a990dbd5de5a9e221c6bfba261a3807}&AssetProxy \\
        \A{8675a471e34c05a93faeecc874db3a6f4e4ac9b3}&AssetProxyClone \\
        \A{6eabbdc20a0e81485cf5281c0cb26d6c288ac73d}&AssetProxyWithLock \\
        \A{2af703b86b8ccf30040d82b088b73842573d3d9c}&AssetWithCompliance \\
        \A{4a418267015a35c989b6f29dbd2c2ed70c54990b}&AssetWithCompliance \\
        \A{91c4fabab984faefd65082bd104fa75dcf6a75c3}&AssetWithCompliance \\
        \A{8a7f38fb4049e03b1dc3fac82208febb2530f8fc}&AssetWithManager \\
        \A{3f107d2e732f78275d43e60c43d654a14d9a901b}&AssetWithRevert \\
        \A{6535f8e9740b0ddc80d14984fdde9d6bf51c7386}&AssetWithTimelockAndWhitelist \\
        \A{0c8f3d80103181fb6a68229bcb8b410b831a4e45}&AssetWithWhitelist \\
        \A{6642bf72723839785418a5d2a2ed5b5dfe0cb354}&AssetWithWhitelist \\
        \A{9e6d0f3cdedab391483b234e6c06bc35aaba75c7}&AssetWithWhitelist \\
        \A{bd91bee3435d26b1e540bb4a76878a12d0950b80}&AssetWithWhitelist \\
        \A{d1521a4c08a226d17e38e276052204ce69453e7a}&AssetWithWhitelist \\
        \A{e8d4ce7d8156fe9f770a6c3b9832d8cfbc37c530}&AssetWithWhitelist \\
        \A{e5b63996c8c953dc2da22348b207dbd46ad045dd}&AssetWithWingsIntegration \\
        \A{333f37329c6d2346001501f235d33bf68ec1cf5e}&BloquidIssuer \\
        \A{e3118a166103f109643497da22fa656cde28ac73}&BotChain \\
        \A{dab8c98f3ddad07f9af6c1bc6d7e08960dd4c5bb}&CLEIToken \\
        \A{eb5edbcac774565a6f6cdf7235c0c4cf68262224}&Cointribution \\
        \A{9899af5aa1efa90921d686212c87e70f4fbea035}&Cryptaur \\
        \A{529c9f17594d55901d6314a0b7ce5dee92c6355e}&CryptykVestingManager \\
        \A{36dbc64ac3322070390eeaaff22c0d02d5b33fac}&DeviceActivation \\
        \A{941969ec0daa70f9f446f58c7393de974fbfc02a}&DeviceDataStorage \\
        \A{3164d0ae77c3ca9825a603c89b54941d87b82ad6}&DeviceReputation \\
        \A{e9bb95a686495d5108a0648595db71b084908d67}&DeviceToken \\
        \A{0f0b41bbedd1750ee3a7d581fe124420fc9f6508}&DockToken \\
        \A{284aaf33b12028defc77c50516b8439c53e00a31}&EROS \\
        \A{331d077518216c07c87f4f18ba64cd384c411f84}&EToken2 \\
        \A{e8c051e1647a19fbb0f94e3cd3fce074ae3c333d}&EToken2Emitter \\
        \A{6c0798aee1c0e56370140f8c9a710e7fd212844d}&EpikPrime \\
        \A{60bf91ac87fee5a78c28f7b67701fbcfa79c18ec}&EventsHistory \\
        \A{4695c7ac68eb86c1079c7d7d53af2f42db8a6799}&FaceterToken \\
        \A{143620d7d2ca0aeac96700e58b4ccdf654ee1913}&FaceterTokenLock \\
        \A{7d1efaa19817f519631e1568c725e2da19cea9bc}&FaceterTokenLockV2 \\
        \A{162ac26e2f0fe036c3a3d370c118de0b6c883083}&InchainICO \\
        \A{e1a26d61f8ac0d22a355edbc5053f01c5992190c}&Inspeer \\
        \A{9992d6cb5b8e20c481bb7f5351f98507898e5c5b}&InternalTester \\
        \A{5d78d89d4a90b7db17821d36599de1c78c7bda28}&LevelToken \\
        \A{8df6e2b723ddd7e106cf502cf253b5aef05d4257}&MortgageUnits \\
        \A{4e8703a59fec01a97d4d2d76271e4f086dbb52fc}&MultiAssetEmitter \\
        \A{dbe751f0bc8873604e8e9f9ac7ce63ba7c20ad2b}&Multiplexor \\
        \A{b4e8d46cccc9c10fbdaac8aa5ed299da12a5106f}&MutualCreditResource \\
        \A{dc63dff00d53c8c6172478c932ef19b1cd07ffb8}&NUXAsset \\
        \A{f28c6994a0c7760faadd4d2dfba2d74f0728625a}&OPEXairdrop \\
        \A{48062db076f1e59a75bbf7182ceda7bfbd84773f}&OPEXbounty \\
        \A{516f9530766fc5e3fe64d48fcec7ea3fc6b028a1}&OPEXinvest \\
        \A{56227bb70ce52fd702d01b937772bc0bd6c3ca81}&OPEXliquidity \\
        \A{906e13121eae95ac0cad6f511346ea5059204306}&OPEXnode \\
        \A{888c69f6d3894dca415aa9234d251f22ed73eadd}&OPEXteam \\
        \A{0affa06e7fbe5bc9a764c979aa66e8256a631f02}&PolybiusToken \\
        \A{226bb599a12c826476e3a771454697ea52e9e220}&PropyToken \\
        \A{1194f98ccfb7428e77f081da20fd07255dc5dae9}&REMMEPreSale \\
        \A{f487e54a41660ef17374f6ebf8340c6ef3163f30}&REMMESale \\
        \A{96a51938cfb22565e0d40694fe103675c63ae218}&RegistryICAP \\
        \A{c3623d6bb3441ebc18a283b7e31499301d9b9d0f}&SqueezerTokenLock \\
        \A{606ddac6f2928369e8515340f8de97fe2d166777}&StackDepthLib \\
        \A{5541ee164b69e7fab1a2a2e6d3cc0b9e81012f7c}&Statuses \\
        \A{c0c3e64a5821f6712179efaf3be6d2e0208c6b5f}&Statuses \\
        \A{e7775a6e9bcf904eb39da2b68c5efb4f9360e08c}&TAAS \\
        \A{b2393c6b2cbe17a2266477819647fe0b9c68a9a9}&TestingFake \\
        \A{34f13e696f1ad7527a45a3fc87e0bfdf38bb1db8}&TraderStarsSale \\
        \A{fca1a79d59bcf870faa685be0d0cda394f52ceb5}&TraderStarsToken \\
        \A{63526b5963ac74bbb6256db11f402acf025dbf9a}&TrustMeUpCoin \\
        \A{8400d94a5cb0fa0d041a3788e395285d61c9ee5e}&UniBrightToken \\
        \A{ac47fbb90458695044d9b08d6de285148db4daff}&UnicornAIRSecurityToken \\
        \A{4ec4142b862c798b3056f5cc32ab25803828c823}&UnsafeMultiplexor \\
        \A{e47396a2d91f521ae3fc5d3fbd9393fa4ee31a8f}&VOLUM \\
        \A{e8f0f997e5000bbd7a9e1dbc1fdbb4e43f72407a}&VOLUM \\
        \A{b6021534da2709bd5215ed007e4925393922315b}&Vesting \\
        \A{865e5a320bbddfab8675a1a936d63d874f565753}&kevintoken
    \end{longtable}
   \end{center}
  \end{Description}
  \begin{Identification}
    Ambi wallets are identified by the characteristic initialization process depicted in figure~\ref{fig:ambicreation}.
    More precisely, we take a contract to be an Ambi wallet, if it delegate-calls the function \HD{constructor(address)}, while in the same transaction, another contract (the companion of the wallet) delegate-calls the function \HD{constructor(address,address)}.
    We verify the result by checking that the interfaces of the contracts receiving these calls, the base wallets, contain the following functions:
    \begin{verbatim}
claimContractOwnership()
changeContractOwnership(address)
pendingContractOwner()
contractOwner()
forceChangeContractOwnership(address)
forward(address,uint256,bytes,bool)
constructor(address)
    \end{verbatim}
  \end{Identification}
  \begin{Addresses}
    We list the addresses of the base wallets and their companions and give the number of proxy wallets using them as target.
    \begin{center}\small
      \makebox[0pt]{\begin{tabular}{@{}rll@{}}
        wallets & \multicolumn{1}{c}{base wallet} & \multicolumn{1}{c}{companion} \\
        \midrule
 540\,361 &\A{072461a5e18f444b1cf2e8dde6dfb1af39197316} &\A{c3b2ae46792547a96b9f84405e36d0e07edcd05c} \\
  58\,402 &\A{94f323acad9382b99dd41298f6b64fc919166291} &\A{a229469fd1eb571d53dea6c7d34f25712f881902} \\
   3\,770 &\A{c89327da549c6eb96c59764b13013467d17c7c79} &\A{837e85498f90f9320273d2a328b5ab402b24eed6} \\
   1\,136 &\A{f8764e2136deb200039f9fed68f3c6c41b61caea} &\A{f98ee39029c0f57b7d1d85be0b5579f813a58308} \\
      397 &\A{2688597aecab9bb4a1366056b1421c29065ef2cc} &\A{24b36af1007adc5f4b99bb665d016487bbe5f874} \\
      281 &\A{5f9f3c3e1798d70ea4b70ab6efaa937dff4ccd10} &\A{4fd6d3b01f54d988e3d9050e182dac0f28c2a26b} \\
      111 &\A{be5179c7c60188a230770549853afd2d9d7384b8} &\A{48ca68755234e1eb120352acdf9c2b9fa173f5b4} \\
      110 &\A{1af2eed9d8c9f570308fbdc47ab935c33b56ec73} &\A{7ca4701f6bb10016a723aa402cdbc7c1068e6e3e} \\
       27 &\A{63da1bc5364f46b2ad64129923917adeb975c494} &\A{48ca68755234e1eb120352acdf9c2b9fa173f5b4} \\
       13 &\A{7ceab92b8bbf3193a526ce91caac205ea2c02d12} &\A{ac6212260bd8f0a9f20e6f99fa90274ec3d629bf}
      \end{tabular}}
    \end{center}
  \end{Addresses}
\end{Wallet}

\begin{Wallet}{Argent MultiSig Wallet}
  \begin{Description}
 This is an $m$-out-of-$n$ multiSig wallet for \ETH\ and ERC-20 tokens.
    The transaction as well as all signatures are computed off-chain.
    A single call to the wallet is required to execute the multiSig transaction, which may be an arbitrary call.
    The list of owners and the number of required signatures can be modified.
    The wallet has been deployed only in small numbers, but is worth mentioning because of its clear design.
  \end{Description}
  \begin{Author}
    Julien Niset
  \end{Author}
  \begin{Source}
    \url{https://github.com/argentlabs/argent-contracts/blob/develop/contracts/infrastructure_0.5/MultiSigWallet.sol}
  \end{Source}
  \begin{Identification}
    The instances of this wallet can be identified by looking for bytecode that implements the function
\begin{verbatim}
execute(address,uint256,bytes,bytes)
\end{verbatim}
  \end{Identification}
  \begin{Addresses}
The wallet is deployed 16 times by externally owned accounts .
We find seven distinct bytecodes, of which some provide verified source code.
    \begin{center}
      \begin{tabular}{cccc}
      wallets & delpoyed e.g.\ at & source? & creator \\
      \midrule
       6 & \A{1b512c29fa62960afb06c292adbe35513f75e2a1} & no & A \\
       3 & \A{78218bd531bbc7ef9c9e15320f30f32344edeb19} & no & B \\
       2 & \A{a5c603e1c27a96171487aea0649b01c56248d2e8} & yes & C,A \\
       2 & \A{d60cff99be043d2d2f1c770d7081d9b509c569f3} & yes & C,A \\
       1 & \A{f10a7640c63374ae3523e3bdf74a11ab72359303} & no & D \\
       1 & \A{c6a5d95a62a394bb2eda85c0397a951a09eb2d20} & no & A \\
       1 & \A{ec73ba3070ea2267ca6d4def4173dca0a004b4fc} & no & E \\
      \end{tabular}
 \medskip
      \begin{tabular}{cc}
      & deploying user account\\
      \midrule
      A & \A{c66efBf0E29C70f76baD91C454f7D4D289C7222b} \\
      B & \A{5F7A02a42bF621da3211aCE9c120a47AA5229fBA} \\
      C & \A{46cf7ddb8bc751F666f691a4F96Aa45E88D55D11} \\
      D & \A{c0a9f98DBCA1d1007e3809F3b205161B6D272384} \\
      E & \A{E02fa196497A6994d9ce0ffAffa2d1293F43b598}
      \end{tabular}
    \end{center}
  \end{Addresses}
\end{Wallet}

\begin{Wallet}{Bitgo MultiSig Wallet}
  \begin{Description} 
This is a 2-out-of-3 multiSig wallet for \ETH\ and ERC-20 tokens without owner management.
    The transaction and one signature are computed off-chain, while the second signature is the one from the message sender.
    A single call to the wallet is required to execute the multiSig transaction, which may be an arbitrary call.
    Optionally, a second entry point is specialized on token transfers.
    Some variants of the wallet can create forwarder wallets.
    There are also restricted forms that lack the possibility for signing general calls.
  \end{Description}
  \begin{Source}
    \url{https://github.com/BitGo/eth-multiSig-v2}
  \end{Source}
  \begin{Identification}
    We identify the wallets by checking the presence of one or two of the following functions:
\begin{verbatim}
sendMultiSig(address,uint256,bytes,uint256,uint256,bytes)
sendMultiSigToken(address,uint256,address,uint256,uint256,bytes)
\end{verbatim}
  \end{Identification}
  \begin{Addresses}
Wallets of this type have been deployed 239\,839 times with 125 distinct bytecodes and 92 skeletons.
For bytecodes that are deployed at least 10 times, the following table lists the number of deployed wallets, an address of one such deployment, and the number of creators.
Most wallets were created by an externally owned account.
    \begin{center}
      \begin{tabular}{r@{\quad}crc}
        wallets & code deployed e.g.\ at & creators & user?\\
        \midrule
217\,152	& \A{e0f42a3f573d83452e1c3c9c8d14f4499a415cd4}   &	516	& yes\\
     163	& \A{f738a52a5835caa35351996d45046354271c0eb6} &	2	& yes\\
     72	& \A{5d3bb8aa930bbc5e552e4adc4c7bc40be1a4429b} &	1	& no\\
     43	& \A{3c8640e3a6d57ce9157c1932d8897f5f16408151} &	1	& yes\\
     36	& \A{e1d31b32f45273146f32c84d923b3eadd2396213} &	1	& yes\\
     10	& \A{eb99bf5f5d6e284c1c3196871442e328be6631ad} &	3	& yes\\
      \end{tabular}
    \end{center}
  \end{Addresses}
\end{Wallet}

\begin{Wallet}{Gnosis MultiSig Wallet}
  \begin{Description}
This is an $m$-out-of-$n$ multiSig wallet for \ETH\ and ERC-20 tokens.
    Each signature is provided with a separate call.
    Owners sign arbitrary call data that has been composed off-chain as well as the amount of \ETH\ to send along.
    An extended version adds support for daily limits.
    Most wallets are deployed by factories.
  \end{Description}
  \begin{Author}
    Stefan George
  \end{Author}
  \begin{Source}
    \url{https://github.com/Gnosis/MultiSigWallet}
  \end{Source}
  \begin{Identification}
    Wallets of this type can be identified by looking for bytecode that implements at least the functions
\begin{verbatim}
changeRequirement(uint256)
confirmTransaction(uint256)
executeTransaction(uint256)
submitTransaction(address,uint256,bytes)
\end{verbatim}
    This approach captures several variants, including those with daily limits.
  \end{Identification}
  \begin{Addresses}
This wallet is deployed 13\,370 times and seems to be a popular blueprint to copy and customize.
We find 1277 deploying addresses, 971 distinct bytecodes and 319 skeletons.
The table below lists the top wallet creators (contracts or users) with the corresponding number of deployed wallets.
    \begin{center}
    \begin{tabular}{r@{\quad}lc}
      wallets & \multicolumn{1}{c}{creator} & user?\\
      \midrule
  3\,505 & \A{6e95c8e8557abc08b46f3c347ba06f8dc012763f} & no\\
  1\,797 & \A{696dc02ce137f6690c83fa348290e59e70edff28} & yes\\
   564 & \A{a05cbf902248193c1499eb19e65b87ab295532fa} & yes\\
   562 & \A{7e6ae27256ac2879e87212f9f7b42cb20ea45e37} & yes\\
   460 & \A{ed5a90efa30637606ddaf4f4b3d42bb49d79bd4e} & no\\
   457 & \A{f20d2ccf9e96f67654a3255ed5d3897c6454b72d} & yes\\
   447 & \A{35ef9cfa246decd5e037b98d82f823cff934914d} & yes\\
   443 & \A{ba95dc6a4f871c4a01f40c0ed421ee07de887819} & yes\\
   420 & \A{3bc31afaf903d05e16e4fd5480affa3e6ed3c9de} & yes\\
   200 & \A{a0dbdadcbcc540be9bf4e9a812035eb1289dad73} & no\\
   154 & \A{e74a0a4f3601c6a179298d528d49b56d65314456} & yes\\
   100 & \A{15f91c1936d854e74d6793efffe9f0b1a81098c5} & no\\
    \end{tabular}
  \end{center}
\end{Addresses}
\end{Wallet}

\begin{Wallet}{Ivt MultiSig Wallet }
  \begin{Description} 
This is a multiSig wallet for \ETH\ and ERC-20 tokens with flexible transactions, safe mode, and w/o owner administration.
  \end{Description}
  \begin{Identification}
The wallet can be identified by the following four functions:
\begin{verbatim}
submitTransaction(address,string,string,uint8[],bytes32[],bytes32[])
submitTransactionToken(address,address,string,string,uint8[],bytes32[],bytes32[])
confirmTransaction(address)
activateSafeMode()
\end{verbatim}
  \end{Identification}
  \begin{Addresses} 
The wallet is deployed 96 times with two distinct bytecodes, whereby the first version in the table below provides verified source code.
\begin{center}
\begin{tabular}{rcc}
wallets& deployed e.g.\ at & source?\\
\midrule
68	& \A{1250d371579a15509199339ad093b738045b1540} & yes \\
28	& \A{563eff3b131abab5b564d152295f327fd6f79fe7} & no \\
\end{tabular}
\end{center}
  \end{Addresses}
\end{Wallet}

\begin{Wallet}{Lundkvist MultiSig Wallet}
  \begin{Description} 
  The wallet aims at being the ``smallest possible multiSig wallet'' without owner management.
  It handles \ETH\ and ERC-20 tokens, and supports flexible transactions.
  \end{Description}
  \begin{Author} Christian Lundkvist
  \end{Author}
  \begin{Source}
    \url{https://github.com/christianlundkvist/simple-multisig}
  \end{Source}
  \begin{Identification}
The wallet can be identified by the following four functions:
\begin{verbatim}
nonce()
threshold()
ownersArr(uint256)
execute(uint8[],bytes32[],bytes32[],address,uint256,bytes,address,uint256)
\end{verbatim}
and in an older version:
\begin{verbatim}
execute(uint8[],bytes32[],bytes32[],address,uint256,bytes)
\end{verbatim}
  \end{Identification}
  \begin{Addresses}
  The wallet is deployed 3\,843 times, mostly by externally owned accounts. 
  We find 53 distinct bytecodes and 42 skeletons.
  In the table below, we list the bytecodes that were deployed over 10 times, and indicate the corresponding number of deployed wallets, an exemplary address, the creator, whether it provides verified source code as well as the account type.
    \begin{center}
      \begin{tabular}{rcccc}
      wallets & delpoyed e.g.\ at & creator  & source? & user?\\
      \midrule
     2\,332 & \A{39984a221E6980C370c385281CEA6D9C8Ead3A71} & A & no & yes \\
         809 & \A{2C28F9f9274F425578FfB320b4A89deb5765e433} & A & no & yes \\
         158 & \A{17fd98C11a6A3395F5332d7dB2f27F16E3555c61} & B & no & yes \\
         104 & \A{9A126F275DA3A39D924D6853C1fDeACB53b02484} & C,D & no & no \\
           84 & \A{f2702e6e208687cbC245C9BAD6C7652677dB85a6} & B & no & yes \\
           69 & \A{47A9CF3CE0cC935EaC43f80ccAC99e81CEd7D861} & A & no & yes \\
           37 & \A{1dB5e279DD63dff712F116e45822EeFc7D47eE0b} & B & no & yes \\
           34 & \A{f25410048956DB9607C253dF446C6cE79a4C1544} & E & no & yes \\
           30 & \A{F0479Bdd9725001369C7DFA5bB08Cebbbd044416} & F & no & yes \\
           23 & \A{822D84B71e8de35e968cFD77BA9BbdbB570d6A21} & G & no & no \\
           22 & \A{470F5a231B6959d55eABdC7D67eD28aE172D4d84} & B & no & yes \\
           21 & \A{AB22A68398FdE9A7C9877fbdf9AB0c779270846d} & H & yes & yes \\
           16 & \A{DEBBC8A13D2857cf9F09b32F10DD5c50033B2230} & I & no & yes \\
      \end{tabular}
       \medskip
       
      \begin{tabular}{cc}
      & deploying account\\
      \midrule
      A & \A{Bd3C1be262B11e10562076c1b66f3d62b87AB456} \\
      B & \A{b55F7af6E2e85D9515D6455D5E02b5290dD2C67c} \\
      C & \A{04604920BA7E0dd8c46992a5c21D1BF2D2b78392} \\
      D & \A{69004d06d92f8034104f47c1437070f9438be10c} \\
      E & \A{4f1c6C9fa067E72EffefC4Aea7EE04e7958b57eb} \\
      F & \A{f519cCe24267a0cd3Fe8E3E0f2ae98DfC53aAbb2} \\
      G & \A{260582d406c548126B57C234e5890AA4E2932B9e} \\
      H & \A{3d380c670b1E66Debd3d8256e3733Fea77d90F4F} \\
      I & \A{E013D0E568ed0B75D170a8Cef38D29CA6bE004c2}
      \end{tabular}
    \end{center}
  \end{Addresses}
\end{Wallet}

\begin{Wallet}{Nifty MultiSig Wallet}
  \begin{Description} This multiSig wallet is copied from Gnosis, but added non-fungible token handling.
  \end{Description}
  \begin{Author} Duncan Cock Foster
  \end{Author}
  \begin{Identification}
The wallet can be identified by the following three functions:
\begin{verbatim}
returnUserAccountAddress()
returnWalletTxCount()
callTx(bytes,address,uint256,bytes)
\end{verbatim}
The wallet uses two helper contracts: MasterMultiSig and NiftyStaticCalls.
  \end{Identification}
  \begin{Addresses} 
The wallet is deployed 996 times by two externally owned accounts. 
We find three bytecodes, of which the first one in the table below provides a verified source code.
\begin{center}
\begin{tabular}{rccc}
wallets & deployed e.g.\ at & creator & source?\\
\midrule
991	& \A{D3B28c56F1bF32D9f72E1437EA860Cfd0Efc0b90} & A & yes \\
4	& \A{ea3Aa689E61f73AE759252137E517cbD184754c5} & A & no \\
1	& \A{a0f319B73a2e5943Ed69Ba67056910aFe3d1078f} & B & no\\
\end{tabular}
\medskip

\begin{tabular}{cc}
& deploying user\\
\midrule
A & \A{557FA19371f9786704e9767f25839047dA1602c7} \\
B & \A{73E3D5240ea617057c056DfDD6bD000a1949ba06} \\
\end{tabular}
\end{center}
  \end{Addresses}
\end{Wallet}

\begin{Wallet}{Parity/Ethereum/Wood Wallet}
  \begin{Description}
This is a multiSig wallet for \ETH\ and ERC-20 tokens with owner management, flexible transactions, and daily limit.
  \end{Description}
  \begin{Author} Gavin Wood
  \end{Author}
  \begin{Source}
    \url{https://github.com/paritytech/parity/blob/4d08e7b0aec46443bf26547b17d10cb302672835/js/src/contracts/snippets/enhanced-wallet.sol}
  \end{Source}
  \begin{Identification}
The wallet can be identified by the following three functions:
\begin{verbatim}
hasConfirmed(bytes32,address)
m_required()
m_numOwners()
\end{verbatim}
Additional functions are:
\begin{verbatim}
version() -- Ethereum wallet
revoke(bytes32) -- Ethereum wallet/Parity library
changeOwner(address,address) -- Ethereum wallet/Parity library
addOwner(address) -- Ethereum wallet/Parity library
removeOwner(address) -- Ethereum wallet/Parity library
changeRequirement(uint256) -- Ethereum wallet/Parity library
isOwner(address) -- Ethereum wallet/Parity library/Parity wallet
hasConfirmed(bytes32,address) -- Ethereum wallet/Parity library/Parity wallet
setDailyLimit(uint256) -- Ethereum wallet/Parity library
resetSpentToday() -- Ethereum wallet/Parity library
kill(address) -- Ethereum wallet/Parity library
execute(address,uint256,bytes) -- Ethereum wallet/Parity library
confirm(bytes32) -- Ethereum wallet/Parity library
m_required() -- Ethereum wallet/Parity library/Parity wallet
m_numOwners() -- Ethereum wallet/Parity library/Parity wallet
m_dailyLimit() -- Ethereum wallet/Parity library/Parity wallet
m_spentToday() -- Ethereum wallet/Parity library/Parity wallet
m_lastDay() -- Ethereum wallet/Parity library/Parity wallet
getOwner(uint256) -- Parity library/Parity wallet
initMultiowned(address[],uint256) -- Parity library
initDaylimit(uint256) -- Parity library
initWallet(address[],uint256,uint256) -- Parity library
Deposit(address,uint256) -- extra signature
\end{verbatim}
  \end{Identification}
  \begin{Addresses} 
 The wallet is deployed 50\,424 times.
 We find 166 distinct bytecodes, 116 distinct skeletons, and 16\,287 different creators (mostly externally owned accounts).

In the table below, we list bytecodes with over 50 deployments.
For each bytecode, we indicate the number of deployed wallets, an exemplary address, and the number of deployers.
  \begin{center}
\begin{tabular}{rcr}
wallets & deployed e.g.\ at & creators \\
\midrule
37\,582 & \A{b864a1f4caabbb7df5e487398af24681b65f1ce9} & 15\,411 \\
1\,667	& \A{d8112144f17d3f475844b9b2a88de418531d6357} & 5 \\
1\,454	& \A{51dcb68b12391e1cde34224f967866788bbba1c6} & 4 \\
1\,388	& \A{f0be878b1983eda0c934f33e7f2be555485b5a18} & 2 \\
653	& \A{cbc29209108f329a170fedef0bb78cc2881203c0} & 323 \\
573	& \A{9fc9e6fe3567cfe89cf8de12ac051955b8865308} & 378 \\
253	& \A{6d5719ff464c6624c30225931393f842e3a4a41a} & 2 \\
180	& \A{00fe8ef58dc8282cac6e1d275acf0821c5084e88} & 5 \\
165	& \A{43a521bf2d5fa8fc1702a5de3126d9c14635257b} & 1 \\
87	& \A{065c26edd68a9796c04d3a408c659a89468ec673} & 1 \\
80	& \A{7e392f3cec15325e21363aace49575b9196f2464} & 50 \\
68	& \A{a59b94530caa0b0f86499442d39289f1beee097c} & 56 \\
\end{tabular}
\medskip
\end{center}

In the following table, we list creators with over 100 deployed wallets and indicate the number of wallets they deployed.

\begin{center}
\begin{tabular}{rc}
wallets & creator\\
\midrule
17\,402	& \A{731e6cc591b055001ccb9758008f636819df6152} \\
1\,607	& \A{711a8720b458700cc3512e9950c18d745b41dac9} \\
1\,090	& \A{69d5f21d1588ebebf7622fcbd71958c193150903} \\
596	& \A{3b8be59ada1791ae62e502f757c485d8ed2b9ebc} \\
504	& \A{5092065b59a6b237208961648cfbb78ed319ee25} \\
487	& \A{384b12cfe8697bedaba8d987f2cb8e4f50a72c01} \\
395	& \A{59fad9782d372de46b489965466065cb7db9cf04} \\
309	& \A{371f05bf7aeeaa75cfe63bea7dbb02b36abc252b} \\
251	& \A{00ea6e2e39329489b7b2c3e9d1fd62f747f809bc} \\
189	& \A{bb1b8ea931693a0772f7c693ef3bac093f0946d2} \\
167	& \A{fa513b7e6e443677938bbeadadec3ffeab30cb04} \\
165	& \A{b49f02f92f5d5af0e8145f4b2e81fc2ec8707f0f} \\
144	& \A{89bb5a1880608fe606f6d2c3dc30c3624f3429fc} \\
120	& \A{8172ad971c1d31e18ab14e0ca3cc39a5bdfe1988} \\
113	& \A{0004121ca746a33c1b0b7c197e881a2d6b1a51f1} \\
\end{tabular}
\end{center} 
  \end{Addresses}
\end{Wallet}

\begin{Wallet}{Teambrella MultiSig Wallet}
  \begin{Description} 
This is a multiSig wallet for \ETH\ only with flexible transactions and owner management. 
In one variant, there is a recovery mechanism.
  \end{Description}
  \begin{Identification}
The wallet can be identified by the following function:
\begin{verbatim}
m_teamId()
\end{verbatim}
  \end{Identification}
  \begin{Addresses} 
The wallet is deployed 852 times.
We find seven distinct bytecodes and skeletons, and 147 different deployers.
In the table below, we list for each bytecode the number of deployed wallets, an exemplary address, the number of deployers and whether it provides verified source code.

  \begin{center}
\begin{tabular}{rccc}
wallets & deployed e.g.\ at & source? & \#creators \\
\midrule
400	& \A{72da082dc5b039a90b82920a02ec1f4a15cd94e9} & yes & 2 \\
182	& \A{77c460dc09d99f4b68ff1ee0f0260cb3b18330c1} & yes & 2 \\
180	& \A{2096acbaaaf96f01268f37b217407c5a234126bd} & no & 164 \\
84	& \A{cdfbd9043c8cb291f008528f5811f1d1c95a5211} & no & 1 \\
3	& \A{5b4a7801badcfe985ac3b5a18d38d6f88e2269ad} & no & 2 \\
2	& \A{be92e39a249affaa13f80508c730cca3759e596e} & no & 1 \\
1	& \A{9307f4893416dce16ffd85171db3e6d3f0c290b8} & no & 1 \\
\end{tabular}
\end{center}
  \end{Addresses}
\end{Wallet}

\begin{Wallet}{Unchained Capital MultiSig Wallet}
  \begin{Description}
This is a 2-out-of-3 multiSig wallet for \ETH\ only with flexible transactions.
  \end{Description}
  \begin{Source}
    \url{https://github.com/unchained-capital/ethereum-multisig}
  \end{Source}
  \begin{Identification}
The wallet can be identified by the following five functions:
\begin{verbatim}
spendNonce()
unchainedMultisigVersionMajor()
unchainedMultisigVersionMinor()
generateMessageToSign(address,uint256)
spend(address,uint256,uint8,bytes32,bytes32,uint8,bytes32,bytes32)
\end{verbatim}
  \end{Identification}
  \begin{Addresses}
 The wallet is deployed 134 times.
 We find four distinct bytecodes and three skeletons, all deployed by externally owned accounts.
In the table below, we list for each bytecode the number of deployed wallets, an exemplary address, whether it provides verified source code, and the number of deployers.
  \begin{center}
\begin{tabular}{rccc}
wallets & deployed e.g.\ at & source & \#creators \\
58	& \A{3584c26fdbac407def37ddbe054f5bfa17aa7362} & yes & 5 \\
52	& \A{11dd0b9eece8a57eba133872c8d6d0ebca149248} & yes & 3 \\
22	& \A{0f6252a414064a13facb36533fda366e5a8de5b6} & no & 1 \\
2	& \A{1b4ce9158a4fbef36ab1a977d61fb4f3735563cc} & yes & 1 \\
\midrule
\end{tabular}
\end{center} 
  \end{Addresses}
\end{Wallet}

\subsection{Forwarder Wallets}\label{app:forwarder}
\begin{Wallet}{BitGo Forwarder Wallet}
  \begin{Description} The wallet can forward its assets to the parent who created it.
  \end{Description}
  \begin{Source}
    \url{https://github.com/BitGo/eth-multisig-v2}
  \end{Source}
  \begin{Identification}
The wallet is characterized by the following three functions:
\begin{verbatim}
parentAddress(),
flushTokens(address),
flush()
\end{verbatim}
Optinal entry points are:
\begin{verbatim}
Forwarder() 
initialize()
logger()
initialize(address)
feeAddress()
0x7cade34a
updateOwner(address)
changeParent(address)
0x0e18b681
newAdmin()
changeAdmin(address)
changeTarget(address)
admin()
0x813f0f59
0xf4069e83
\end{verbatim}

  \end{Identification}
  \begin{Addresses}
 The wallet is deployed 2\,003\,429 times.
 We find 113 distinct bytecodes, 60 distinct skeletons, and 957 different creators (mostly externally owned accounts).

In the table below, we list bytecodes with over 100 deployments.
For each bytecode, we indicate the number of deployed wallets, an exemplary address, and the number of deployers.
  \begin{center}
\begin{tabular}{rcr}
wallets & deployed e.g.\ at & \#creators \\
\midrule
1\,010\,331	& \A{234d0ea487a3f7d58dbf930e443f252126d1a5fc} & 819 \\
    957\,349	& \A{db8ffb9ec317627036ede95e1ab503ab764996ac} & 14 \\
      13\,924	& \A{491d26f172e961af962a851229a96ebf03e64968} & 2 \\
6\,714	& \A{5202df5290e56216c7f96da8c8a0801b26340c4c} & 5 \\
4\,840	& \A{d233e670342ec181155dfd58e23aa54e5c6506f9} & 9 \\
2\,886	& \A{ba39910c7df7b36fe5494f350acbeae2b3f0516b} & 3 \\
1\,015	& \A{d124b0a8e0da91a212f95e19aae4e4ef16cd5857} & 25 \\
1\,010	& \A{7e32c59afab1ad8809a1a72eb051489e79661d86} & 1 \\
785	& \A{290e168c494cfc74d89f6f4718e2cbbed1b9aa9c} & 4 \\
701	& \A{5680942f6ae431558b1b66f39b0a96ed48eede7c} & 2 \\
646	& \A{94babef745501090e1e41d023d2ac32b6e8be0ae} & 1 \\
622	& \A{7e9ab9ae383048b4456ec2e91873d0e309915a47} & 3 \\
606	& \A{644579a22a2198d1790e8c75a7d04505255cd390} & 1 \\
255	& \A{6de659271348306c1638c2caa7e98a81758c8aad} & 2 \\
175	& \A{ece0db88c0479b07e63ca9f8c3d32258214453f9} & 3 \\
153	& \A{ab885acd53b3a6b7a7197b7dce49abe0200f4c5d} & 9 \\
128	& \A{908efc4c5563cd3e864e3ad6216b02cc52ae1d45} & 6 \\
120	& \A{95188daad0aaccaae71bc67599c5e8cd6b1399f0} & 1 \\
102	& \A{a62a1c50488f7cb84ef4fa64871ec20b75ed5988} & 6 \\
\end{tabular}
\medskip
\end{center}

In the following table, we list creators with over 100\,000 deployed wallets and indicate the number of wallets they deployed.

\begin{center}
\begin{tabular}{rc}
wallets & creator\\
\midrule
957\,336	& \A{5b9e8728e316bbeb692d22daaab74f6cbf2c4691} \\
137\,925	& \A{bf0c5d82748ed81b5794e59055725579911e3e4e} \\
101\,331	& \A{2a549b4af9ec39b03142da6dc32221fc390b5533} \\
\end{tabular}
\end{center} 
\end{Addresses}
\end{Wallet}

\begin{Wallet}{IntermediateWallet}
  \begin{Description} 
This is a wallet for \ETH\ and ERC-20 tokens with forwarding of assets to hard-coded address.
  \end{Description}
  \begin{Identification}
The wallet is characterized by the following five functions:
\begin{verbatim}
owner()
wallet()
transferOwnership(address)
setWallet(address)
retrieveTokens(address,address)
\end{verbatim}
A variant also contains the function:
\begin{verbatim}
tokenFallback(address,uint256)
\end{verbatim}
  \end{Identification}
  \begin{Addresses} 
This wallet is deployed 2\,520 times (2\,315 of the first variant and 205 of the second variant).
We find five bytecodes, three skeletons and two creators for the first variant, and one bytecode for the second variant.
In the table below, we list for each bytecode the number of deployed wallets, an exemplary address, whether it provides verified source code, and the number of deployers.
\begin{center}
\begin{tabular}{rccc}
wallets & deployed e.g.\ at & source?& \#creators \\
\midrule
1\,260	& \A{94ce13129781a9f1ea7b454ff73382fbc6dd59b0} & yes & 1 \\
660	& \A{9e12478f75c8181593608c177010b94c6d63718e} & no & 1 \\
390	& \A{dddbadae98c512d4307f5a74bbc5a5be0f03b95f} & yes & 1 \\
4	& \A{8e3b6673b88a722d6be7df05ca9fe94264fdd1a4} & yes & 2 \\
1	& \A{6d5d8397976536f5ff8da66f4e6342021e6754f5} & no & 1 \\
\midrule
205	& \A{61a61b4c1c47675d6465e1853dad7cfb02f855d3} & yes & 1 \\
\end{tabular}
\medskip
\end{center}

In the following table, we list creators and indicate the number of wallets they deployed.

\begin{center}
\begin{tabular}{rc}
wallets & creator\\
\midrule
2\,313	& \A{23ccb865ac268927ccb36de108bf2a10fd69c0d3} \\
2	& \A{ea15adb66dc92a4bbccc8bf32fd25e2e86a2a770} \\
\midrule
205	& \A{ea15adb66dc92a4bbccc8bf32fd25e2e86a2a770} \\
\end{tabular}
\end{center} 
  \end{Addresses}
\end{Wallet}

\begin{Wallet}{Poloniex2}
  \begin{Description}
This is a forwarder wallet for \ETH\ and ERC-20 tokens with owner administration that facilitates flexible transactions.
  \end{Description}
  \begin{Identification}
    The wallet can be identified by the creation history.
  \end{Identification}
  \begin{Addresses}
The wallet is deployed  401\,549 times.
We find one byrecode and two creators.
In the table below, we list for the bytecode the number of deployed wallets, an exemplary address, whether it provides verified source code, and the number of deployers.
\begin{center}
\begin{tabular}{rccc}
wallets & deployed e.g.\ at & source?& \#creators \\
\midrule
401\,549	& \A{e32d19b68388b013b25b592d388b5bc81e10adb1} & no & 2 \\
\end{tabular}
\medskip
\end{center}

In the following table, we list creators and indicate the number of wallets they deployed.

\begin{center}
\begin{tabular}{rc}
wallets & creator\\
\midrule
401\,548	& \A{b42b20ddbeabdc2a288be7ff847ff94fb48d2579} \\
1	& \A{0da57faf4c86d17f4586e0159981fa756747256e} \\
\end{tabular}
\end{center} 
  \end{Addresses}
\end{Wallet}

\begin{Wallet}{SimpleWallet3}
  \begin{Description}
    This is a wallet for \ETH\ and ERC-20 tokens with ownership management.
    Figure~\ref{fig:SimpleWallet3} describes its functionality as Solidity source code.
    The wallet mimics a token interface by using the functions \HD{balanceOf} and \HD{transferFrom} as well as the event \HD{Transfer} (but with the semantics differing from token contracts).
    The wallets are created by a small number of wallet factories.
    By design, each wallet may have a separate user.
    However, none of the wallets deployed so far has ever changed its owner.
    All wallets deployed by a factory share the same owner (see the address section below). 
    Therefore, this type of wallet resembles forwarder wallets under the control of a central owner.
    \begin{figure}
\begin{lstlisting}
pragma solidity ^0.5.0;

contract SimpleWallet3 {
    address public owner;

    event OwnershipTransferred(address indexed prevOwner,
        address indexed newOwner);
    event @Transfer@(address indexed token, address indexed recipient,
        uint256 amount);

    constructor (address _owner) public { owner = _owner; }

    function() external payable {}

    function transferOwnership(address _newOwner) public {
        require (msg.sender == owner && _newOwner != address(0));
        emit OwnershipTransferred(owner, _newOwner);
        owner = _newOwner;
    }

    function balanceOf(address _token) public view returns (uint256) {
        if (_token == address(this)) {
            return address(this).balance;
        } else {
            return ERC20(_token).balanceOf(address(this));
        }
    }

    function transferFrom(address _token,
                 address payable _recipient, uint256 _amount) public {
        require (msg.sender == owner || _recipient == owner);
        if (_token == address(this)) {
            require (address(this).balance >= _amount);
            _recipient.transfer(_amount);
        } else {
            require (ERC20(_token).balanceOf(address(this)) >= _amount);
            ERC20(_token).@transfer@(_recipient,_amount);
            emit @Transfer@(_token, _recipient, _amount);
        }
    }
}
\end{lstlisting}
      \caption{Approximate Solidity code of \emph{SimpleWallet3}}
      \label{fig:SimpleWallet3}
    \end{figure}
  \end{Description}
  \begin{Identification}
    The wallets in this group can be (almost) uniquely identified by the four signatures of their interface.
    \begin{flushleft}
      \HD{transferFrom(address,address,uint256)} \\
      \HD{balanceOf(address)} \\
      \HD{owner()} \\
      \HD{transferOwnership(address)}
    \end{flushleft}
    Moreover, all wallets are created by calling the function \HD{create(address)} of a wallet factory.
    This distinguishes them from the single contract with the same interface that in fact is a token contract.
  \end{Identification}
  \begin{Addresses}
    Each line in the table corresponds to a distinct deployed wallet code, all but one sharing the same skeleton.
    The owner column gives the initial owner; none has been changed so far.
    \begin{center}
      \begin{tabular}{rcccc}
         count & skel & sample address & factory & owner \\
        \midrule
        1\,209 &  C   & \A{f6e8d71069841518052c9f8c9ac461958e2fca5f} & [a] & [1] \\
            67 &  C   & \A{d75967e9ddba86b45f94457fb23270ab5cb950a5} & [b] & [2] \\
             9 &  C   & \A{ccbc4985a451ad7c89a80ec0404117b14d2bd2dc} & [c] & [3] \\
             3 &  D   & \A{2701916e72806a7f7f4771042c81219677bbda5e} & [d] & [2] 
      \end{tabular}
    \end{center}
    The factories are contracts uniquely identified by their interface
    \begin{flushleft}
      \HD{owner()} \\
      \HD{create(address)} \\
      \HD{get(address)} \\
      \HD{transferOwnership(address)}
    \end{flushleft}
    The factories create the wallets with an owner identical to their own.
    List of factories:
    \begin{center}
      \begin{tabular}{cccl}
        ref id & skel & factory address & owner \\
        \midrule\relax
        [a] & A & \A{ec65b1d69135eea775135e56bab572766d379e14} & [1] \\\relax
        [b] & A & \A{b5ce40f9b98c6499f0e165e2512e4f1d0573e0de} & [2], later [1]\\\relax
        [c] & A & \A{f59737e4907d193d2b2ed97dd9c876ccd75a4b41} & [3] \\\relax
        [d] & B & \A{69b42ddfc73a2320329a3e9acf6c94176a168ec4} & [2]  
      \end{tabular}
    \end{center}
    Owner addresses:
    \begin{center}
      \begin{tabular}{cc}
        ref id & owner address \\
        \midrule\relax
        [1] & \A{e2b628a82b8f6e061f6739b7cdfa547bf9615a3e} \\\relax
        [2] & \A{949b917578b88af67c4e4302190208fdd7c96d85} \\\relax
        [3] & \A{68d5aed7a804f7d7085a99b1844e0abefa715169}
      \end{tabular}
    \end{center}
  \end{Addresses}
\end{Wallet}

\begin{Wallet}{Wallet2}
  \begin{Description}
    This is a wallet for \ETH\ and ERC-20 tokens that accepts \ETH\ via the fallback function.
    The interface offers a single function that transfers the given amount of \ETH\ (if the address is 0) or the amount of tokens (token contract identified by the address) to a hard-coded destination.
  \end{Description}
  \begin{Identification}
    The wallets are identified by their (three) skeletons or by their (six) bytecodes.
    Alternatively, the wallets can be characterized as the contracts with the single entry point
    \begin{flushleft}
      \HD{transfer(address,uint256)}
    \end{flushleft}
    that have been deployed by wallet factories with the single entry point \HD{createWallet()}.
  \end{Identification}
  \begin{Addresses}
    The wallet is deployed 12\,499 times.
    It does not provide verified source code.

    Each line in the table below corresponds to a distinct deployed wallet code, some of which have identical skeletons.
    When calling the \HD{transfer} function of a wallet, the assets are forwarded to a hard-coded destination address.
    All but two wallets have been deployed by contracts (`factories').
    \begin{center}
      \tabcolsep=3pt
      \begin{tabular}{rcccc}
         count & skel & sample address & destination & factory \\
        \midrule
        11\,076& A   & \A{3ee9b1fda4ee7a080156ea54371c46eb4751c316} & [1] & [a] \\
         1\,021& A   & \A{231114d49808df31b53d76abfe7f684e92f122f6} & [2] & [b] \\
            240& B   & \A{e86db80c1c2d9b41193264c27ca99c7399a73e95} & [3] & [c] \\
            160& C   & \A{7e3cd718ce1072da84e0460c73ec59420f2ec6bf} & [3] & [d] \\
              1& B   & \A{cf1f62d7ee46a9590ce34c5abd645ce1f1d54f27} & [4] &     \\
              1& A   & \A{80854f4f959c0e82d73bcfce38ba4b06c49ede41} & [4] &    
      \end{tabular}
    \end{center}
    Destination addresses to which the assets are sent:
    \begin{center}
      \begin{tabular}{cc}
        ref id & destination address \\
        \midrule\relax
        {[1]} & \A{e7b1ec34d65f8c9a1b02694bc0ff5b3f7d2c1bc9} \\
        {[2]} & \A{96ed53794f48b9c652189558ba147041ea743fe8} \\
        {[3]} & \A{c2a923d6938d828c6f08d0964891d6126fdf7625} \\
        {[4]} & \A{54dbe9c116f92021f2b28756f59d2c1649a92f56}
      \end{tabular}
    \end{center}
    The factories possess the single entry point \HD{createWallet()}.
    Two of the four factory contracts possess the same skeleton.
    \begin{center}
      \begin{tabular}{ccc}
        ref id & skel & factory address \\
        \midrule\relax
        [a] & D & \A{484eca64c063643d315b598f20ab45b8e91cc2db} \\\relax
        [b] & D & \A{978039119e31886bddf790a50d3b0681b8e73872} \\\relax
        [c] & E & \A{0ffa490933ebdd05d783a790a1f47fded54dc0a7} \\\relax
        [d] & F & \A{f2a8f8670482d5cbbc78fc85842ac0801f145699}
      \end{tabular}
    \end{center}
  \end{Addresses}
\end{Wallet}

\subsection{Controlled Wallets}\label{app:controlled}
\begin{Wallet}{Bittrex}
  \begin{Description}
    This controlled wallet involves three types of contracts: controller, sweeper, and user wallets.
    The controller creates wallets and maintains a list of sweepers that are used by the wallets to perform transfers.
    A separate sweeper can be set for each kind of token or Ether.
    In practice the so-called default sweeper is used most of the time; it is deployed at the same time as the controller.
    With a total of 14 \texttt{addSweeper} call, only eight controllers have set a new sweeper so far: three times for each of the OMG, SONM, Tether token, twice for the Binance token and for Ether, and once for the Mithril token.

    Figure~\ref{fig:controlledCalls} shows the interaction of controller, (default) sweeper, and user wallets.
    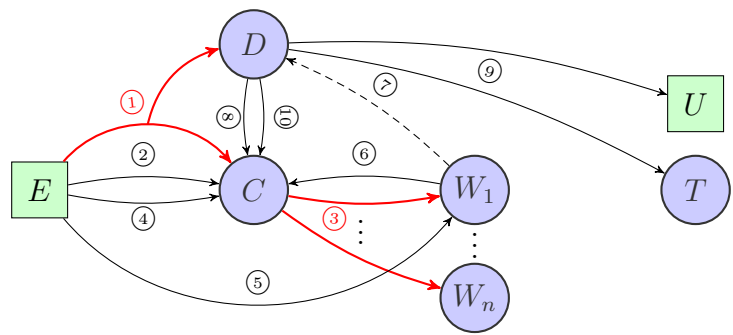
\begin{figure}[!h]\centering
\begin{tikzpicture}[node distance=2cm]
  \node[contract](c){$C$};
  \node[contract](w1)[right=of c]{$W_1$};
  \node[contract](wn)[below=5mm of w1]{$W_n$};
  \path (w1) -- node{\raisebox{-1ex}[0mm][0mm]{$\vdots$}} (wn);
  \node[contract](ds)[above=10mm of c]{$D$};
  \node[user](u1)[left=of c]{$E$};
  \path[create] (u1) edge[bend left=50]
                     node[action,above,pos=0.43]{\action1}
                     coordinate(cds)
                (c);
  \path[create] (cds) edge[bend left=30]
                (ds);
  \path[call]   (u1) edge[bend left=10] node[action,above]{\action2} (c);
  \path[create] (c) edge[bend right=10] coordinate(cw1) node[action,below,pos=0.3]{\action3} (w1);
  \path[create] (c) edge[bend right=10] coordinate(cwn) (wn);
  \path (cw1) -- node{\raisebox{-1ex}[0mm][0mm]{$\vdots$}} (cwn);
  \path[call] (u1) edge[bend right=10] node[action,below]{\action4} (c);
  \path[call] (u1) edge[bend right=50] node[action,above]{\action5} (w1);
  \path[call] (w1) edge[bend right=10] node[action,above]{\action6} (c);
  \path[delegatecall] (w1) edge[bend right=10] node[action,above]{\action7} (ds);
  \path[call] (ds) edge[bend right=10] node[action,below]{\action8} (c);
  \node[contract](t)[right=of w1]{$T$};
  \node[user](u4)[above=3mm of t]{$U$};
  \path[call] (ds) edge[bend left=10] (u4);
  \path[call] (ds) edge[bend left=10] node[action,above]{\action9} (t);
  \path[call] (ds) edge[bend left=10] node[action,above]{\action{10}} (c);
\end{tikzpicture}
\caption{Bittrex controlled wallet: interactions between controller, sweeper, and wallets.}
\label{fig:controlledCalls}
\end{figure}
    A user or contract representing e.g.\ an exchange, $E$, deploys two companion contracts in a single transaction: a controller~$C$ and a default sweeper~$D$~(1).
    In further transactions, $E$ calls $C$ with \texttt{makeWallet}~(2).
    This prompts $C$ to deploy wallets~$W_i$~(3).
    In total, 2.5\,M wallets have been deployed this way.
    The wallet addresses can be given to customers, who publish them to receive \ETH{} and tokens.
    The control over the wallets, however, stays with $E$, similar to accounts of traditional banks.

    Now suppose \ETH{} or tokens owned by $W_1$ are to be transferred to the destination address~$d$.
    First, $E$ calls $C$ with \texttt{changeDestination} to store $d$ in the controller~(4).
    Destination addresses are changed rarely (so far 65 times for all controllers, in total).
    Then $E$ initiates the transfer by calling the wallet with \texttt{sweep}, passing $W_1$ the token address (0 for \ETH{}) and the amount~(5). The wallet asks the controller for the sweeper in charge of the token~(6).
    Next, the wallet performs a \emph{delegate call} to the sweeper, $D$, handing over token address and the amount~(7).
    The delegate call has the effect that all actions by the sweeper will seem to originate from the wallet even though the code is actually contained in the sweeper.
    The sweeper calls the controller several times to learn the destination address and to check permissions~(8).
    Then it performs the token transfer or sends the \ETH{}~(9).
    Finally, on successful completion, the sweeper prompts the controller to issue the event \texttt{LogSweep}~(10).
  \end{Description}
  \begin{Identification}
The controllers and default sweepers are detected by a combination of signature and message analysis.
A contract,~$C$, is a controller if it satisfies the following criteria.
\begin{itemize}
\item During the deployment of~$C$, another contract ($D$) is created.
\item The only signature of~$D$ is \texttt{sweep(address,uint256)} or \texttt{sweepAll(address)}, plus optionally \texttt{controller()}.
\item Contract~$C$ is able to create contracts of a single type, the wallets~($W$).
\item The only signature of~$W$ is \texttt{sweep(address,uint256)} or \texttt{sweepAll(address)}, plus optionally \texttt{tokenFallback(address,uint256,bytes)}.
\end{itemize}
\end{Identification}
\begin{Addresses}
  The table below lists the addresses of all controllers that have deployed at least one wallet, together with their default sweepers.
  Even though 116 different pairs, the wallets show only 31 different skeletons.
  In the lines with a mark in column `src', the controller, the sweeper, or both have a verified source code on Etherscan.
\begin{center}\scriptsize
\begin{longtable}{@{}r@{\ }c@{\ }ll@{\ }c@{}}
        wallets&skel&\multicolumn{1}{c}{controller}&\multicolumn{1}{c}{default sweeper}&src\\\midrule 
        \endhead
        \multicolumn{5}{@{}l@{}}{{\ldots continued on next page}} \\
        \endfoot
        \endlastfoot
1\,578\,470 & A & \A{a3c1e324ca1ce40db73ed6026c4a177f099b5770} & \A{b2233fcec42c588ee71a594d9a25aa695345426c} & \YES \\
633\,646 & B & \A{4f01001cf69785d4c37f03fd87398849411ccbba} & \A{3105d1027fdd1cf6b2d67056b61956249f6fc861} & \\
142\,580 & B & \A{edce883162179d4ed5eb9bb2e7dccf494d75b3a0} & \A{88de41a34871e239c52926f920efb4eefa0f3de3} & \YES \\
69\,391 & A & \A{0bd9bf737f70c339db6b87f6a9fa8f6862b30dd6} & \A{4fff49c5b4691775a13d2bea90b60d1876b6d1c9} &  \\
56\,000 & C & \A{2754b28227f041a66c46509d5620782bfc4766ef} & \A{5e85f6cdb466771b870757658593d073b8f3f9c8} \\
10\,697 & D & \A{d8b7c1e952cb3b4ba3ab993096bb8c6be26a17a0} & \A{f9dd79eef74db6de9efbe9715bc256f76f138005} \\
9\,250 & A & \A{e91eb1fb4b0eae01afa35b99a2c84409f353c49b} & \A{7460c1d30154c06e2253e6ead49a1bc5f4353bcd} & \YES \\
7\,563 & E & \A{4746809ecb3a8449a61d77246992a33fdd73f61b} & \A{27304c0dc0d1ade22c315ec633136d1e249df9d4} \\
6\,572 & F & \A{7142eb34d2220152dedc5868745079bc6ffa0fdd} & \A{a928e01614a4d746ec4acaebbdd4f8239ae6739c} & \YES \\
4\,644 & G & \A{a7e307d1e4181cdf9885fbde941f571698521c45} & \A{f35e698d0eb03bfc9d6147b45d95b1242bf0f123} \\
3\,999 & H & \A{e8dbeeb48ba67bae3d4761b10dbcc0960750c2b5} & \A{83010351ff829c2b93ff46641d616b6396821318} & \\
3\,693 & I & \A{ecba8b19e058891f581badd5697606097300ab60} & \A{fcc5b3023797254e9904ea62c9140fe9fb306f10} \\
3\,350 & J & \A{267353a2cc8b329d8ed3a13cee17adf2427fa680} & \A{3bc8271be786a28c7fc3b731bfd46181eda0b741} \\
3\,265 & H & \A{c2acad3578019495ed7814a612aa0866b7b733e0} & \A{1f4f8ee88e732f2c1731931d0456ed7ad37478de} & \\
2\,580 & K & \A{f6014ba24ddfeb009699ffd5b753487f0b2eddd1} & \A{a5c283dc4b7457318def37c7254832344ce2f0a7} \\
2\,400 & L & \A{0f81e92e206316e9e6d4d9c57f3921ee60817900} & \A{1db3d98054b23a2f740a8ae07bbe07d8a8a3ef26} & \YES \\
1\,764 & M & \A{95e76bc353d4d146bf6e3a6fd1d9db46acacbfca} & \A{b23dac47d88ceebfd233d17ab5d226de07e67444} & \YES \\
1\,755 & N & \A{6933127b497c83466bf2995bd4e22b1536eec5ad} & \A{6825398a9d0c8661e5477c01e6185ac56e7f7276} \\
1\,496 & O & \A{e01c5efcc674a72eb690670ed3f63624542ffbc3} & \A{ed61189fa8d62b29d3f5c12c96a1e9672b05df3f} \\
1\,058 & J & \A{357f9afb22a2bb56f8f75ffa36c4d12f3b62ecc5} & \A{bfe86824d5f428f1efccc6ca4142136e25ef543c} \\
977 & P & \A{c8fc145491a928a24f1da88e31e6ff9771e5b59d} & \A{48ab755bff0a4f82037e339b1cc2c167ddb64cec} \\
887 & J & \A{cfdcc1a98172c13ecd625bad384ab6f578f1bb3c} & \A{f0c85a8e8db7e1e7c550f61e1891a5e0b2cf37c5} \\
870 & L & \A{4202b62990c763860ffaf5e4ee935b1459890e25} & \A{8f16c452e74c8688cd2795f9ab7623afea7accae} & \YES \\
824 & H & \A{f577a6e6a0b5b8677208ab2a301941153374d796} & \A{902ce56f67519c2f1c4e4ab9563f42a405779b2c} & \\
420 & G & \A{a7ab11e268bf2852f103641133044cea9fe75114} & \A{d0932b2c8a854565947316080f243ddc4013c74e} \\
394 & G & \A{9d44d0855aec425c4f9e82cc58c5c0955e335aeb} & \A{d0547cbc83b386ca23506e9ae8936614ed7664b4} \\
383 & O & \A{38db3c2244f5de9e102cd844d374ba40f3818af3} & \A{a9a2514366e3d8b5f13e878d3befb13215c26006} \\
340 & Q & \A{714853895516168212703896401a343699ea9b1a} & \A{271e7d6f3e1a8e876521d4c324b7eef890a30a82} \\
325 & R & \A{d909ace851399e1d92af958ba57875353504e289} & \A{48cd9c3eab67c535291250b8e785fb6d43b1dba9} \\
305 & S & \A{4d352b1338ea8c7cc70b0e669b99b35db8498a54} & \A{ceb40183b1cb52ed0a7203cf9c468d6e38a3c98b} & \YES \\
300 & S & \A{b4cbc0c87ab21bc02e37c1432e771f70cd6db5db} & \A{f55af536a52420a1df21ecff768bab18d80c58b2} & \YES \\
298 & P & \A{f0833b48c6b342437da4a2017fbbf4fd1a3c9682} & \A{16cd93bc5752a490c7d12443d50a2b9ae21904e1} \\
275 & P & \A{aa35a2bf8b7cf512e0f6dfe20d969062a41eaa91} & \A{30a514e48a3c1e395ab2d0a19006e3c28b8ada57} \\
270 & L & \A{2c164b1e756744183c14aa683e316a61e597ff52} & \A{183b0087bc7b85cc393b844b0432c97d37da695e} & \\
223 & H & \A{6658d2c231232d1f945380c18009de4f58dcb690} & \A{ca6d9f429fe7d85bd27e8c1b1d7f30082127fa35} & \\
213 & T & \A{eb818c6a48ccd60a8078aaa20997cc3cb2538c9e} & \A{8e7abaf1316a0edb985e494f572fdf148e8a7e93} \\
212 & U & \A{adbe6eacc402ffd3ce0b20a45fa30a1cd7eaa142} & \A{f5eac2e3b4a3b18ed37b840cdf374f67cc284b7b} \\
209 & B & \A{00e8c228e75995eea114631f1e27f458071b70fb} & \A{f76956a5dba85d8d913ebc382e49d5aa8ea5aaa2} & \\
199 & H & \A{e4460cc335dda80b22e21510a63de14bf46b098a} & \A{b98b445e9054aeab56e387e84dde7659bdcd6972} & \\
152 & H & \A{deea90d523606c52cdf1fcf01a06f732a3dfc355} & \A{c73514758a4ce50ac66d1f081c90adc7eb1427f2} & \\
141 & C & \A{69e6030616d9b1fdaab8c05ab147382b2ac23af0} & \A{7e90d82fd51e81fbb8e94a977273733fd53e49f5} \\
140 & G & \A{435ad629145bb01547fc3269020e04d3b8034b48} & \A{38a21ba7fca8a3bee7c98a3cbc5369358755cfea} \\
117 & V & \A{6f68736ede3b8263d2f769b148a4d410cf44d392} & \A{af51ebbf24042ef8928ce72911ee7214318e1bf5} \\
107 & W & \A{b824643b483bb0726401072b0e66e29f9e70e8cf} & \A{78f523a649114e0cf7991b3d267c344a490a5eea} \\
93 & P & \A{a1df420345bda29622281ad713bb22fa8b03c97b} & \A{1c8cf8b563fa6ea9e82c78798bce69691db04753} \\
71 & A & \A{c227fc68da6efa32cd6710fcd4bc5ae3041d1e86} & \A{286b0bad77b3b817998b3040c79d2864744ff010} & \\
62 & J & \A{2081ca8a6156b4b7e0a9473754a8d97a18a71f88} & \A{68578231c5c4c431dc847e8ee8456148990137f2} \\
62 & C & \A{7c9338bc675fd0c513b31438f5418be947911f8c} & \A{5f0375956b17b236b3208a86dc40dd20a69ef9cb} \\
60 & L & \A{721a6f3e7f66efe60645a7135a6f5d5fd1ab523b} & \A{8640f5c3726f86fe14d3a7b1a7ae2c9ba48c5afc} & \\
50 & U & \A{945ef56694081994cd0a0275b39542a2521a4f67} & \A{3eb119d8c52fae8947e1b8440c97eb98150b7fa5} \\
40 & S & \A{e030fc55a27b2d1d84f34336fca3a276a6ddf0d4} & \A{d7fe8940357eb32341a5e89d0aad49f65ee9b965} & \YES \\
40 & X & \A{037bfb02e53877b9e0ba62705e69ecd1ae1f5f01} & \A{f77b5f4752b47a4bc4116505b4869611541a94e1} & \YES \\
37 & B & \A{6cb0754da2843d3bea24638cb32dcd92b73b2eda} & \A{2208fcba966b93b47211faf6ac44840b228395ee} & \\
36 & J & \A{1e73119561eb749f65b227acd16f5d9b64eddda6} & \A{1bce24e05eb30dd118a1f7a958b52d78100aa304} \\
32 & Y & \A{a001e29c80cbae6aa47bdb69d215752a48874258} & \A{35d1be55f4ec8d114b6552062997fa552fc48ff2} \\
31 & T & \A{127925fec603720409ce8fad7688abbc859598f9} & \A{c99ba12c5a7c0a71ae22c156dccf021880c8a3a9} \\
28 & H & \A{896324fdbcf9fedda6641147a771f53ea6700064} & \A{ea05b97d5ddb61deb3dff6d06894ece0c52c310d} &  \\
25 & J & \A{26fb44ddad061d7cd5dd2f7b2b3a38145e6db0af} & \A{c29fb1d8c40804ac5c2b20e2c471c2902f9715de} \\
24 & H & \A{ad77050da7c4a6cecb6b0eed3e28e434c4a6308f} & \A{6c709216cb6a85226781bcef5db91d65a7b40b63} &  \\
24 & Y & \A{f33dd1caeca4fdce684720a1930f324542707aad} & \A{38af5deb8b0913016674fbf7d445ef1bbc8c3318} \\
20 & Y & \A{e1d65a902a8349f217d6864e456408cd351923d3} & \A{61e6b6500bd934eeb8b86250c3a6ab6816c30d42} \\
19 & B & \A{3c861c1db7bffd094a1f8dc9646860eaf91ad5d9} & \A{fd0c3dbfb9972303d38b445e84327ccb550dd000} & \\
16 & J & \A{b0db51bb47305a496c0757348c893cbc4559d8b4} & \A{d729dbef8ca6ac183b32e864657310765c56c799} \\
15 & A & \A{539faf9c0a455d12e8a6334344be270e679f0269} & \A{60b8f00ae2a5eb41341791c41ef7247505fd3944} & \\
15 & H & \A{c110311376064e3b1ba267db6bc5466e54ac5224} & \A{c5df48507eac52fa8a289676c38f7881860db489} & \\
11 & C & \A{d7d4449e2439eaca8a3d78324fb520d8cb7e22a3} & \A{dc93b68926b626e36d56550079084f765ff127d0} \\
11 & Z & \A{bdab70f6a4ce05b20f9274f540494532784f10ad} & \A{4d5e1fe7d3cee644483d847c3c99c3150c110f3a} & \YES \\
9 & G & \A{87521a991ac4091ba683171895ee2e9a856c8dea} & \A{71683ae4a06c09d3aa69ad5157184cb78da0d4ad} \\
8 & W & \A{e9dcc223c5ec3628967115c9c73c589d59cba2fc} & \A{c1600aaa7c3ac47c26e7a697d9a4c3ac55ed5822} \\
8 & H & \A{cee8ed1342409f987424fa6617bc9862551fba5c} & \A{aaa6623e894a97b37f5eaa5b068dc920588e0639} & \\
6 & H & \A{ee5baa445fe45a6d7bc99e48ce896ed1d604b8a0} & \A{674ade690ada2db1bd19325f80fc98973366ea06} & \\
5 & H & \A{52f19bf189667c85f16b4949dd9770eba9cc8a0a} & \A{525954771f266a96976213827ef1e68360e95466} & \\
4 & U & \A{edd7211e10c4d99785ac6c98ed240ce9c35df94e} & \A{d3ddccdb3b9fe5133e3b3e4dc2814840e90f55f1} \\
4 & W & \A{be25de6336c40bb8335cfc968e4b7763a25be4e6} & \A{89638dd9190223174dade7b7f313e4e8cbe7fee3} \\
4 & X & \A{84b6241ed47e935597bc5a14d5e9ef9fb53e77de} & \A{e2c43804f37b808fb37162dd99860e236a90c371} & \YES \\
3 & W & \A{fe3f6bb09320517517061f8062ba1a4466cc9929} & \A{79866c97bc7f4c0c674a593a484e7d31594de6ba} \\
3 & H & \A{625c5b4f21f938c7167f1d8046e32bc31ddc17e9} & \A{8fcf8126dd357189f9b7007b9687ece6594c6f3a} & \\
3 & H & \A{bf1c8a0fb650f15688104153baeb55fdd03c4232} & \A{cf1daab1364d1f03d5d63f5fd68588cc5a193b6f} & \\
2 & H & \A{f9bee1eee68fff5d284cb3f9ccce27355d944e82} & \A{d69a83a915154d98099ecd1ece38f3e2f277ef71} & \\
2 & H & \A{288b71fcc995ad2071f81cfe6c776192b54e58d2} & \A{8a7111271223d329d2b4300ac336f117644904fd} & \\
2 & H & \A{98940a17b0f12233d31d5f38bfcf4ac57ba16a78} & \A{8686f0ba04b4574bbeb2ac88bc94dada7931d1b4} & \\
2 & H & \A{034b01e0fcea72ec9c79b2a00f603a8370a32caf} & \A{3592dc41c5e57cb114f8c6a775402ca4801490b4} & \\
2 & W & \A{a20342c6f3687066f55faf86aafdbc0fceabcb24} & \A{1a1bab5640eb324c4370e1da9869da8940fa07ae} \\
2 & AA & \A{94d14692551a7cdc544675519c3ca7ff0303e49e} & \A{1310df319b1f4efcd3fd3c7f263d7523980ef696} \\
2 & W & \A{a2ffb93a21beda8ce7f4fcb5b72652a8ddacd5b8} & \A{59deb9156b6fefac513f6c8abdd09d6b40a6d1ba} \\
2 & W & \A{3914df0a52d60b9f90645f011bd9da4fa62a0ec3} & \A{37459f96ac4341ce06e263d9557caf3d06c9a257} \\
2 & W & \A{cc3e11a88bee01f25ae335c9c38b491ed3d0d5eb} & \A{cff34e3b2788845e1c2eb672bed627cc21f55014} \\
2 & K & \A{973668f9b70dd6879475b222adbf0dcd9b61c5dd} & \A{1da1239d8d490c67c5a3bae1c44495314f000628} \\
2 & H & \A{f9017c753072db7b795cd4b3461185bf5e1e2eb2} & \A{4d6b06581a452b259eab851903a1f725012a741f} & \\
2 & H & \A{c1f20ea713da6c7414da00309ace2e397c85dd5e} & \A{d60b90b37679381f1040501c4b773995205bd9bb} & \\
2 & H & \A{896859f53afe8e90936f4cff432d14e98c1846cb} & \A{28a7e8e78d8b1b47e7d01123183e22678829d6a8} & \\
2 & M & \A{d6fa1019203bbcac47345c669fa357075830b6c5} & \A{ee711be03773edc0910e770a6b32a595f88adc8f} & \YES \\
2 & S & \A{2af14b457e6fe815d0b33a1e2791f895ab7741f9} & \A{99f4e80f8e5f443d1164e288b94302dd6e556282} & \YES \\
1 & AB & \A{dae7df6f3f4a5b6c80348071d8f01ecf3e779f10} & \A{9f6f7febd337dd595112a500202f0d446f1bf76e} \\
1 & B & \A{8e9dd6c58165ac8e2aa6e7da3d39743e60003c57} & \A{31f69112aab8ca22658e417ff7ee8bc0aecefb48} & \YES \\
1 & A & \A{a370aad6389d6b2e5cef6fec3c805a7b64020121} & \A{29ab40a597e37efd48ce04b8882c1eadc3e8da08} & \\
1 & A & \A{314ea887007f30121ed587d9d6b51f759abb8c1b} & \A{0878aa875c8e0e4a086d2baa4aa7c585cd4e662e} & \\
1 & AC & \A{dba1a2bf05508b0b53cff2001e7a568241cb7d84} & \A{f1a6ec7ead6bdd6b2cbac4846a843b049ad911f7} \\
1 & AD & \A{b45558158ec811380d6f154e19b4e4c95d0033dd} & \A{42245ece47ff3fdf6eea5a82f38e095ae2403d40} \\
1 & P & \A{a13c1233e51504b3789b6a8d7ddd78d17e9b8f19} & \A{9ca1a09deef43cad39170ae94302103df2e94ecf} \\
1 & H & \A{ce4383e7309090ac0a7ead826991094c852c68ff} & \A{41f532da2fe94448933d09b40b974a61331244f8} & \YES \\
1 & H & \A{7ea5b903204b93b7e2624457919886bfb99e5512} & \A{edd17030d5df34b7b1e8cc6cb6179cf08f7436b2} & \\
1 & H & \A{0eaf2d408b55ac69fc8e94e0ab276602356f0f75} & \A{d10c46330c2a3918803c1e8e0732ae0ab3322e3a} & \\
1 & H & \A{a61edd21e302141631806723a0c70766131cc5ff} & \A{29a6aa66cc9d93d3f9ef688e0ee5abc8f36656f4} & \\
1 & H & \A{56f3d94475226ece7d77cae4f9041ef4bc360b76} & \A{4b3a26cc8fe3aa7ba35445a1abff5a649a1b3c32} & \\
1 & H & \A{6ca022aa46ae12cfd1ef5d12724042b321ce35c6} & \A{2fef37758723585e906897cf50da4bdc55267c3a} & \\
1 & H & \A{162419f95b85cd7beb8bd0354037c9ce3001b2b5} & \A{12be1536978590d61182c5dbf601c70c6069857c} & \\
1 & H & \A{68e86e6463cd3acb6280ae3dbe05096577c3ba6e} & \A{d58206f9131832114012634ec8c338a188826b3f} & \\
1 & H & \A{b05f2f3f0868543ad3f0a8cc6234920fe57036b9} & \A{a0d0ebc1cf6f638a7343218ba9adbccfdf4a0f34} & \\
1 & H & \A{7c17de5abb920871982d464010b3b868e7e4a8c1} & \A{d859c398b275cdf44a6d8476d7d232cba5895723} & \\
1 & U & \A{734606e5cd919f2194361915bda4035939c9f626} & \A{7994c4076eff697e20b3cd12ca214682787ac887} \\
1 & AA & \A{1fe3523cb60c26fc61859cb10349d3436db066dc} & \A{3d17d9bc29c60ade3dd67e3f152e793c9d57c9cc} \\
1 & L & \A{a2d3c535f3a31fc624654cbaaae7d899a3731342} & \A{c66797e1807aa244dcd327d541be05aa6af35af6} & \YES \\
1 & L & \A{2a2af451d6b33e6595d5b3c4672dd24b4c2010d6} & \A{d55e2e2dc961836209eb1a96595a82f181cd85ef} & \YES \\
1 & L & \A{455542507d3c9975538f13f65d6a79b669584d21} & \A{a99721e320c1bb2fad16500c286670ffeb702340} & \YES \\
1 & AE & \A{2977d7f1ea7bcd532443163dc72967031ce3bdce} & \A{2a250b6d8d3f450db1462f411c343f79a2d4dba3} & \YES \\

      \end{longtable}
\end{center}
\end{Addresses}
\end{Wallet}

\begin{Wallet}{ICTlock}
  \begin{Description}
This wallet is only for time-locking ICT (intelligence chain token at address \A{283640b9a2bba66a294c9b19cc4404cafd35c7cc}).
It has owner administration and can forward the token.
  \end{Description}
  \begin{Identification}
  It can be identified via the creation history.
  \end{Identification}
  \begin{Addresses}
The wallet was deployed 309 times by two creators
We find 278 bytecodes that collapse to a single skeleton.
The wallet does not provide verified source code.

In the table below, we list for each creator the number of deployed wallets, the address of an exemplary deployment, and its address.
\begin{center}\scriptsize
\begin{tabular}{rcc}
wallets & deployed e.g.\ at & creator \\
\midrule
305	& \A{6b0c2ae673ab031952312282becdcbe0bbf773c3} & \A{c6d7cd473add4ebfbb4fb1cfdd6ad310313a61c9} \\
4	& \A{a14cbca711d35ebf9be6f78db22bfe7df994126b} & \A{2814a7f5cc45174a05917d18ab6bd801adb713a9} \\
\end{tabular}
\end{center}
  \end{Addresses}
\end{Wallet}

\begin{Wallet}{SimpleWallet4}
  \begin{Description}
This is a basic wallet for \ETH\ and ERC-20 tokens with third-party control.
  \end{Description}
  \begin{Identification}
The wallet is identified via the creation history.
  \end{Identification}
  \begin{Addresses} 
The wallet was deployed once by the externally owned account \A{33b397DAdB08513C26fF4984902f542B703d179E} and can be found at the address \A{71d2edc7888dd67dff650cfb3da4f203aa026518},  where it provides verified source code.
  \end{Addresses}
\end{Wallet}

\subsection{Update Wallets}\label{app:update}
\begin{Wallet}{Eidoo}
  \begin{Description} 
The wallet manages \ETH\ and ERC-20 tokens, can place orders, and can be updated,.
  \end{Description}
  \begin{Identification}
The wallet is characterized by the following functions:
\begin{verbatim}
owner_()
exchange_()
tokenBalances_(address)
logic_()
birthBlock_()
depositEther()
depositERC20Token(address,uint256)
updateBalance(address,uint256,bool)
updateExchange(address)
updateLogic(uint256)
verifyOrder(address,uint256,uint256,address)
withdraw(address,uint256)
balanceOf(address)
\end{verbatim}
The helper contract WalletLogic contains the functions:
\begin{verbatim}
owner_()
exchange_()
tokenBalances_(address)
deposit(address,uint256)
updateBalance(address,uint256,bool)
verifyOrder(address,uint256,uint256,address)
withdraw(address,uint256)
safeTransfer(address,address,uint256)
safeTransferFrom(address,address,address,uint256)
safeApprove(address,address,uint256)
\end{verbatim}
  \end{Identification}
  \begin{Addresses} 
The wallet is deployed 3\,916 times.
We find three bytecodes, two skeletons, and three creators.
In the table below, we list for each bytecode the number of deployments, an exemplary address and the creating contract.
While only the last bytecode provides verified source code, all three factories do.
    \begin{center}\scriptsize
      \begin{tabular}{@{}rcc@{}}
        wallets & code deployed e.g.\ at & factory \\
        \midrule
2\,321	& \A{0a23143f7f8abfb0d14f990fe1e778bd00148cf4}	& \A{f1c525a488a848b58b95d79da48c21ce434290f7} \\
1\,079	& \A{d5c263ae5d11eb0c613372666a7413649e13f564}	& \A{da9d0543311acf95df039d7b0e6fe45996f42382} \\
   516	& \A{e14ddda4eb53aa1056b500e5d5efd470668698af}	& \A{df72b12a5f7f5a02e9949c475a8d90694d10f198} \\
      \end{tabular}
    \end{center}
  \end{Addresses}
\end{Wallet}

\begin{Wallet}{LogicProxyWallet}
  \begin{Description} 
This wallet can execute flexible transactions, as well as change the owner and the version of the implementation.
It has a helper contract that handles the wallet logic and also the proxies.
It is used by \texttt{instadapp.io}.
  \end{Description}
  \begin{Identification}
The wallet is identified by the following two functions:
\begin{verbatim}
execute(address,bytes,uint256,uint256)
isAuth(address)
\end{verbatim}
  \end{Identification}
  \begin{Addresses} 
The wallet is deployed 9\,359 times.
We find three bytecodes, three skeletons, and three creators.
In the table below, we list for each bytecode the number of deployments, an exemplary address and the creating contract.
Only the first bytecode and factory provide verified source code.
     \begin{center}\scriptsize
      \begin{tabular}{@{}rcc@{}}
        wallets & code deployed e.g.\ at & creator \\
        \midrule
9\,357	& \A{5d4a945271fb3e16481bf6ce0bad5f6b2e9d13db}	& \A{498b3bfabe9f73db90d252bcd4fa9548cd0fd981} \\
1	& \A{aa609969b5cadbec199cfb4f7d5f906eac616880}	& \A{5f92b1a2b854a7626cc5dcbbec2212dc9ef99f30} \\
1	& \A{1ce2db6d8cd8fcea3f02c66203144c43cb552929}	& \A{4c0aeb88a7bf3aad6f2523093b87551b394e90f1} \\      
    \end{tabular}
   \end{center}
  \end{Addresses}
\end{Wallet}

\begin{Wallet}{LoopringWallet}
  \begin{Description}
    Loopring wallets are deployed by factory contracts as generic proxies that delegate virtually all calls to a base wallet.
    As the proxies are deployed via \op{create2}, their addresses are known in advance.
    Hence Loopring wallets can be passively used before actually deploying the wallet code.
    The basic functionality of a wallet consists of the management of modules that can be added or removed, and of a table controlling which calls are forwarded to what module.
    Modules may call the wallet function \HD{transact} to perform an arbitrary transaction under the address of the wallet.
  \end{Description}
  \begin{Identification}
    The base wallets can be identified as the contracts, whose interface contains the function
    \begin{flushleft}
      \HD{transact(uint8,address,uint256,bytes)}
    \end{flushleft}
    The wallets are implemented as proxies delegating calls to the base wallets.
    Over time, three different types of proxies have been deployed.
    The codes of the first two are unique to the Loopring wallet, while the more recent deployments use a standardized proxy code.
    The proxies can be detected via the contracts deploying them.
    These proxy factories can be identified as the contracts, whose interface contains either
    \begin{flushleft}
      \HD{computeWalletAddress(address,uint256)}
    \end{flushleft}
    or
    \begin{flushleft}
      \HD{computeWalletAddress(address)}
    \end{flushleft}
    Modules are harder to detect in a uniform manner. A necessary, but not sufficient condition is that their interface has to contain the signatures
    \begin{flushleft}
      \HD{activate()}\\
      \HD{deactivate()}
    \end{flushleft}
    Modules in use can be detected by their address appearing in a call to the function \HD{addModule(address)} of base wallets. 
  \end{Identification}
  \begin{Addresses}
    The following table lists the base wallets in chronological order, giving their address, the creator, and the availability of source code.
    The base wallets do not share skeletons.
    \begin{center}
      \begin{tabular}{ccc}
        base wallet address & creator & src\\
        \midrule
 \A{55ef274c0286410184859490f59645011be8d779} & [1] &      \\
 \A{c054befa7401ef8df61c5fba56d9b5d4b9059a49} & [1] & \YES \\
 \A{f8c1f5848969bac54a5cc0178e0a36504b818db9} & [1] & \YES \\
 \A{433e04b573d7c0b105586970e70e6ea612e7c4ce} & [1] & \YES \\
 \A{a7c03d39082b54e8aac266fcf9a9b56d0892edff} & [1] & \YES \\
 \A{303baa149efc0b3b47136177f27637f2c491e457} & [1] & \YES \\
 \A{e5857440bbff64c98ceb70d650805e1e96adde7a} & [2] & \YES
      \end{tabular}
    \end{center}
    The following table lists the factory contracts in chronological order, giving the number of deployed proxy wallets, their type, the factory address, an identifier for the factory skeleton, the factory creator, and the availability of source code on Etherscan.
    \begin{center}
      \begin{tabular}{@{}rcc@{}c@{\ }c@{\ }c@{}}
        proxies & type & factory address & skeleton & creator & src\\
        \midrule
       0 & [a] & \A{29a27d44e129d0460ecd1e33a1fe0135f157860c} & A & [1] &      \\
       0 & [a] & \A{d081b3537ac710a537048f2db0913c1fe957802b} & A & [1] &      \\
       2 & [b] & \A{52eec07daf9176688695eda3481a52c8ff17b15d} & B & [1] & \YES \\
      15 & [b] & \A{23a732ace185a2b62efea22320875bb823e0d97b} & B & [1] & \YES \\
      44 & [b] & \A{d49da05aef8077cc6824e3b84493d6491f452b0a} & B & [1] & \YES \\
       0 & [c] & \A{dd867ac32d8fd56a2829135fc865543e2132e795} & C & [1] \\
       0 & [c] & \A{ec0e26ec06a2f78053c220f40838301cea80bd27} & C & [1] \\
       3 & [c] & \A{0c8dbfb40a324674ad26fe7aeccd4c654036aa82} & C & [1] & \YES \\
       0 & [c] & \A{bac254cc0146d1c11d97f1155e64692daa4d2c28} & D & [1] \\
       0 & [c] & \A{64aee7c9a23d6e4f4dd033205c23833e62a83b0e} & D & [1] & \YES \\
       0 & [c] & \A{1b71b5820c871ffc6733bb7859866df05827dc94} & E & [1] & \YES \\
       0 & [c] & \A{7eec46d5914fb345d163778fa5cc6fa989e22951} & E & [1] & \YES \\
     432 & [c] & \A{339703fb41df4049b02dfce624fa516fcfb31c46} & E & [1] & \YES \\
       0 & [d] & \A{08afa2375eae0398fb420dfc696fbcc35ac9e361} & F & [2] \\
       0 & [d] & \A{262f27480ccb98fa0b91d7a9f11bb82e3547ada1} & F & [2] & \YES \\
    3133 & [d] & \A{9fad9ffcea95c345d41055a63bd099e1a0576109} & F & [1] & \YES
      \end{tabular}
    \end{center}
    Code samples for the proxies of type b, c and d can be found at the addresses below, while their source code is available as part of the factory sources.
    Proxies of type~a have not been deployed so far.
    \begin{center}
      \begin{tabular}{cc}
        type & proxy deployed e.g.\ at \\
        \midrule
        {[b]} & \A{a534cb53a11e223c8595c30cc7f56c156d3dd890} \\
        {[c]} & \A{92b6e56acff470f377fe19fee5db6a9ac16a7ca8} \\
        {[d]} & \A{98435da87a12c2f2d7b068eec4e445a9cff9a75c}
      \end{tabular}
    \end{center}
    The following table lists, in chronological order, the modules that have been added to any of the wallets.
    \begin{center}
      \begin{longtable}{@{}cl@{}c@{\ }c@{}}
        module address & name & creator & src \\
        \midrule
        \endhead
        \multicolumn{4}{@{}l@{}}{{\ldots continued on next page}} \\
        \endfoot
        \endlastfoot
        \A{62de6ffc1d000503cffbb28402337515e7fa26e4} &                     &  [1] &      \\
        \A{f272176232d7598dd7b942adf91899c2b9534feb} & GuardianModule      &  [1] & \YES \\
        \A{e4b780dd298cdbc1af8206744382636f69aac1d9} & RecoveryModule      &  [1] & \YES \\
        \A{6de7b657e2f1fb3fc6433e37ab1ba937229e59dd} & LockModule          &  [1] & \YES \\
        \A{66d6877bff54288abe1bc94085d411833c6391db} & WhitelistModule     &  [1] & \YES \\
        \A{a0dfd5356adb9a2dcf6fdd004f5f38dbe61877ea} & QuotaModule         &  [1] & \YES \\
        \A{5ca59166b1d97be05ef54d25d69c873ced3cd414} & QuotaTransfers      &  [1] & \YES \\
        \A{63a3c58d843782adbe3398fbe0bd0062ba9bc066} & ApprovedTransfers   &  [1] & \YES \\
        \A{89b8c7319265fa102f9114dfc70114159973085e} & DappTransfers       &  [1] & \YES \\
        \A{8451569e9a9c7f8784e80241e3d224d6cc5d1e4a} &                     &  [1] &      \\
        \A{b94b9e42b0c9404e716d960f0cd3b6b7a826aeb6} &                     &  [1] &      \\
        \A{afbeb85219308c11eaca8885a9f421b86f257c63} &                     &  [1] &      \\
        \A{1c2e4096ad21bcf0c6fbea37e17c283a28399b8f} &                     &  [1] &      \\
        \A{39a0163e4193609977a9107a1dce1744a4a3e9d3} &                     &  [1] &      \\
        \A{90f0630d3b5dbeb498ff32ba4ef5e51f7475895a} &                     &  [1] &      \\
        \A{bdae72a749214b6318f1fefb97f38de1ec4137d0} &                     &  [1] &      \\
        \A{e7b1c5c0dec98bed1b1bcba5065d0031c2115e02} &                     &  [1] &      \\
        \A{e1b59bf1da5ade6b58ff780d8868d82ff70fd1e7} & ERC1271Module       &  [1] & \YES \\
        \A{d49da05aef8077cc6824e3b84493d6491f452b0a} & WalletFactoryModule &  [1] & \YES \\
        \A{01ddf96fd2130fb1a1e72b4f241ef402ca4b43bd} & GuardianModule      &  [1] & \YES \\
        \A{0535ccf8e3e6940b82a9d502701fc21c33d47e00} & WhitelistModule     &  [1] & \YES \\
        \A{cc4fa8d650bb137515272f12276e259537e53242} & QuotaTransfers      &  [1] & \YES \\
        \A{18ca1f3ca8f467696aefb9d0599116854f6e2e99} & ApprovedTransfers   &  [1] & \YES \\
        \A{dd6d0747d686fbef23e3567f48ba538132477711} & DappTransfers       &  [1] & \YES \\
        \A{6ae6c5bd9a41b978ee71f4b59bc0f6ad87837e03} & UpgraderModule      &  [1] & \YES \\
        \A{f9eaa41280469fb184f845dc48516a28bf5fb01f} &                     &  [1] &      \\
        \A{b70bdad4567e5529f5afc7df5ebbe7dbdf0550aa} &                     &  [1] &      \\
        \A{dc145e984a01266027ce9ebd481e1a4a7691a755} &                     &  [1] &      \\
        \A{53cc3eda2369ad61e1a2a9cf34d72e60f584d920} & GuardianModule      &  [1] & \YES \\
        \A{ef4a0a588c574f0561b95be80d130340b6974b56} &                     &  [1] &      \\
        \A{166df39721b2fd26d4907f91df74f6ae064a38ae} & FinalCoreModule     &  [1] & \YES \\
        \A{eed7818b55121d9901ae9e09aaf32d28952d7b14} & FinalSecurityModule &  [1] & \YES \\
        \A{62ed952ba331137629f144e0f61cd176cee5bcec} & FinalTransferModule &  [1] & \YES \\
        \A{23033e1babff3ab65e25f644a18ae6ebf601eb70} &                     &  [1] &      \\
        \A{21a357e5723608f24c2ea0c6d79d9f9073a12c85} &                     &  [1] &      \\
        \A{750f41b0abaf33fa79287a80efa714171feedde2} &                     &  [1] &      \\
        \A{af25b62b1aff8fe3fc72d0d7f3cf1cc6941eafb2} &                     &  [1] &      \\
        \A{c8a7a320b247d9353a98e93b8a3001bea16af683} &                     &  [1] &      \\
        \A{8207d121df7aee10d47fb5852ad50b7eab0bb216} &                     &  [1] &      \\
        \A{0947e1d932ee0806fde3e7a8ba855efe0b5bbf70} &                     &  [1] &      \\
        \A{6546c62a31771c307a205018dc5c87c6072b01de} &                     &  [1] &      \\
        \A{218559ef0bdeccdbb4fa79aaeb23ac7cae01b559} &                     &  [1] &      \\
        \A{2041507e1e860bef75a77508304cc37e021ddcc8} &                     &  [1] &      \\
        \A{e6b09c48e1893358557a42fcac328083c6f7cf52} &                     &  [1] &      \\
        \A{1aafc323b4711c6278105a653fac56abc58804a6} & FinalSecurityModule &  [1] & \YES \\
        \A{c475c56388b2ae3d82d71224c8e2eb433e484bae} & FinalTransferModule &  [1] & \YES \\
        \A{06eaabfe640e2c5202eeba9a4ad7be66050ddffc} &                     &  [1] &      \\
        \A{75ba2e145bf6b02dd013fbe9decacc3abe71737b} &                     &  [1] &      \\
        \A{e915058df18e7efe92af5c44df3f575fba061b64} & FinalCoreModule     &  [1] & \YES \\
        \A{b684849f3a7bd53fbad882302b5f7b9276c9b491} & FinalSecurityModule &  [1] & \YES \\
        \A{5693e9ef54f7b78ddef14997c1fbc51aa1d2fac9} & FinalTransferModule &  [1] & \YES \\
        \A{829ce855af88e3f6c6306de25c04fbc01394f680} & UpgraderModule      &  [1] & \YES \\
        \A{9bd708ba2e187ed4540310e01bfb8e347528b434} & UpgraderModule      &  [2] & \YES \\
      \end{longtable}
    \end{center}
    The base wallets, proxy factories and modules have been deployed by two user addresses.
    \begin{center}
      \begin{tabular}{cc}
        {[1]} & \A{fbbdec9bd33324b960195d9ff951377d41a35980}\\
        {[2]} & \A{4374d3d032b3c96785094ec9f384f07077792768}
      \end{tabular}
    \end{center}
  \end{Addresses}
\end{Wallet}

\subsection{Smart Wallets}\label{app:smart}
\begin{Wallet}{Argent Smart Wallet}
  \begin{Description}
    This is a modular wallet.
    The base wallet implements ownership management, generic transactions, and module authorization.
    All other functionality is outsourced to modules that can be added and removed flexibly.
    Authorized modules may invoke a function that performs generic transactions on behalf of the wallet.
    The base wallet is deployed in small numbers, but is the target of numerous proxy wallets that delegate all calls to the base wallet. 
  \end{Description}
  \begin{Author} Julien Niset
  \end{Author}
  \begin{Source}
    \url{https://github.com/argentlabs/argent-contracts/tree/develop/contracts/wallet/}
  \end{Source}
  \begin{Identification}
    Base wallets can be identified by the following two (of a total of 10) functions.
    \begin{flushleft}
      \HD{init(address,address[])}\\
      \HD{authoriseModule(address,bool)}
    \end{flushleft}
    The factory contracts that deploy the proxies can be identified by the
    function:
    \begin{flushleft}
      \HD{createWallet(address,address[],string)}
    \end{flushleft}
    The proxies of this wallet type are the contracts created by one of the factories.
  \end{Identification}
  \begin{Addresses}
    The following table lists the 11 bytecodes of the base wallet.
    The first column gives the number of deployments (42 in total), the second one the number of active wallets (only 12 were called so far), the address of one deployment where the code may be found, a column indicating which bytecodes share skeleton, a column identifying contracts with source code on Etherscan, and finally references to the external users that deployed the base wallets (see below for the addresses).
    \begin{center}
      \tabcolsep=2pt
      \begin{tabular}{rccccc}
        wallets & active & sample deployment & skel & src & creator \\
        \midrule
      13 &0&\A{cc28b303982b241b51b2b5356ddcbbf6a5e8a847} &   A&      & [1]  \\
      13 &0&\A{33b2ed2ae137c45d0c37438053ffa75bc2c71ff3} &   B&      & [2]  \\
       3 &2&\A{8ca6057ba18ed534e1c8de5330fb0841619a78e8} &   C&      & [3]  \\
       3 &2&\A{b6d64221451edbac7736d4c3da7fc827457dec03} &   D& \YES & [1,4]\\
       2 &1&\A{5588b1ad36bf5c0848831896ef5ef6d01c01b818} &   E&      & [1,4]\\
       2 &2&\A{b1dd690cc9af7bb1a906a9b5a94f94191cc553ce} &   F& \YES & [1,4]\\
       2 &2&\A{bc0b5970a01ba7c22104c85a4e2b05e01157638d} &   G& \YES & [1,4]\\
       1 &0&\A{377e2e723fd05b1bd79a14d25ba1a0f96d9d0a05} &   A&      & [1]  \\
       1 &1&\A{6850808054499dc9c3ef220a0305e72530914620} &   H&      & [5]  \\
       1 &1&\A{609282d2d8f9ba4bb87ac9c38de20ed5de86596b} &   C&      & [1]  \\
       1 &1&\A{811a7f70d12fbd29ec494edc75645e66f5fcccfc} &   I& \YES & [6]
      \end{tabular}
    \end{center}
    The table below lists the addresses of the proxy factories, together with the number of proxies they created, the type of proxy (see below), a column indicating which bytecodes share skeleton, a column identifying contracts with source code on Etherscan, and finally references to the external users that deployed the factories (see below for the addresses).
    \begin{center}
      \tabcolsep=2pt
      \begin{tabular}{rccccc}
      proxies & ptype & deployed at & skel & src & creator \\
        \midrule
   17.770 & [a] & \A{40c84310ef15b0c0e5c69d25138e0e16e8000fe9} & A & \YES & [4] \\
   17.202 & [b] & \A{851cc731ce1613ae4fd8ec7f61f4b350f9ce1020} & B & \YES & [4] \\
      823 & [c] & \A{2fe1f9bbc9ce8ea4e00f89fc1a8936de6934b63d} & C & \YES & [6] \\
      146 & [b] & \A{cf106b9644eb97deb5b78ab22da160ffca67a448} & B &      & [1] \\
      104 & [a] & \A{618ba96a418d24288a3b11d42600e2ff40cd6c59} & A &      & [1] \\
       29 & [d] & \A{3c7089d58b44ec8083c665503482900872230ed0} & D &      & [3] \\
       13 & [e] & \A{96114e2c7371bf51876c2c04cd5cd866db5e289a} & E &      & [1] \\
        8 & [d] & \A{0cd62a685307b9f11105c1005f560c661fe537bb} & D &      & [3] \\
        1 & [f] & \A{d8c6c2bd977d9533d616fd3dce5e3e99ce9cdf0b} & F &      & [5] \\
        0 &     & \A{46ef78645b59e9e77da150d2fdcf030ad41257aa} & G &      & [1] \\
        0 & [e] & \A{5a33a52ed9c82845d6aebb1c5a1283c12e070b32} & E &      & [4] \\
        0 & [a] & \A{95e5872048a5309e370e77248cbbffc1c3b106e3} & A &      & [4]
      \end{tabular}
    \end{center}
    The factories deploy 6 different proxy codes.
    The table below lists one address for each code.
    For the first three, source code can be found on Etherscan.
    \begin{center}
      \begin{tabular}{cccc}
              & sample deployment & skel & src \\
        \midrule
        {[a]} & \A{e1c7fe723752bada5075ca8ee5d53eb04b8910a6} & A & \YES \\ 
        {[b]} & \A{0364c42a15c2cc3073eba1e11ee5ab0c6a1b5b40} & B & \YES \\ 
        {[c]} & \A{7540bc4c7eab8507ab67c5c070e88e560793e746} & A & \YES \\ 
        {[d]} & \A{b40cd073018948b275efbf25d5261f4be7da7254} & A &      \\ 
        {[e]} & \A{1a40d7eaeb5f8084811f864f959b1dfed4ed6286} & C &      \\ 
        {[f]} & \A{38eef97ebf22cbc9936b858226f346f18238504c} & D &      \\ 
      \end{tabular}
    \end{center}
    The following table lists, in chronological order, the modules that have been added to any of the wallets (identified from the calls to the \HD{authoriseModule} function).
    \begin{center}
      \begin{longtable}{@{}cl@{}r@{\ }c@{}}
        module address & name & \llap{creator} & src \\
        \midrule
        \endhead
        \multicolumn{4}{@{}l@{}}{{\ldots continued on next page}} \\
        \endfoot
        \endlastfoot
 \A{96114e2c7371bf51876c2c04cd5cd866db5e289a} &                          & [1] &      \\
 \A{cf106b9644eb97deb5b78ab22da160ffca67a448} &                          & [1] &      \\
 \A{58e719254f1564ed29a86db7554c47fab778f3fe} &                          & [1] &      \\
 \A{9abb5db4b23a866ffd649716c6ce2674b2c28c17} &                          & [1] &      \\
 \A{76fe1ecb4a94f1b88e8b75de11445160a492ea5a} &                          & [1] &      \\
 \A{6c764fac2ed1c5fabf8bcd86bae68d8cdbe8290e} &                          & [1] &      \\
 \A{4556d522453633cfc6962cbde7cc4da840eb6707} &                          & [1] &      \\
 \A{69c90605f5a3224ac54f23bb7923462e0630603a} &                          & [1] &      \\
 \A{5cd15b0960de93b6ed9df6012163cb88bf45d7bb} &                          & [1] &      \\
 \A{642a28247b2b91cfb852b01c0e1f76dbf48b0f14} &                          & [1] &      \\
 \A{851cc731ce1613ae4fd8ec7f61f4b350f9ce1020} & WalletFactory            & [4] & \YES \\
 \A{4dd68a6c27359e5640fa6dcaf13631398c5613f1} & ModuleManager            & [4] & \YES \\
 \A{ff5a7299ff6f0fbaad9b38906b77d08c0fbdc9a7} & GuardianManager          & [4] & \YES \\
 \A{0bc693480d447ab97aff7aa215d1586f1868cb01} & LockManager              & [4] & \YES \\
 \A{dfa1468d07fc86840a6eb53e0e65cebde81d1af9} & RecoveryManager          & [4] & \YES \\
 \A{0045684552109f8551cc5c8aa7b1f52085adff47} & ApprovedTransfer         & [4] & \YES \\
 \A{df6767a7715381867738cf211290f61697ecd938} & TokenTransfer            & [4] & \YES \\
 \A{d5e1dff5f039b2c42978d98ed60c0ac5c8f6a266} & DappManager              & [4] & \YES \\
 \A{ed0da07aab7257df53efc4dfc076745744138ed9} & TokenExchanger           & [4] & \YES \\
 \A{c0e95b6d572ed7165cce2702203be05765cb8d19} &                          & [1] &      \\
 \A{2e278c93f95827c18437e4ea7324d705f7fcfeaa} &                          & [1] &      \\
 \A{711dab2333d815341ad1cb94d473f40e5749bb8f} &                          & [1] &      \\
 \A{1848e646bba45174f4044443719db6e5e6cf5d66} & NftTransfer              & [4] & \YES \\
 \A{bd75d3693301b6971b9b3294d853130b42dccca4} &                          & [1] &      \\
 \A{275a7d7c267793fa3df3f91027d83bdb193134a5} &                          & [1] &      \\
 \A{963f86da34cf2ce619d4b8e5ce96577943f95b6b} & MakerManager             & [4] & \YES \\
 \A{50d9174dd1e494823f4fadc4de2531aa81c1140a} &                          & [1] &      \\
 \A{2301d711ae627ba2f1cb76788c64b35eef869915} &                          & [1] &      \\
 \A{a5d7d68d7975e89feb240f42fed1d77bb71b1caf} & CompoundManager          & [4] & \YES \\
 \A{5388b0f8106bde37dc6982b4ba5771d2e8d9dc42} &                          & [4] &      \\
 \A{359b4287b69fb67096250c5f00e507b735f8e1bf} &                          & [1] &      \\
 \A{dc6a3775d618c143f7037c1693b50e8f36d290bd} &                          & [1] &      \\
 \A{945b093da7cc6f532f7d6f2841dce1cb693dc142} &                          & [1] &      \\
 \A{528cda0f6aa74da824659808c46d748c4a5a9403} &                          & [1] &      \\
 \A{d8c6c2bd977d9533d616fd3dce5e3e99ce9cdf0b} &                          & [5] &      \\
 \A{69922e74e7251e67bbeccac206591bf6b7d5424b} &                          & [1] &      \\
 \A{04d5a499ee1b436e6e942a1549c09aab03d5ee45} &                          & [1] &      \\
 \A{5a054b4d428cfe3a3c54197da32ebfa8c10e4bd4} &                          & [1] &      \\
 \A{5e63e80f59ea6cd7aef05721ca72f83ef7b04c42} &                          & [1] &      \\
 \A{28ea57c1b2d5ef040a5b9e2694f69fb43c35b67e} &                          & [1] &      \\
 \A{c2c40bc13df1239a0f5c0d783628debae7b0d8cb} &                          & [1] &      \\
 \A{fbeb18c80df7667175165614c8de55fda8da61ac} &                          & [1] &      \\
 \A{453bf02f901fb2b078ff6c963b648bc5c2eea006} &                          & [1] &      \\
 \A{1b394ca195f6f418f8d6dfcbaeb554bc5d03b704} &                          & [1] &      \\
 \A{c2c2d6c0173893ecbb5979387b1ad1e65ae786dd} &                          & [1] &      \\
 \A{2b6d87f12b106e1d3fa7137494751566329d1045} & TransferManager          & [4] & \YES \\
 \A{cd23f51912ea8fff38815f628277731c25c7fb02} & ApprovedTransfer         & [4] & \YES \\
 \A{7557f4199aa99e5396330bac3b7bdaa262cb1913} & MakerV2Manager           & [4] & \YES \\
 \A{0cd62a685307b9f11105c1005f560c661fe537bb} &                          & [3] &      \\
 \A{3c7089d58b44ec8083c665503482900872230ed0} &                          & [3] &      \\
 \A{2fe1f9bbc9ce8ea4e00f89fc1a8936de6934b63d} & WalletFactory            & [6] & \YES \\
 \A{618ba96a418d24288a3b11d42600e2ff40cd6c59} &                          & [1] &      \\
 \A{40c84310ef15b0c0e5c69d25138e0e16e8000fe9} & WalletFactory            & [4] & \YES \\
 \A{eccc82ea51e23185affb8c00598d644914184844} &                          & [1] &      \\
 \A{c3fc9249f52e6592e2f4be1366d286e5eed1222a} &                          & [1] &      \\
 \A{5cdff52846831ad32fe1e9e690d310deca11c7dc} &                          & [1] &      \\
 \A{eea3f122f8ea035eb3783fb4f56a4e7cc84bf410} &                          & [1] &      \\
 \A{275446211af16cfc071c296c0cf27b04180d21b6} &                          & [1] &      \\
 \A{90dfe5050db825eb818ddb9dcdbe43238173ec60} &                          & [1] &      \\
 \A{201041f7486abb71b0b363eff78aeeeff26e7ed6} &                          & [1] &      \\
 \A{f1d3bed1445193d05083581d7f2533b0ff8d918c} &                          & [1] &      \\
 \A{e39cf675defc983d7e6fb4848e15560e3065d702} &                          & [1] &      \\
 \A{de6c0867658c92808eaf53ba5b766b50994fb3ae} &                          & [1] &      \\
 \A{c927a1e4c431babb798d705988ab036e2bdcf640} &                          & [1] &      \\
 \A{6cd7edbb0de4979e667c9d28540ebe96015fb621} &                          & [1] &      \\
 \A{2070538c8efdd3c003170f6fc6d26b4a58ab599e} &                          & [1] &      \\
 \A{af1c9fc0588f3caf654c796b3ce83d74f477419c} &                          & [1] &      \\
 \A{c1395f0b1fe125369a955145b1775b9ed585c889} &                          & [1] &      \\
 \A{67933cccf3f6b40148271f5e7b5408fd1d5b2837} & ApprovedTransfer         & [4] & \YES \\
 \A{fcfab7cdc0613fea78982ebd52559b7362db7976} & RecoveryManager          & [4] & \YES \\
 \A{8a7da5fe29f33f5c053d46cb79086cdcc24ac610} & MakerV2Manager           & [4] & \YES \\
 \A{103675510a219bd84ce91d1bcb82ca194d665a09} & TransferManager          & [4] & \YES \\
 \A{a24783c07cdd995c56c1d33ce485e6ac39e0c018} & SimpleUpgrader           & [4] & \YES \\
 \A{94b447babc7e4c445ed241ca8e52c547275fd27e} &                          & [1] &      \\
 \A{80ef91092bf35b886ae7cfe0011913ea10add5f6} &                          & [1] &      \\
 \A{4b0e6ccb42471d9f044fbad5acf5cdaed687d116} &                          & [1] &      \\
 \A{dc350d09f71c48c5d22fbe2741e4d6a03970e192} & RecoveryManager          & [4] & \YES \\
 \A{4a8c4b77221ca1e82a61ce91ca603e4c9cb7181f} & SimpleUpgrader           & [4] & \YES \\
 \A{e305ed426ad28a583e3e5f9ce10027729edf039e} &                          & [4] &      \\
 \A{0d1ca6ec20b19cf30d7e32df62ff166a47ac5ecd} & WalletFactory            & [1] & \YES \\
 \A{3de91553d67f7003ed9e4a1f073416c5d6a58d6a} &                          & [1] &      \\
 \A{ebafd3eaba87a40024dc4a656b4b6f91c40bea98} &                          & [1] &      \\
 \A{9ae0acdb750bfcf694675f46b580847fc49a48bf} & WalletFactory            & [4] & \YES \\
 \A{4b3fbe6d554c540c2672eb7a501018a1a39f7f53} & VersionManager           & [4] & \YES \\
 \A{fa2abb20bc59d5b57cec3ff4348743ceeef5369f} &                          & [4] &      \\
 \A{3a938f6dff602874ea67ab87fcddb892b3bf897a} &                          & [4] &      \\
 \A{e4dfc679f21e47609238c7f92d9416d5b726005f} & UpgraderToVersionManager & [4] & \YES \\
 \A{3fe5125e4b28062ea5be32acedb37bffaa256456} &                          & [4] &      \\
 \A{4b22f66f26a3d961511def26ef267b4ce41a5b5d} &                          & [1] &      \\
 \A{0971700a1b520d4ab6571e65a0fe375e95a551a0} &                          & [1] &      \\
 \A{b73e133283792d6cadbb0a8cfb57481a30bf8555} &                          & [1] &      \\
 \A{cd9ad3b54581640db1d2452bc69af9c83f6e0a95} &                          & [1] &      \\
 \A{dec1c7a7b5930871ddff751a0e07247a67f4a707} & VersionManager           & [1] & \YES \\
 \A{58a2e91f01efeccf8f61b356d1d7a2cde61e49a6} &                          & [1] &      \\
 \A{155124ddeb6b87b15f188ed8d3bc14375b3c6372} &                          & [1] &      \\
 \A{22ef27955fd2e49e25fd034e4b847ab6d870f770} &                          & [1] &      \\
 \A{3c95bb792fe4c28b7dc1e3eb5bcf4edae6df1936} &                          & [1] &      \\
 \A{645ba45dbe3c6942c812a46f9ee8115c89b524ec} &                          & [4] &      \\
 \A{1b05623800526da9c03de7583917344d5d30e139} &                          & [4] &      \\
 \A{e7f618212cf739d5f209453a2a91bfc198b065e0} &                          & [4] &      \\
 \A{908842ed4b885318711dcdb16d0503d6a000251e} &                          & [4] &      \\
 \A{bcdf8dba8760fd1a94b12a8233385e9c98d1e399} &                          & [4] &        
      \end{longtable}
    \end{center}    
    The user accounts mentioned above that created the base wallets, modules and factories:
    \begin{center}
      \begin{tabular}{cc}
        {[1]} & \A{c66efbf0e29c70f76bad91c454f7d4d289c7222b} \\
        {[2]} & \A{b61c85f9ce87c89c012e06fcdb501ed229a3776f} \\
        {[3]} & \A{5f7a02a42bf621da3211ace9c120a47aa5229fba} \\
        {[4]} & \A{46cf7ddb8bc751f666f691a4f96aa45e88d55d11} \\
        {[5]} & \A{c0a9f98dbca1d1007e3809f3b205161b6d272384} \\
        {[6]} & \A{e02fa196497a6994d9ce0ffaffa2d1293f43b598}   
      \end{tabular}
    \end{center}
  \end{Addresses}
\end{Wallet}

\begin{Wallet}{Dapper Smart Wallet}
  \begin{Description}
This wallet is suited for \ETH\ , ERC-20 and advanced tokens and has cosigner functionality, flexible transactions, and a recovery mechanism.
The multiSig functionality employs signed messages based on ERC-191 for confirmation support.
  \end{Description}
  \begin{Source}
    \url{https://github.com/dapperlabs/dapper-contracts}
  \end{Source}
  \begin{Identification}
The wallet is characterized by the following function:
\begin{verbatim}
invoke0(bytes)
\end{verbatim}
Its factory is identified by:
\begin{verbatim}
deployFullWallet(address,address,uint256)
\end{verbatim}
  \end{Identification}
  \begin{Addresses}
  The wallet is deployed 46\,474 times, mostly by factories.
  We find 9 bytecodes, seven skeletons, and 10 creators.
   \begin{center}
    \begin{tabular}{rcc}
    wallets & deployed e.g.\ at & creator \\ \midrule
45\,960	& \A{c917434e55463a8ed222c20a90be82841cdf9143} & A, B, C, D, E \\
352	& \A{65c7e22422e9a1ebcc37bbcb694014d6776267ae}	& F, B \\
157	& \A{bb1a73afc9a9856624a232e675605aaca08f7f3d}	& G, H \\
2	& \A{989a2ad9acaa8c4e50b2fc6b650d6e1809b9195b}	& G , I \\
1	& \A{1d06375a7c74c47d21183fbf330f752ab3657ead}	& G \\
1	& \A{cbc2d66110784f7131b7ff583838f12ab96bab4c}	& G \\
1	& \A{ed69ac1cab88cc82ff417131bdc69d93427107b4}	& J \\
    \end{tabular}
    
    \begin{tabular}{cc}
    creator & address \\ \midrule
    A & \A{839245f153c27efb75c28762c5daf676a4fc205f} \\
    B & \A{a78bbf97033e534c54b0a4fa62aa77b652ae4097} \\
    C & \A{ab76c6d00c603a7615d5459132c1745eb1fb4f6c} \\
    D & \A{bd4ae80f258ba3e75dd9894d0d697a3e330b9483} \\
    E & \A{c889e895e73771e0e0386623fec2ba8f7721327d} \\
    F & \A{9ef182a63b3109b77dc4026c5f8605dc8730fadd} \\
    G & \A{6b9ca9f0fbea46ecf0efd71262ae576fc523cb3b} \\
    H & \A{7820c32b9b74929f1f12d508253ff66444c61571} \\
    I & \A{f078305e5faecbe2ba0045197798ef94e0cd5225} \\
    J & \A{4585b1986089c75e338498575125a70c7c92141c} \\
    \end{tabular}
   \end{center}
  \end{Addresses}
\end{Wallet}

\begin{Wallet}{Gnosis Smart Wallet}
  \begin{Description}
This is a modular multiSig wallet with flexible transactions for \ETH\ as well as ERC-20 and advanced tokens. It has a daily limit and recovery mechanism.
  \end{Description}
  \begin{Author}
  Stefan George, Richard Meissner, Ricardo Guilherme Schmidt
  \end{Author}
  \begin{Source}
    \url{https://github.com/gnosis/safe-contracts/blob/development/contracts/GnosisSafe.sol}
  \end{Source}
  \begin{Identification}
The wallet is characterized by the following function:
\begin{verbatim}
approvedHashes(address,bytes32)
\end{verbatim}
  \end{Identification}
  \begin{Addresses} 
The wallet is deployed 12\,078 times.
We find 35 bytecodes, 22 skeletons, and 712 creators.
In the follwoing table, we list for each skeleton with more than 10 deployments the number of deployed wallets and an exemplary address.
   \begin{center}
    \begin{tabular}{rc}
wallets & deployed e.g.\ at \\ \midrule
10\,118	& \A{9ff80c034f165bb25c041080634845e064d5599a}  \\
1\,178	& \A{7f2722741f997c63133e656a70ae5ae0614ad7f5}  \\
673	& \A{567725581c7518d86c7d163dd579b2c4258337d0} \\
64	& \A{059af59a40f06cccc8332203f3e01d8f828eed7d} \\
15	& \A{fa29d3b68c5d99e44f363581b0ebaa023a7a8432} \\
    \end{tabular}
   \end{center}
Regarding creators, we list the ones with over 10 deployments in the table below.
   \begin{center}
    \begin{tabular}{rcc}
wallets & creator & note \\ \midrule
9\,480	& \A{76e2cfc1f5fa8f6a5b3fc4c8f4788f0116861f9b} & ProxyFactory 1.1.1\\
1\,172	& \A{12302fe9c02ff50939baaaaf415fc226c078613c} & ProxyFactory 1.0.0 \\
631	& \A{0fb4340432e56c014fa96286de17222822a9281b} & CPKFactory \\
51	& \A{1eda606967a97522b432d39d19a2bf4daf2229de} & CPKFactoryCustom\\
13	& \A{02a97f7ff6739086025d2f40190ebceb18750da0} \\
12	& \A{88cd603a5dc47857d02865bbc7941b588c533263} & ProxyFactory 0.1.0\\
    \end{tabular}
   \end{center}

  \end{Addresses}
\end{Wallet}

\end{document}